\documentclass[iop]{emulateapj}
\newcommand{\cha}{\textit{Chandra\/}}
\def\xmm{{XMM-{\it Newton\/}}}

\def\swi{{{\it Swift}-BAT}\/}
\def\xrt{{{\it Swift}-XRT}\/}

\def\flu{{ erg s$^{-1}$} cm$^{-2}$}

\def\nustar{{\it NuSTAR}}

\shorttitle{Compton-thick AGN in the \nustar\ era}
\shortauthors{Marchesi et al.}

\usepackage{natbib}
\usepackage{float}
\usepackage{color}
\usepackage{graphicx}
\usepackage{gensymb}
\usepackage{array}
\usepackage{enumitem}
\usepackage{longtable}
\usepackage{hyperref}

\begin{document}


\title{Compton-thick AGN in the \nustar\ era}

\author{S. Marchesi\altaffilmark{1}, M. Ajello\altaffilmark{1}, L. Marcotulli\altaffilmark{1}, A. Comastri\altaffilmark{2}, G. Lanzuisi\altaffilmark{2,3}, C. Vignali\altaffilmark{3,2}} %G. Cusumano\altaffilmark{3}, V. La Parola\altaffilmark{3}, A. Segreto\altaffilmark{3}}
%\affil{Astronomy Department, University of California}å
%\today

\altaffiltext{1}{Department of Physics \& Astronomy, Clemson University, Clemson, SC 29634, USA}
\altaffiltext{2}{INAF--Osservatorio Astronomico di Bologna, Via Piero Gobetti, 93/3, 40129, Bologna, Italy}
\altaffiltext{3}{Dipartimento di Fisica e Astronomia, Alma Mater Studiorum, Universit\`a di Bologna, Via Piero Gobetti, 93/2, 40129, Bologna, Italy}
%\altaffiltext{4}{INAF - Istituto di Astrofisica Spaziale e Fisica Cosmica, Via U. La Malfa 153, I-90146 Palermo, Italy}

\begin{abstract}
We present the 2--100\,keV spectral analysis of 30 candidate Compton thick (CT-) active galactic nuclei (AGN) selected in the \swi\ 100-month survey. The average redshift of these objects is $\langle z\rangle\sim$0.03 and they all lie within $\sim$500\,Mpc. 
We used the \texttt{MyTorus} \citep{murphy09} model to  perform X-ray spectral fitting both without and with the contribution of the \nustar\ data in the 3--50\,keV energy range. When the \nustar\ data are added to the fit, 14 out of 30 of these objects (47\% of the whole sample) have intrinsic absorption N$_{\rm H}<10^{24}$\,cm$^{-2}$ at the $>$3$\sigma$ confidence level, i.e., they are re-classified from Compton thick to Compton thin. Consequently, we infer an overall observed fraction of CT-AGN with respect to the whole AGN population lower than the one reported in previous works, and as low as $\sim$4\%. We find evidence that this over-estimation of N$_{\rm H}$ is likely due  to the low quality of a subsample of spectra, either in the 2-10 keV band or in the \swi\ one.
%Furthermore, the high quality of the \nustar\ data at $>$10\,keV allows us to get a direct insight on the geometry of the obscuring material surrounding the accreting supermassive black hole (SMBH). Our measurements show a preference for small covering factors ($f_c$) in heavily obscured sources, the average covering factor for the 13 sources for which we have a significant $f_c$ measurement being $\langle f_c\rangle$=0.30$_{-0.20}^{+0.18}$. We also find evidence of an anti-correlation between $f_c$ and the AGN 2-10\,keV luminosity, suggesting that more luminous sources are capable to blow away a fraction of the obscuring material surrounding the SMBH.
\end{abstract}


\section{Introduction}
According to the different models of Cosmic X-ray Background (CXB), the diffuse X-ray emission observed in the  1 to $\sim$200--300\,keV band, is mainly caused by accreting supermassive black holes (SMBH), the so-called active Galactic Nuclei \citep[AGN; e.g.,][]{alexander03,gilli07,treister09}.
Particularly, at the peak of the CXB \citep[$\sim$30\,keV,][]{ajello08a} a significant fraction of emission (10--25\%) is expected to be produced by a numerous population of heavily obscured, Compton thick (CT-) AGN \citep[e.g.,][]{risaliti99}, having intrinsic column density N$_{\rm H}\geq10^{24}$\,cm$^{-2}$. Nonetheless, in the nearby Universe ($z\leq$0.1) the observed fraction of CT-AGN with respect to the total population appears to be lower than the one expected on the basis of the majority of CXB model predictions \citep[$\sim$20--30\%; see, e.g., ][and references therein]{ueda14}, being between 5 and 10\% \citep{comastri04,dellaceca08,vasudevan13,ricci15}, although observational biases against detecting CT-AGN can at least partially explain this discrepancy \citep[see, e.g.,][]{burlon11}.

Entering the Compton thick regime, the fraction of emission directly produced by the AGN at energies $\geq10$\,keV significantly decreases \citep[][]{burlon11,ricci15}, while at lower energies only the emission scattered, rather than absorbed, by the obscuring material is detectable \citep[see, e.g.,][]{matt99,yaqoob10,koss16}. As a consequence, the detection and characterization of the CT-AGN population in the nearby Universe is possible only using instruments that can map the $>$10\,keV band with deep observations. The wide-field (120$\times$90 deg$^2$) Burst Alert Telescope \citep[BAT; ][]{barthelmy05}, one of the instruments mounted on the \textit{Swift} satellite \citep{gehrels04}, partially fulfills these requirements. \swi\ is an all-sky instrument which continuously scans and images the whole sky in the 15-150\,keV band. Combining good sensitivity and all-sky coverage, \swi\ is a strategical instrument to create a census of the hard X-ray, low luminosity AGN in the nearby Universe. Several works based on BAT--selected objects and on the joint analysis of the BAT spectra with the spectra collected with different 0.3--10\,keV instruments such as \xmm, \cha, \xrt\ and \textit{Suzaku}, have in fact been able to discover several tens of new candidate CT-AGN  \citep[see, e.g., ][]{burlon11,vasudevan13,ricci15,marchesi17,marchesi17b}.

Obscuration in AGN is commonly explained with the presence of a so-called ``dusty torus'', i.e., gas and dust distributed around the SMBH and in proximity to the accretion disk. The actual shape and composition of this material is however an open topic, although several works suggest that a clumpy distribution of optically thick clouds may be preferred to a more homogeneous structure \citep[e.g.,][]{jaffe04,elitzur06,risaliti07,honig07,nenkova08,burtscher13}. In the last years, several tori models, based on Monte Carlo simulations, have been developed to properly treat the complex X-ray spectra of Compton thick AGN \citep[e.g.,][]{ikeda09,murphy09,yaqoob10,brightman11,yaqoob12,liu14,furui16}. Each of these models is based on different assumptions on the obscuring material geometry and chemical composition, while all the models assume a homogeneous distribution of obscuring material. Moreover, both \citet{ikeda09} and \citet{brightman11} allow to measure the torus half-opening angle, which is directly related to the torus covering factor, $f_c$. It has been shown that the CT-AGN population may have a wide variety of covering factors, since there are both objects for which the obscuring material has been measured to be spherically distributed around the SMBH ($f_c$=1) as well as sources with a geometrically thin ($f_c\sim0.1$) torus \citep[see, e.g.,][]{brightman15}.
A proper use of these different models, however, requires excellent spectral statistics in the 2--50\,keV band, which can be provided neither by one of the several 0.3--10\,keV facilities nor by BAT.

The launch of the Nuclear Spectroscopic Telescope Array \citep[hereafter \nustar, ][]{harrison13}, the first telescope with focusing optics at $>$10 keV, represented a major breakthrough in the actual characterization of obscured AGN, providing an improvement on sensitivity of about two orders of magnitude with respect to previous facilities at these energies. Consequently, several works have already been published on heavily obscured AGN as seen by \nustar\ \citep[e.g.,][]{balokovic14,puccetti14,annuar15,bauer15,brightman15,koss15,rivers15,masini16,puccetti16}. Nonetheless, the majority of these works focused on single or few sources, and the largest sample of heavily obscured AGN analyzed with \nustar\ contains only 11 objects \citep{masini16}, which are radio-selected megamasers, i.e., sources well known to host a large fraction of CT-AGN \citep{greenhill08}. Therefore, a systematic analysis of the role of \nustar\ in the characterization of heavily obscured AGN is so far absent in the literature, or limited to small samples of objects. To fill this gap, in this work we present the analysis of the 30 candidate CT-AGN in the BAT 100-month catalog for which an archival \nustar\ observation exists. Notably, for 17 out of 30 sources this is the first time the \nustar\ data analysis is published.

This work is organized as follows: in Section \ref{sec:sample} we present the sample of 30 candidate CT-AGN with available \nustar\ observations and we describe the data reduction and spectral extraction process for both \nustar\ and the 0.3--10\,keV observations. In Section \ref{sec:model} we describe the model used to perform the spectral fitting. In Section \ref{sec:results} we present the main results of the analysis, with a particular focus on peculiar sources, while in Section \ref{sec:nus_role} we highlight the fundamental role played by \nustar\ in characterizing CT-AGN, analyzing the differences in the spectral fit results obtained with and without the addition of the \nustar\ data. 
%In Section \ref{sec:flux_ratio} we discuss how the \nustar\ data can help us in characterizing the shape of the obscuring material surrounding the SMBH, both indirectly, using the observed to intrinsic flux ratio in the 3--75\,keV and in the 15--55\,keV bands, and directly. 
In Section \ref{sec:sc} we test the Spectral Curvature method \citep{koss16} developed to select candidate CT-AGN. %In Section \ref{sec:combine_spec} we show the combined 3--80\,keV spectrum of the CT-AGN population analyzed in this work, and its relation with the CXB shape.  
Finally, we report our conclusions in Section \ref{sec:concl}. All reported errors are at a 90\% confidence level, if not otherwise stated. The errors have been obtained with the XSPEC error command.

\section{Sample selection and data reduction}\label{sec:sample}
In this work, we take advantage of the most recent catalog developed using the BAT survey data, i.e., the Palermo BAT 100-month catalog\footnote{\url{http://bat.ifc.inaf.it/100m\_bat\_catalog/100m\_bat\_catalog\_v0.0.htm}}, which reaches a flux limit $f\sim$3.3 $\times$ 10$^{-12}$\,\flu\ in the 15-150 keV band. The public data used in this work have been downloaded from the HEASARC public archive and processed using the BAT\_IMAGER code \citep{segreto10}. With BAT\_IMAGER it is possible to analyze data obtained using coded mask instruments: the software screens and combines all the available observations and performs the source detection process. In our analysis we use background subtracted, exposure-averaged spectra; the spectral redistribution matrix we use is the official BAT one\footnote{\url{http://heasarc.gsfc.nasa.gov/docs/heasarc/caldb/data/swift/\\bat/index.html}}. %We point out that the 100-month BAT catalog is yet to be published (Segreto et al. in prep.), but all the data used in this work have been fully reprocessed.

The Palermo \swi\ 100-month catalog contains 911 AGN. Based on previous results from the literature and an independent spectral analysis lead by our group (Kadan et al. 2017 in prep.), 50 of these objects are candidate CT-AGN, i.e., their best-fit intrinsic absorption value is $N_{\rm H, z}\geq10^{24}$\,cm$^{-2}$. Out of these 50 sources, 30 have archival \nustar\ data available as of December 1, 2017. %\footnote{Another source, 2MASXJ09235371--3141305, has an archival \nustar\ observation available (obsid: 60061339002), but we have not been able to reduce the data. Therefore, we exclude this object from our sample.}. 
These 30 objects are reported in Table \ref{tab:sample}, where the information on the work that first classified the object as Compton thick is also provided.

\begin{figure}%[!t]
  \centering
  \includegraphics[width=1.\linewidth]{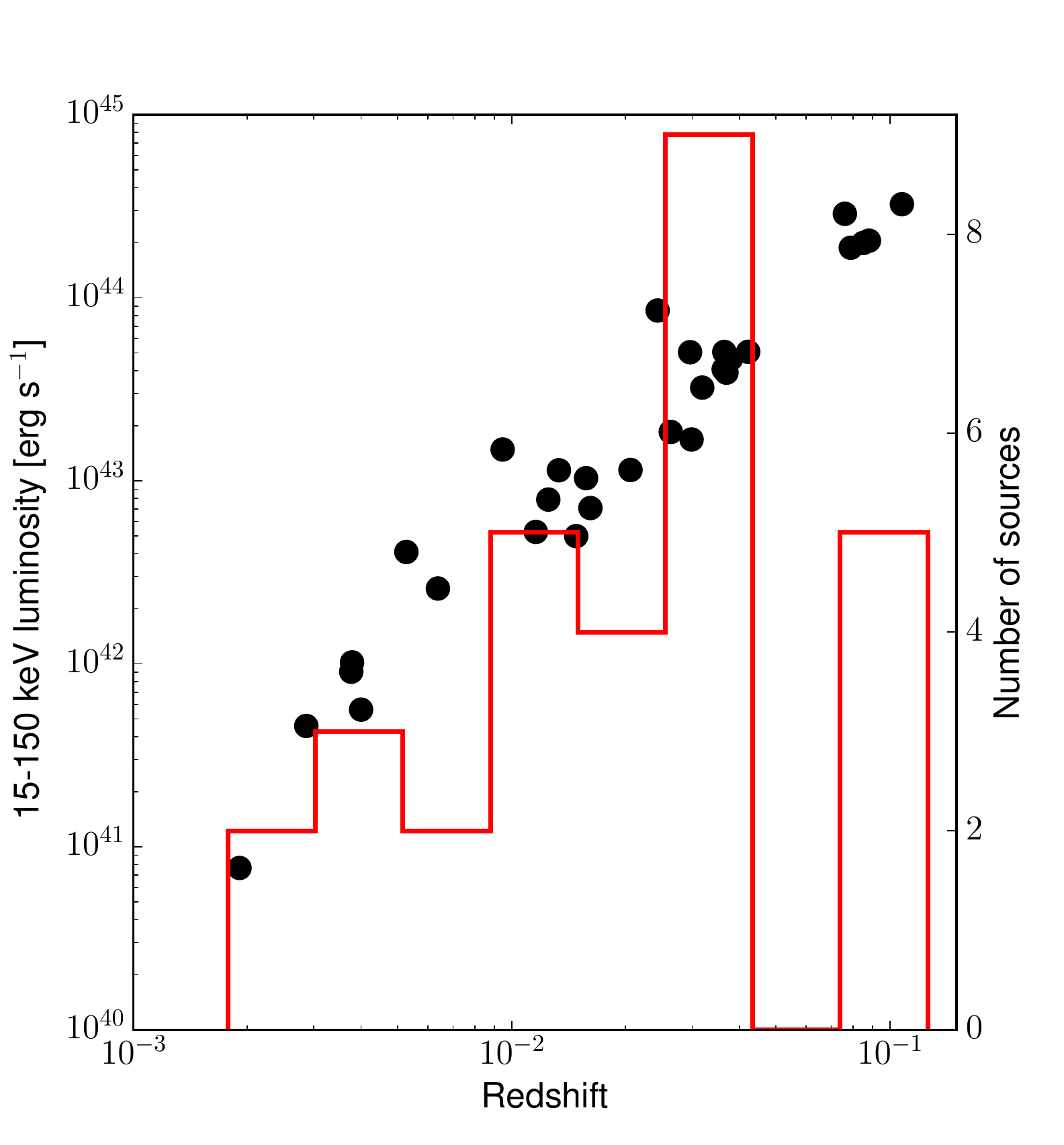}
\caption{\normalsize 15--150\,keV luminosity, measured with \swi, as a function of $z$ for the 30 candidate CT-AGN studied in this work. The redshift histogram is also shown (red solid line) All objects have $z$$\leq$0.108 (d$_L\leq$500\,Mpc).}\label{fig:z_vs_lx}
\end{figure}

In Figure \ref{fig:z_vs_lx} we show the distribution of the 30 objects in the 15--150\,keV luminosity versus redshift ($z$) plane. As can be seen, since we apply a \swi\ selection and the Palermo BAT 100-month catalog samples the bright, nearby AGN population, we are studying the low-$z$ CT-AGN population. More in detail, the average redshift of our sample is $\langle z\rangle$=0.03 (corresponding to an average luminosity distance $\langle d_L\rangle$=135\,Mpc) and the farthest object has redshift $z$=0.108 (d$_L$=500\,Mpc). 15 (50\% of the sample) and 25 (83\%) of the sources are located at distances $<$100\,Mpc and $<$200\,Mpc, respectively. The luminosity distances are computed assuming a cosmology with H$_0$ = 69.6 km s$^{-1}$ Mpc$^{-1}$, $\Omega_M$ = 0.29, and $\Omega_{\Lambda}$= 0.71.

The data retrieved for both  \nustar\  Focal Plane Modules \citep[FPMA and FPMB;][]{harrison13} were processed using the  \nustar\  Data Analysis Software (NUSTARDAS) v1.5.1. 
The event data files were calibrated running the {\tt nupipeline} task using the response file from the Calibration Database (CALDB) v. 20100101. With the {\tt nuproducts} script we generated both the source and background spectra, and the ancillary and response matrix files. 
For both focal planes, we used a circular source extraction region with a 30$^{\prime\prime}$ diameter centered on the target source; for the background we used the same extraction region positioned far from any source contamination in the same frame. The \nustar\ spectra have then been grouped with at least 15 counts per bin. 

\begingroup
\renewcommand*{\arraystretch}{1.15}
\begin{table*}
\centering
\scalebox{0.83}{
\begin{tabular}{lccccccccccc}
\hline
\hline
  \multicolumn{1}{c}{4PBC name} & \multicolumn{1}{c}{Source name} &   \multicolumn{1}{c}{R.A.} & \multicolumn{1}{c}{Decl} &  \multicolumn{1}{c}{Type} & \multicolumn{1}{c}{$z$} & \multicolumn{1}{c}{Telescope} & \multicolumn{1}{c}{ObsID} & \multicolumn{1}{c}{Date} &  \multicolumn{1}{c}{Exposure} &  \multicolumn{1}{c}{Rate} &  \multicolumn{1}{c}{Ref.}\\
  & & deg & deg & & & & & & ks & cts s$^{-1}$\\
  \multicolumn{1}{c}{(1)} & (2) & (3) & (4) & (5) & (6) & (7) & (8) & (9) & (10) & (11) & (12)\\
\hline
  J0111.5--3804 & NGC 424 & 17.86511 & --38.08347 & 1.9 & 0.0118 & \xmm & 550950101 & 2008--12--07 & 309.3 & 0.023 & (a)\\
  ...                              & ...             &  ...              &      ...   & ...    &      ...       & \nustar & 60061007002 & 2013--01--26 & 31.0 & 0.035 & (b)*\\
  J0122.5+5004 & MCG+08-03-018 & 20.64346 & 50.05500 & 2 & 0.0204 & \xrt & 38011, 80019 & 2008--11--19 & 5.1 & 0.011 & (c)*\\
    ...                              & ...             &  ...              &      ...   & ...    &      ...       & \nustar & 60061010002 & 2014--01--27 & 63.3 & 0.031 & --\\
  J0242.6+0000 & NGC 1068 & 40.66963 & --0.01328 & 2 & 0.0038 & \xmm & 740060401 & 2014--08--19 & 138.4 & 0.046 & (a)\\
   ...                              & ...             &  ...              &      ...   & ...    &      ...       & \nustar & 60002030002 & 2012--12--18 & 115.7 & 0.107 & (d)*\\
  J0303.8--0106 & NGC 1194 & 45.954621 & --1.10374 & 1.9 & 0.0136 & \xmm & 307000701 & 2006--02--19 & 43.6  & 0.019 & (e)\\
   ...                              & ...             &  ...        &       ...   & ...     &      ...       & \nustar & 60061035002 & 2015--02--28 & 63.0 & 0.043 & (f)*\\
  J0308.1--2256 & NGC 1229 & 47.04494 & --22.96080 & 2 & 0.0363 & \xrt & 41743, 80534 & 2010--10--19 & 15.5 & 0.002 & (c)*\\
   ...                              & ...             &  ...      & ...        &    ...   &      ...       & \nustar & 60061325002 & 2013--07--05 & 49.8 & 0.024 & --\\
  J0350.5--5019 & ESO 201-IG 004 & 57.59567 & --50.30261 & 2 & 0.0359 & \xmm & 501210401 & 2007--07--15 & 40.8 & 0.012 & (c)*\\
  ...                              & ...             &  ...         &      ...    & ...   &      ...       & \nustar & 60061331002 & 2014--08--09 & 46.3 & 0.015 & --\\
  J0356.2--6251 & 2MASXJ03561995--6251391 & 59.08321 & -62.86076 & 1.9 & 0.1076 & \xrt & 37304, 81865 & 2008-08-05 & 12.7 & 0.003 & (c)*\\
   ...                              & ...             &  ...     & ...         &      ...            &      ...       & \nustar & 60201034002 & 2016--05--06 & 53.1 & 0.035 & --\\
  J0453.3+0403 & CGCG 420-15 & 73.35729 & 4.06158 & 2 & 0.0294 & \xmm & 307000401 & 2005--08--30 & 29.8 & 0.025 & (g)*\\
  ...                              & ...             &  ...      & ...        &      ...            &      ...       & \nustar & 60061053004 & 2014--08--13 & 36.6 & 0.062 & --\\
  J0605.5--8638 & ESO 005-G 004 & 91.42346 & --86.63186 & 2 &  0.0062 & \xrt & 35254, 80367 & 2005--12--14 & 24.1 & 0.003 & (h)\\
  ...                              & ...             &  ...      & ...        &      ...            &      ...       & \nustar & 60061063002 & 2015--11--10 & 49.4 & 0.018 & --\\
  J0714.0+3518 & MCG+06-16-028 & 108.51608 & 35.27928 & 1.9 &  0.0157 & \xrt & 40931, 80381 & 2010--05--12 & 5.0 & 0.003 & (c)*\\
  ...                              & ...             &  ...     & ...         &      ...            &      ...       & \nustar & 60061072002 & 2013--12--03 & 47.1 & 0.024 & (i)\\
  J0924.0--3141 & 2MASXJ09235371--3141305 & 140.97388 & --31.69186 & 2 & 0.0424 & \xrt & 33661, 80674, 91688 & 2013--04--07 & 12.3 & 0.005 & (c)*\\
 ...                              & ...             &  ...      & ...        &      ...            &      ...       & \nustar & 60061339002 & 2014--04--19 & 42.5 & 0.048 & --\\
  J1001.9+5540 & NGC 3079 & 150.49085 & 55.67979 & 1.9 & 0.0037 & \xmm & 110930201 & 2001--04--13 & 36.8 & 0.010 & (j)\\
  ...                              & ...             &  ...      & ...        &      ...            &      ...       & \nustar & 60061072002 & 2013--11--12 & 43.0 & 0.034 & (k)*\\
  J1048.3--2509 & NGC 3393 & 162.09775 & --25.16206 & 2 & 0.0125 & \xmm & 140950601 & 2003--07--05 & 36.0 & 0.007 & (l)\\
  ...                              & ...             &  ...       & ...       &      ...            &      ...       & \nustar & 60061205002 & 2013--01--28 & 31.4 & 0.033 & (m)*\\
  J1052.6+1036 & 2MASXJ10523297+1036205 & 163.13745 & 10.60558 & 1 & 0.0878 & \xmm & 693430401 & 2012--11--21 & 75.1 & 0.051 & (n)\\
  ...                              & ...             &  ...       & ...       &      ...            &      ...       & \nustar & 60160414002 & 2017--01--30 & 81.3 & 0.065 & --\\
  J1149.1--0416 & RBS 1037 & 177.32783 & --4.28083 & 1 & 0.0845 & \xrt & 38057, 80061 & 2010--11--08 & 21.1 & 0.032 & (n)\\
  ...                              & ...             &  ...        & ...      &      ...            &      ...       & \nustar & 60061215002 & 2017--02--02 & 81.3 & 0.049 & --\\
  J1206.3+5243 & NGC 4102 & 181.59580 & 52.71108 & 2 & 0.0028 & \xmm & 601780701 & 2009--10--30 & 26.5 & 0.013 & (o)\\
  ...                              & ...             &  ...       & ...       &      ...            &      ...       & \nustar & 60061215002 & 2015--11--19 & 41.2 & 0.059 & --\\
  J1207.5+3352 & B2 1204+34 & 181.88711 & 33.87778 & 2 & 0.0791 & \xrt & 37315, 80691 & 2008--02--17 & 21.7 & 0.022 & (n)\\
  ...                              & ...             &  ...       & ...       &      ...            &      ...       & \nustar & 60160472002 & 2014--12--16 & 43.8 & 0.076 & --\\
  J1305.4--4928 & NGC 4945 & 196.36449 & --49.46821 & 2 & 0.0019 & \xmm & 204870101 & 2004--01--11 & 27.3 & 0.097 & (p)\\
  ...                              & ...             &  ...      & ...        &      ...            &      ...       & \nustar & 60061356002 & 2013--06--15 & 109.1 & 0.274 & (q)*\\
  J1321.0+0859 & NGC 5100 & 200.24417 & 8.98194 & AGN & 0.0319 & \xrt & 38063, 81140  & 2009--11--23 & 16.1 & 0.012 & (n)\\
   ...                              & ...             &  ...     & ...         &      ...            &      ...       & \nustar & 60160536002 & 2016--01--09 & 42.1 & 0.068 & --\\
  J1416.9--4641 & IGR J14175-4641 & 214.26526 & --46.69478 & 2 & 0.0766 & \xrt & 36108, 81864 & 2006-12-26 & 6.4 & 0.003  & (c)*\\
   ...                              & ...             &  ...      & ...        &      ...            &      ...       & \nustar & 60201033002 & 2016--05--25 & 43.0 & 0.025 & --\\
  J1432.7--4409 & NGC 5643 & 218.16977 & --44.17441 & 2 & 0.0040 & \xmm & 0601420101 & 2009--07--25 & 120.8 & 0.010 & (r)\\
   ...                              & ...             &  ...      & ...        &      ...            &      ...       & \nustar & 60160536002 & 2014--05--24 & 44.9 & 0.022 & (s)*\\
  J1440.7+5330 & Mrk 477 & 220.15874 & 53.50441 & 1 & 0.0377 & \xmm & 651100301 & 2010-07-21 & 32.1 & 0.037 & (t)\\
  ...                              & ...             &  ...     & ...         &      ...            &      ...       & \nustar & 60061255002 & 2014--05--15 & 36.1 & 0.065& --\\
  J1442.4--1714 & NGC 5728 & 220.59957 & --17.25308 & 2 & 0.0094 & \cha & 4077 & 2003--06--27  & 18.7 & 0.020 & (u)\\
  ...                              & ...             &  ...     & ...         &      ...            &      ...       & \nustar & 60061256002 & 2013--01--02 & 48.7 & 0.100 & --\\
  J1445.6+2702 & CGCG 164--019 & 221.40351 & 27.03478 & 1.9 & 0.0299 & \xrt & 37385, 49742,  80536 & 2010--11--26 & 18.0 & 0.004  & (c)*\\
  ...                              & ...             &  ...      & ...        &      ...            &      ...       & \nustar & 60061255002 & 2013--09--13 & 48.5 & 0.014 & (i)\\
  J1643.3+7038 & NGC 6232 & 250.83433 & 70.63253 & 2 & 0.0148 & \xrt & 45377, 80537 & 2011-05-01 & 19.7 & 0.003 & (c)*\\
  ...                              & ...             &  ...       & ...       &      ...            &      ...       & \nustar & 60061328002 & 2013--08--17 & 36.2 & 0.006 & (i)\\
  J1653.0+0223 & NGC 6240 & 253.24530 & 2.40093 & 1.9 & 0.0245 & \xmm & 101640101 & 2000--09--22 & 49.5 & 0.049 & (v)\\
  ...                              & ...             &  ...        & ...      &      ...            &      ...       & \nustar & 60002040002 & 2014--03--30 & 61.7 & 0.092 & (w)*\\
  J2102.5--2809 & ESO 464-G016 & 315.59901 & --28.17486 & 2 & 0.0364 & \xrt & 41113, 41906 & 2011--06--23 & 9.5 & 0.003 & (c)*\\
  ...                              & ...             &  ...      & ...        &      ...            &      ...       & \nustar & 60101013002 & 2016--04--13 & 44.2 & 0.022 & --\\
  J2148.3--3456 & NGC 7130 & 327.08133 & --34.95124 & 1.9 & 0.0162 & \cha & 2188 & 2001--10--23 & 38.6 & 0.006 & (x)\\
   ...                              & ...             &  ...     & ...         &      ...            &      ...       & \nustar & 60261006002 & 2016--12--15 & 84.1 & 0.010 & --\\
  J2207.0+1013 & NGC 7212 & 331.75542 & 10.23111 & 2 & 0.0266 & \xmm & 200430201 & 2004--05--20 & 36.7 & 0.018 & (y)\\
  ...                              & ...             &  ...     & ...         &      ...            &      ...       & \nustar & 60061310002 & 2013--09--01 & 49.1 & 0.019 & (i)\\
  J2318.3--4222 & NGC 7582 & 349.59792 & --42.37056 & 2 & 0.0053 & \xrt & 32534, 91915 & 2012--09--01 & 6.5 & 0.021 & (z)\\
  ...                              & ...             &  ...     & ...         &      ...            &      ...       & \nustar & 60061318002 & 2012--08--31 & 32.9 & 0.015 & (a2)*\\
\hline
\hline
\end{tabular}}\caption{\normalsize Sample of candidate CT-AGN analyzed in this work. Column (1): ID from the Palermo BAT 100-month catalog (Cusumano et al. 2017 in prep.). (2): source name. (3) and (4): right ascension and declination (J2000 epoch). (5): optical classification (1.9: Seyfert 1.9 galaxy; 2: Seyfert 2; AGN: active galactic nucleus), as reported in \citet{koss17}. (6): redshift. (7): telescope used in the analysis. (8): observation ID. (9): observation date. For \swi, this is the date of the first observation taken. (10): total exposure, in ks. For \xmm\ and \nustar, this is the sum of the exposures of each camera. (11): average count rate (in cts s$^{-1}$), weighted by the exposure for \xmm\ and \nustar, where observations from multiple instruments are combined. Count rates are computed in the 3--70\,keV band for \nustar\ and in the 2--10\,keV band otherwise. (12): reference for previous assessments of CT nature for the source, as follows. When \nustar\ data were used, the reference is reported on the \nustar\ observation line. Sources previously fitted with a torus model are flagged with a *. a) \citet{matt00}; b) \citet{balokovic14}; c) \citet{ricci15}; d) \citet{bauer15}; e) \citet{greenhill08}; f) \citet{masini16}; g) \citet{severgnini11}; h) \citet{ueda07}; i) \citet{koss16}; j) \citet{iyomoto01}; k) \citet{brightman15}; l) \citet{levenson06}; m) \citet{koss15}; n) \citet{vasudevan13}; o) \citet{gonzalez11}; p) \citet{guainazzi00}; q) \citet{puccetti14}; r) \citet{matt13}; s) \citet{annuar15}; t) \citet{shu07}; u) \citet{markwardt05}; v) \citet{vignati99}; w) \citet{puccetti16}; x) \citet{levenson05}; y) \citet{guainazzi05}; z) \citet{turner00}; a2) \citet{rivers15}.}\label{tab:sample}
\end{table*}
\endgroup
%this is the sum of the exposures of or different observations (for \xrt), in ks

\subsection{0.3--10\,keV data selection}
To each \swi\ 100-month candidate CT-AGN we associate a 0.3--10\,keV archival observation, selected using the following criteria:
\begin{enumerate}
\item When available, we use a \xmm\ observation; if no \xmm\ observation is available, we use a \cha\ one. Finally, we use \xrt\ observations when neither \xmm\ nor \cha\ data are available.
\item When multiple \xmm\ or \cha\ observations are available, we choose the longest one. For \xrt, instead, we combine all the available observations (see Section \ref{sec:xrt}).
\end{enumerate}

The only source for which we do not follow these rules is NGC 7582. This object is well known for its complex, highly variable spectrum \citep[see, e.g.,][]{piconcelli07,bianchi09,rivers15}; therefore, we select a NGC 7582 \xrt\ observation taken simultaneously to the \nustar\ one, instead of a longer \xmm\ one.

Following these criteria, 14 sources have an \xmm\ counterpart, 2 have a \cha\ counterpart and 14 have a \xrt\ counterpart. A summary of these observations is reported in Table \ref{tab:sample}. All the \xmm\ and \cha\ observations were made targeting specifically the sources in our sample, which are therefore imaged on axis. Most of the sources targeted by \xrt, instead, have been observed slightly off-axis: in 31 (41, 49) out of 54 \xrt\ observations used in this work, the source analyzed in this work lies within 3 (4, 5)$^\prime$ from the observation center. One source, 2MASXJ03561995--6251391, has been only observed serendipitously while targeting another object (SWIFT J0357.5--6255), at a distance of $\sim$9$^\prime$ from the pointing position.

\subsection{\xmm\ data reduction}
We reduced the \xmm\ data using the SAS v16.0.0\footnote{\url{http://xmm.esa.int/sas}} packages and adopting standard procedures. The source spectra were extracted from a 15$^{\prime\prime}$ circular region, while the background spectra were obtained from a circle having radius 45$^{\prime\prime}$ located near the source and not contaminated by nearby objects. Each spectrum has been binned with at least 15 counts per bin.

\subsection{\cha\ data reduction}
The \cha\ data have been reduced using the CIAO \citep{fruscione06} 4.7 software and the \textit{Chandra} Calibration Data Base (\texttt{caldb}) 4.6.9, adopting standard procedures; no source shows significant pile-up, as measured by the CIAO \textsc{pileup\_map} tool. We used the CIAO \texttt{specextract} tool to extract both the source and the background spectra. Source spectra have been extracted in circular regions of 4$^{\prime\prime}$, while background spectra have been extracted from annuli having inner radius $r_{\rm int}$=10$^{\prime\prime}$ and outer radius $r_{\rm out}$=25$^{\prime\prime}$: regions inside the background area have been visually inspected to avoid contamination from nearby sources. Finally, point-source aperture correction has been taken into account when extracting the spectra. Each spectrum has been binned with at least 15 counts per bin.

\subsection{\xrt}\label{sec:xrt}
All the sources in our sample have been observed multiple times by \xrt: given the \xrt\ smaller effective area with respect to \xmm\ and \cha, to maximize the spectral statistics we combined all the observations to produce a single spectrum. To do so, we used the \xrt\ data products generator available online \citep[\url{http://www.swift.ac.uk/user\_objects/}; see also][]{evans09}. %\footnote{http://www.swift.ac.uk/user\_objects/} 
Sources with \xrt\ 0.3--10\,keV data generally have lower count rates than those with \xmm\ or \cha\ data (see Table \ref{tab:sample}): consequently, we binned the \xrt\ spectra with 10 counts per bin when possible, and with 7 counts per bin for those sources with less than 50 net counts (namely NGC 1229, 2MASX J03561995--6251391, MCG+06-16-028, IGR J14175-4641 and ESO 464--G016).

\section{Spectral fitting procedure}\label{sec:model}
We fitted our spectra using the XSPEC software \citep{arnaud96}, taking into account the Galactic absorption measured by \citet{kalberla05}. We used \citet{anders89} cosmic abundances, fixed to the solar value, and the \citet{verner96} photoelectric absorption cross-section. In heavily obscured AGN the emission produced by the accreting SMBH is completely suppressed in the 0.5--2\,keV band \citep[see, e.g.,][]{gilli07,treister09}, where the emission is instead dominated by processes such as star-formation and/or diffuse gas emission \citep[see, e.g.,][]{koss15}. This is particularly true in \xmm\ and \xrt\ data, since the PSFs of these instruments are not sharp enough (5-15$^{\prime\prime}$ on axis) to avoid contamination from non-nuclear regions. Modeling this soft emission may be difficult, particularly in low-statistics spectra, such as all spectra from \xrt\ in our sample, and affect the final measurement of the AGN photon index. For these reasons, we choose to fit our data in the 2--150\,keV regime.

The X-ray spectral characterization of heavily obscured AGN presents a level of complexity that cannot be easily treated by simple XSPEC absorption models (e.g., \texttt{zwabs}, \texttt{ztbabs}, \texttt{zpcfabs}) without potentially introducing biases \citep[see, e.g.,][]{murphy09}. Consequently several models, based on Monte Carlo simulations, have been developed to analyze these complex spectra in a more self-consistent way. 
In this work we fit our spectra using the \texttt{MyTorus} \citep{murphy09} model.  %two of them:

\texttt{MyTorus} includes three distinct and separable components. The first one is a multiplicative component containing photoelectric absorption and Compton scattering attenuation (with associated equivalent neutral hydrogen column density denoted by $N_{\rm H,z}$), and it is applied to the main continuum (usually a power law): notably, in heavily obscured AGN, i.e., sources with $N_{\rm H, z}\gtrsim$5$\times$10$^{23}$ cm$^{-2}$, basically no flux from the AGN continuum is detected below 3--4\,keV (see Figure 5.1 in \texttt{MyTorus} manual\footnote{\url{http://mytorus.com/mytorus-manual-v0p0.pdf}}). This supports our choice to fit the spectra only above 2\,keV.

The second \texttt{MyTorus} component is the scattered continuum, also known as the ``reflected component'', i.e., those photons that reach the observer after interacting with the material surrounding the SMBH. The relative normalization between the reflected component and the main one is hereby denoted as A$_{\rm S}$. Finally, the third component models the neutral Fe fluorescent emission lines: while these are not the only elements responsible for the presence of emission lines in AGN X-ray spectra, they nonetheless are those that produce the most prominent lines. Particularly, \texttt{MyTorus} models the Fe K lines, both the K$\alpha$ at 6.4\,keV and the K$\beta$ at 7.06\,keV. In our analysis, we always started fitting the data assuming the relative normalization of the Fe K lines, A$_{\rm L}$=A$_{\rm S}$. However, we find that in four objects (namely NGC 424, NGC 1068, NGC 4945 and NGC 7130) the fit significantly improves when A$_{\rm L}$ is allowed to vary with respect to A$_{\rm S}$, although the physical interpretation of this is unclear. In our model, both A$_{\rm S}$ and A$_{\rm L}$ are computed within XSPEC adding a multiplicative constant component before the reflected continuum and the fluorescent lines components.%, and the Ni K$\alpha$ line at 7.47\,keV. 

Since in \texttt{MyTorus} the iron emission lines are produced self-consistently, one cannot use the XSPEC \texttt{eqwidth} task to compute the iron K$\alpha$ equivalent width \citep[EW; see, e.g.,][]{yaqoob15}. Therefore, to compute EW we first measure the monochromatic continuum flux, without the emission line contribution, at E$_{K\alpha}$=6.4\,keV rest-frame (i.e., the iron K$\alpha$ centroid); then we measure the flux of the emission line component alone in the 6.08--6.72\,keV rest-frame energy range, i.e., between 0.95\,E$_{K\alpha}$ and 1.05\,E$_{K\alpha}$. We choose this energy range to avoid contaminations from the iron K$\beta$ line at 7\,keV. The rest-frame EW is then computed multiplying by (1+$z$) the ratio between the line flux and the monochromatic continuum flux. Finally, the uncertainty on EW is derived estimating the uncertainty on A$_{\rm L}$ and then recomputing the line flux using as new A$_{\rm L}$ value the lower and upper boundaries of this parameter.

In \texttt{MyTorus}, the obscuring material surrounding the SMBH is assumed to have a toroidal, azimuthally simmetric shape. The torus has a fixed half-opening angle $\theta_{\rm OA}$=60$\degree$, i.e., a covering factor $f_c$=cos($\theta_{\rm OA}$)=0.5. While the torus half-opening angle is fixed, the angle  between the observer and the torus axis is free to vary in the range $\theta_{\rm obs}$=[0--90]$\degree$. In this work, we follow the approach detailed in \citet[][i.e., decoupled mode with column densities tied]{yaqoob15}: we first fix $\theta_{\rm obs}$=90$\degree$ for the main continuum, then we fit each spectrum twice, once assuming a reflection component viewing angle $\theta_{\rm obs, AS, AL}$=90$\degree$, the other with $\theta_{\rm obs, AS, AL}$=0$\degree$. Sources best-fitted with $\theta_{\rm obs, AS, AL}$=90$\degree$  correspond to a scenario where the dense obscuring torus is observed ``edge-on'' and the obscuring material lies between the AGN and the observer. In sources best-fitted with $\theta_{\rm obs, AS, AL}$=0$\degree$, instead, the reflection component comes from the back-side of a patchy, rather than uniform obscuring torus.

While in principle a more complex decoupled configuration of these three components can be used, with column densities untied, in the so-called ``\texttt{MyTorus} decoupled'' configuration \citep{yaqoob12}, which can be used to mimic different obscuring material geometries, we do not make use of this configuration in our analysis. In fact, in this work we are particularly interested in studying how the additional \nustar\ data affects the measurements of the main spectral parameters (particularly $\Gamma$ and $N_{\rm H, z}$). To do so, we choose to use a relatively simple model, such as the simple decoupled MyTorus one, in order to reduce parameters degeneracies and have a more consistent comparison between the results obtained without and with the \nustar\ data.
%\item The \citet{brightman11} \texttt{Torus} (hereafter BN \texttt{Torus}) model. In this model, the half-opening angle ($\theta_{\rm OA}$) of the obscuring material is left as a free parameter, and can assume values in the range $\theta_{\rm OA}$=[25.8--84.3]. In this model, the torus is geometrically assumed to be a sphere, with two cones interjecting it and representing the torus opening. We fix the observing view angle to the maximum allowed in this model, i.e., $\theta_{\rm obs}$=87$\degree$. The three different components which are part of \texttt{MyTorus} are all embedded in a single one in BN \texttt{Torus}.
%\end{enumerate}

%For the vast majority of the sources in our sample the two models produced results that are consistent within the uncertainties: the best-fit results that we report in Table \ref{tab:results} are those obtained with the model having the smallest reduced $\chi^2$. 16 sources have been fitted better with \texttt{BN Torus}, the remaining 14 with \texttt{MyTorus}.

%In both cases, 
We also added to the model a second power law, with photon index $\Gamma_2$=$\Gamma_1$, where $\Gamma_1$ is the photon index of the primary power law. This second power law models the fraction of emission (usually $\leq$1\% of the main component) which is scattered, rather than absorbed, by the gas surrounding the SMBH. The fractional contribution of the scattered component is calculated using the normalizations of the two power laws: in XSPEC, we measure the intensity of the scattered component using a constant to multiply the second power law component. We assume this second power law to be unabsorbed.

To take into account cross-calibration offsets between the 2-10\,keV and the \nustar\ data, as well as variability between different observations, we introduce in the model a constant $C_{NuS-2-10}$: the constant is fixed to 1 in the 2--10\,keV and \swi\ datasets and is free to vary in the \nustar\ ones. Therefore, $C_{NuS-2-10}$$>$1 indicates a scenario where the \nustar\ flux is higher than the 2--10\,keV one.
For the majority of the sources, the best-fit model is consistent with a lack of variability, C$_{2-10-NuS}\sim$1, within the uncertainties. We discuss in Section \ref{sec:variable} those sources for which we instead measure significant flux variability between the 2-10\,keV and the \nustar\ observations.

\subsection{Complex spectra modelling}\label{sec:peculiar}
A few objects in our sample are not properly fitted by the basic model described above and require a more complex modelling. We report here the additional components introduced in the fit, as well as the sources for which these components are required.
\begin{enumerate}
\item Spectra with deep ($>$100\,ks) \xmm\ observations required additional emission lines in the model. Particularly, following the work of \citet{bauer15} on NGC 1068, we add to the spectrum of this source three broad Gaussian lines, one to model a complex Si XIII and XIV feature at 2.38\,keV, the others to model the Fe He-like and H-like features at 6.69\,keV and $\sim$7\,keV, respectively. Similar features are also present in the spectra of NGC 424 and NGC 4945  \citep[see also][]{puccetti14}.
\item The soft X-ray spectrum of NGC 7130 is dominated by the emission produced by star-forming processes \citep{levenson05}. While we fit the \cha\ data only in the 2--10\,keV, where the AGN emission is dominant, the presence of significant residuals in the 2--3\,keV band require us to add a thermal, phenomenological component to the model. The best-fit temperature is $kT$=0.32$^{+0.08}_{-0.10}$.
\item Recent studies of the complex and highly variable source NGC 7582 showed that the \nustar\ spectrum of this object is likely obscured by a patchy torus with high covering factor \citep[80--90\%;][]{rivers15}. In this characterization, the fractional second power law is obscured by a Compton thin medium ($N_{\rm H, z}$$\sim$3$\times$10$^{23}$ cm$^{-2}$) and its strength is higher ($\sim$20\% of the main component) than the one usually observed in CT-AGN ($<$5\%). Consequently, in our fit to NGC 7582 we assume that the second power-law is absorbed: we find an intrinsic absorption value $N_{\rm H, z}$$\sim$(3.1$\pm$0.7)$\times$10$^{23}$ cm$^{-2}$, in excellent agreement with the one reported by \citet{rivers15}.
\item NGC 424 is a widely studied reflection-dominated CT-AGN, with possibly only a fractional contribution to the observed emission coming from the main power law \citep[see, e.g.,][]{collinge00,iwasawa01,matt03,balokovic14}. Consequently, for this source we allowed the reflected component normalization to be A$_{\rm S}\gg$1; we also let A$_{\rm L}$ free to vary independently from A$_{\rm S}$.
\end{enumerate}

\section{Fitting results}\label{sec:results}
In Table \ref{tab:results} we report the best-fit values of the main spectral parameters ($N_{\rm H, z}$, $\Gamma$, iron K$\alpha$ EW) obtained first by fitting only the 2--10\,keV and the \swi\ data, then adding to the fit also the \nustar\ data. In Table \ref{tab:results2} we report the other best-fit parameters: the 2--10\,keV to \nustar\ cross-normalization constant, $C_{NuS-2-10}$; the main power law component normalization, norm$_{\rm 1}$; the reflection and and iron lines relative normalizations, A$_{\rm S}$ and A$_{\rm L}$; the fraction of scattered emission, $f_{\rm scatt}$. The parameters reported in this second table are those obtained from the joint 2--10\,keV--\nustar--\swi\ fit.

Preliminarily, we point out that in the rest of this work we will refer to a subsample of 26 out of 30 sources. Based on our analysis, the remaining four objects (namely 2MASX J10523297+1036205, B2 1204+34, NGC 5100 and Mrk 477) have best-fit parameters that are not consistent with a CT-AGN origin, both including and excluding the \nustar\ data from the fit. Since the main purpose of this paper is to study a population of bona-fide CT-AGN, we exclude these objects from the following analysis: however, we will present their spectra in Appendix \ref{sec:app_no-ct}, where we also investigate the discrepancy between our results and those of previous works.

Another object, RBS 1037, presents two possible solutions with similar statistics when the \nustar\ data are not included in the analysis: ($i$) a slightly favoured CT-AGN scenario, with $N_{\rm H, z}$=1.35$_{-0.67}^{+8.37}$$\times$10$^{24}$ cm$^{-2}$ and reduced $\chi^2$,  $\chi^2_\nu$=$\chi^2$/dof=89.1/79=1.13; ($ii$) an unobscured AGN scenario, with $N_{\rm H, z}<$10$^{22}$ cm$^{-2}$ and $\chi^2$/dof=92.4/79=1.17. Notably, the CT-AGN result is in good agreement with the one reported in \citet[][$N_{\rm H, z}$=1.70$_{-0.63}^{+3.92}$$\times$10$^{24}$ cm$^{-2}$]{vasudevan13}. 
However, when the \nustar\ data are added to the fit the unobscured AGN solution becomes the statistically favoured one, with $\chi^2_\nu$=$\chi^2$/dof=315.6/311=1.01. The CT-AGN solution cannot be ruled out, since fixing  the intrinsic absorption value to $N_{\rm H, z}$=1.35$\times$10$^{24}$ cm$^{-2}$ leads to an only marginally worse $\chi^2_\nu$=$\chi^2$/dof=322.7/312=1.03. However, this solution is not stable, since computing the uncertainty on $N_{\rm H, z}$ using XSPEC always causes the best-fit value to move from the CT local minimum to the unobscured AGN best-fit value.
Furthermore, the iron K$\alpha$ line equivalent width (EW=0.17$_{-0.08}^{+0.08}$\,keV) is relatively small, while in CT-AGN these lines are expected to have larger EW values, EW$\sim$1--2\,keV \citep[e.g.,][]{matt96}. 
Finally, it is worth noticing that this source is optically classified as a Seyfert 1 galaxy, and at the present day no evidence of Compton thick Seyfert 1 galaxy has been reported \citep[see, e.g.,][]{ricci15}. Taking into account all these factors, we believe that the unobscured scenario for RBS 1037 is more likely than a CT one.

\subsection{Source variability}\label{sec:variable}
While the majority of the objects are well fitted using the \texttt{MyTorus} model combined with the additional components described in the previous section, a minority of sources presented more complex spectra, that required additional components to obtain a reliable fit. We report here these sources, and the components added to the fit.
\begin{enumerate}
\item We find significant flux variability (i.e., $C_{NuS-2-10}$$\neq$1 at the $>$3\,$\sigma$ level) in nine sources. In five of these objects, the \nustar\ flux is higher than the 2--10\,keV one: these sources are NGC 1068 ($C_{NuS-2-10}$=1.51$_{-0.05}^{+0.05}$), CGCG 420--15 ($C_{NuS-2-10}$=1.34$_{-0.11}^{+0.13}$), 2MASXJ09235371--3141305 ($C_{NuS-2-10}$=1.46$_{-0.23}^{+0.30}$), NGC 3393 ($C_{NuS-2-10}$=1.60$_{-0.21}^{+0.23}$) and NGC 4945 ($C_{NuS-2-10}$=16.14$_{-4.35}^{+7.23}$). The 2--10\,keV flux is instead higher than the \nustar\ one in the following four objects: ESO  201-IG 004 ($C_{NuS-2-10}$=0.62$_{-0.08}^{+0.09}$), ESO 005-G 004 ($C_{NuS-2-10}$=0.51$_{-0.07}^{+0.09}$) IGR J14175--4641 ($C_{NuS-2-10}$=0.63$_{-0.10}^{+0.12}$) and NGC 6232 ($C_{NuS-2-10}$=0.46$_{-0.17}^{+0.21}$). For three of these objects, a comparison between 2--10\,keV and \nustar\ data already exists in the literature: NGC 3393 has been studied by \citet{koss15} using both \xmm\ and \cha\ data in the 2--10\,keV band: they found $C_{NuS-2-10}$=1.70$_{-0.25}^{+0.25}$, in excellent agreement with our best-fit result, $C_{NuS-2-10}$=1.60$_{-0.21}^{+0.23}$. Similarly,  NGC 4945 is known to be highly variable above 10\,keV \citep[see, e.g.,][and references therein]{yaqoob12,puccetti14}, the normalization of the main continuum varying by a factor $\sim$6--8 in less than a month \citep[][]{puccetti14}. NGC 1068, instead, was found to lack any significant variability on a timespan of $\sim$15\,yrs \citep{bauer15}, a result in agreement with the fully reflection dominated nature of this source (N$_{\rm H, z}>$10$^{25}$ cm$^{-2}$). However, \citet{bauer15} show that fitting the complex NGC 1068 spectrum with a single reflector component, like the one we are using in our analysis, leads to statistically inaccurate modeling. The best-fit model they propose contains instead a multi-component reflector, which more accurately describes the \nustar\ excess that we model with a phenomenologically effective, but physically inaccurate cross-normalization constant.
\item In a few sources we find a significant improvement in the fit when leaving $N_{\rm H, z}$ free to change between the 2--10\,keV dataset(s) and the \nustar\ ones. A first case is that of NGC 4102, where the \xmm\ observation was taken six years before the \nustar\ one (October 2009 versus November 2015, see Table \ref{tab:sample}).  For this source, the best-fit requires two different intrinsic absorption values, $N_{\rm H, z, XMM}$=1.44$_{-0.20}^{+0.25}$$\times$10$^{23}$ cm$^{-2}$ and  $N_{\rm H, z, NuS}$=7.78$_{-0.95}^{+0.88}$$\times$10$^{23}$ cm$^{-2}$, therefore suggesting a possible ``changing look'' behavior for this source. Interestingly, the additional \nustar\ data allow us to properly measure the photon index ($\Gamma_{\rm NuS}$=1.67$_{-0.12}^{+0.12}$), which had a significantly steeper best fit value ($\Gamma_{\rm NoNuS}$=1.96$_{-0.17}^{+0.16}$) when the \xmm\ and \swi\ data only were taken into account. Consequently, the $N_{\rm H, z, XMM}$ best-fit value significantly decreases with the addition of the \nustar\ data to the fit (it was $N_{\rm H, z, NoNuS}$=1.91$_{-0.26}^{+0.44}$$\times$10$^{24}$ cm$^{-2}$). Two out of four objects for which our 2--10\,keV+BAT analysis does not lead to a CT-AGN result (2MASX J10523297+1036205, Mrk 477), in disagreement with previous works, are also best fitted with two different $N_{\rm H, z}$ values: we describe them in Appendix \ref{sec:app_no-ct}.
\item In NGC 4945 the photon indexes are $\Gamma_{\rm XMM}$=1.83$_{-0.05}^{+0.04}$ and $\Gamma_{\rm NuS}$=1.97$_{-0.06}^{+0.06}$. NGC 4945 is a particularly complex source, and evidence of variability in the photon index, both between different observations and within the same \nustar\ observation, has been reported in \citet[][]{puccetti14}. Notably, their measurements of both $\Gamma_{\rm NuS}$ and $N_{\rm H, z}$ are in excellent agreement with ours, although the 2--10\,keV data used in their work come from \cha\ and \textit{Suzaku}, while we used \xmm\ data.
\end{enumerate}

\begingroup
\renewcommand*{\arraystretch}{1.5}
\begin{table*}
\centering
\scalebox{0.84}{
\begin{tabular}{ccc|cccc|cccc}
\hline
\hline
&  & & \multicolumn{4}{c|}{Without \nustar} & \multicolumn{4}{c}{With \nustar}\\ 
Source & $N_{\rm H, gal}$ & $\theta_{\rm obs, AS, AL}$ & $N_{\rm H, z}$ & $\Gamma$ & EW & $\chi^2$/DOF & $N_{\rm H, z}$ & $\Gamma$ & EW & $\chi^2$/DOF\\% & Model \\ % & Facility\\
& 10$^{20}$ cm$^{-2}$ & \degree &10$^{22}$ cm$^{-2}$ & & keV & & 10$^{22}$ cm$^{-2}$ & & keV &\\
\hline
NGC 424 & 1.6 & 0 & 319.2$_{-39.8}^{+50.3}$ & 2.00$_{-0.03}^{+0.04}$ & 0.40$_{-0.05}^{+0.06}$ & 329.3/265 & 244.2$_{-21.5}^{+23.1}$ & 1.92$_{-0.03}^{+0.03}$ & 0.39$_{-0.12}^{+0.65}$ & 446.9/336\\% \\ % & T \\ % & XMM\\  0 % Updated referee
MCG +08--03--018 & 13.6 & 90 & 99.7$_{-34.3}^{+60.6}$ & 2.38$_{-0.34}^{+0.22 u}$ & $<$1.75 & 9.9/21 & 47.9$_{-7.2}^{+7.4}$ & 1.84$_{-0.12}^{+0.10}$ & 0.34$_{-0.15}^{+0.17}$ & 176.9/151 \\ % & M \\ % & XRT\\ % 90 % Updated referee
NGC 1068 & 0.9 & 0 & 1000.0$_{-279.3}^{+0.0u}$ & 2.33$_{-0.06}^{+0.07}$ & 0.45$_{-0.06}^{+0.07}$ & 282.0/187 & 1000.0$_{-89.1}^{+0.0u}$ & 1.73$_{-0.07}^{+x0.07}$ & 1.25$_{-0.19}^{+0.18}$ & 1056.2/740 \\ % & T \\ % & XMM\\  0 % Updated referee
NGC 1194 & 6.0 & 90 & 101.3$_{-14.7}^{+17.0}$ & 1.79$_{-0.17}^{+0.18}$ & 0.60$_{-0.16}^{+0.20}$ & 90.8/65 & 81.1$_{-7.9}^{+8.6}$ & 1.50$_{-0.09}^{+0.10}$ &  0.78$_{-0.14}^{+0.16}$ & 307.5/243 \\ % & M \\ % & XMM\\ % 90 % Updated referee
NGC 1229 & 1.7 & 90 & 820.6$_{-528.4}^{+179.4u}$ & 1.96$_{-0.14}^{+0.12}$ & -- & 6.5/8 & 43.0$_{-6.2}^{+6.8}$ & 1.4$^f$ & 0.25$_{-0.23}^{+0.25}$ & 98.6/96\\ % & T \\ % & XRT\\ 0 noNus, 90 Nus % Updated referee
ESO 201-IG 004 & 1.2 & 90 & 115.6$_{-23.3}^{+23.3}$ & 2.03$_{-0.24}^{+0.23}$ & 0.40$_{-0.27}^{+0.86}$ & 47.5/30 & 71.3$_{-13.1}^{+15.4}$ & 1.51$_{-0.11}^{+0.14}$ & 0.47$_{-0.32}^{+0.48}$ &108.7/83 \\ % & T \\ % & XMM\\  90 % Updated referee
2MASXJ03561995--6251391 & 3.1 & 90 &  177.4$_{-95.5}^{+99.5}$ & 2.29$_{-0.38}^{+0.31 u}$ & $<$1.13 & 7.4/9 & 83.9$_{-10.5}^{+9.4}$ & 1.98$_{-0.16}^{+0.06}$ & $<$0.37 &  134.5/137 \\ % & T \\ % & XRT\\  % 90 % Updated referee
CGCG 420--15 & 6.6 & 90 & 123.9$_{-40.6}^{+26.5}$ & 2.19$_{-0.34}^{+0.21}$ & 0.83$_{-0.38}^{+0.61}$ & 94.5/70 &  71.5$_{-9.7}^{+8.5}$ & 1.66$_{-0.12}^{+0.11}$ & 0.41$_{-0.12}^{+0.14}$ & 268.6/218 \\ % & M \\ % & XMM\\ 90 % Updated referee
ESO 005--G 004 & 10.2 & 90 & 101.9$_{-27.1}^{+49.2}$ & 1.58$_{-0.19}^{+0.19}$ & 1.32$_{-0.83}^{+1.04}$ & 17.6/12 & 106.9$_{-21.2}^{+24.7}$ & 1.54$_{-0.16}^{+0.17}$ & 1.89$_{-0.44}^{+0.43}$ & 77.4/72 \\ % & M \\ % & XRT\\ 90 % Updated referee
MCG +06--16--028 & 5.6 & 90 & 199.6$_{-130.3}^{+455.4}$ & 2.06$_{-0.31}^{+0.52}$ & -- & 9.0/7 & 104.7$_{-17.3}^{+17.0}$ & 1.56$_{-0.14}^{+0.13}$ & 0.37$_{-0.35}^{+0.35}$ & 82.3/86 \\ % & M \\ % & XRT\\ 90 % Updated referee
2MASXJ09235371--3141305 & 13.3 & 90 & 160.4$_{-24.5}^{+21.1}$ & 2.60$_{-0.16}^{+0.00 u}$ & $<$1.93 & 17.6/11 & 67.3$_{-9.6}^{+9.6}$ & 1.76$_{-0.13}^{+0.09}$ & 0.09$_{-0.05}^{+0.07}$ & 194.6/144 \\ % & T \\ % & XRT\\ 90 % Updated referee
NGC 3079 & 0.9 & 90 & 253.9$_{-49.8}^{+527.2}$ & 2.13$_{-0.14}^{+0.18}$ & $<$0.94 & 93.0/76 & 246.7$_{-23.5}^{+23.5}$ & 1.94$_{-0.10}^{+0.10}$ & 0.77$_{-0.30}^{+0.32}$ & 206.9/182 \\ % & M \\ % & XMM\\ 90 % Updated referee
NGC 3393 & 6.2 & 90 & 194.5$_{-28.3}^{+72.5}$ & 1.87$_{-0.22}^{+0.33}$ & 3.08$_{-1.30}^{+2.64}$ & 26.4/21 & 189.7$_{-16.5}^{+40.5}$ & 1.78$_{-0.12}^{+0.22}$ & 1.75$_{-0.46}^{+0.51}$ & 66.7/92 \\ % & M \\ % & XMM\\ 90 % Updated referee
RBS 1037 & 2.2 & 90 & 135.0$_{-66.9}^{+837.1}$ & 1.73$_{-0.08}^{+0.08}$ & -- & 89.1/79 & $<$1.0 & 1.75$_{-0.05}^{+0.05}$ & 0.17$_{-0.08}^{+0.08}$ & 315.6/311 \\ % & M \\ % & XRT\\ % 90 % Updated referee
NGC 4102 & 1.7 & 90 & 190.9$_{-25.7}^{+43.8}$ & 1.96$_{-0.16}^{+0.17}$ & 0.46$_{-0.37}^{+0.49}$ & 32.3/29 & 77.8$_{-8.8}^{+9.5}$ & 1.67$_{-0.12}^{+0.12}$ & 0.25$_{-0.18}^{+0.27}$ & 217.5/190 \\ % & M \\ % & XMM\\ 90 % Updated referee
NGC 4945 & 15.7 & 0 & 497.8$_{-58.4}^{+51.3}$ & 1.76$_{-0.03}^{+0.03}$ & 0.48$_{-0.30}^{+0.30}$ & 117.5/113 & 377.0$_{-15.7}^{+16.6}$ & 1.97$_{-0.06}^{+0.06}$ & 0.42$_{-0.12}^{+0.09}$ & 1508.8/1496 \\ % & T \\ % & XMM\\ % 
IGR J14175-4641 & 8.2 & 90 &  138.4$_{-48.8}^{+131.9}$ & 2.01$_{-0.27}^{+0.29}$ & $<$4.09 & 5.8/7 & 80.1$_{-12.9}^{+14.0}$ & 1.79$_{-0.14}^{+0.15}$ & $<$0.65 & 84.3/79 \\ % & T \\ % & XRT\\ %90 % Updated referee
NGC 5643 & 8.0 & 0 & 193.7$_{-44.2}^{+280.7}$ & 2.04$_{-0.12}^{+0.15}$ & 1.41$_{-0.23}^{+0.41}$ & 57.1/66 & 159.4$_{-29.8}^{+40.2}$ & 1.58$_{-0.15}^{+0.11}$ & 1.41$_{-0.19}^{+0.35}$  & 154.0/137 \\ % & T \\ % & XMM\\  0 % Updated referee
NGC 5728 & 7.4 & 90 & 149.2$_{-15.3}^{+14.3}$ & 1.94$_{-0.08}^{+0.09}$ & -- & 18.8/22 & 142.3$_{-8.7}^{+8.6}$ & 1.88$_{-0.06}^{+0.06}$ & 0.57$_{-0.22}^{+0.29}$ & 362.4/329 \\ % & M \\ % & Cha\\ %90 % Updated referee
CGCG 164--019 & 2.5 & 0 &  111.3$_{-86.9}^{+389.4}$ & 1.66$_{-0.26l}^{+0.74}$ & $<$1.16 & 8.4/11 & 119.5$_{-36.2}^{+50.0}$ & 1.78$_{-0.26}^{+0.29}$ & 0.41$_{-0.33}^{+0.33}$ & 59.9/57 \\ % & M \\ % & XRT\\ 0 % Updated referee
NGC 6232 & 5.7 & 90 & 405.4$_{-168.0}^{+167.5}$ & 2.19$_{-0.24}^{+0.22}$ & -- & 10.7/13 & 59.3$_{-17.5}^{+34.1}$ & 1.44$_{-0.04l}^{+0.34}$ & $<$0.44 & 35.9/34 \\ % & T \\ % & XRT\\  90 % Updated referee
NGC 6240 & 4.9 & 90 & 149.0$_{-8.9}^{+9.6}$ & 1.95$_{-0.07}^{+0.07}$ & 0.31$_{-0.16}^{+4.58}$ & 203.5/153 & 135.5$_{-6.4}^{+6.5}$ & 1.80$_{-0.05}^{+0.06}$ &  0.34$_{-0.22}^{+0.26}$ & 533.8/495 \\ % & T \\ % & XMM\\  90 % Updated referee
ESO 464--G016 & 7.3 & 90 & 162.4$_{-71.7}^{+197.6}$ & 2.29$_{-0.59}^{+0.31u}$ & $<$5.13 & 5.6/7 & 84.8$_{-15.6}^{+17.3}$ & 1.88$_{-0.22}^{+0.24}$ & $<$0.58 & 67.9/76 \\ % & T \\ % & XRT\\ % 90 % Updated referee
NGC 7130 & 1.9 & 90 & 154.2$_{-64.6}^{+34.6}$ & 1.72$_{-0.30}^{+0.25}$ & 1.09$_{-0.70}^{+1.73}$ & 14.5/18 & 221.8$_{-29.4}^{+42.4}$ & 1.50$_{-0.10l}^{+0.19}$ & 2.30$_{-0.55}^{+0.61}$ & 61.3/83 \\ % & M \\ % & Cha\\ 90 % Updated referee
NGC 7212 & 5.0 & 0 & 132.5$_{-32.6}^{+54.6}$ & 1.93$_{-0.19}^{+0.20}$ & 0.81$_{-0.25}^{+0.49}$ & 54.7/54 & 126.9$_{-24.5}^{+31.4}$ & 1.92$_{-0.17}^{+0.16}$ & 0.72$_{-0.23}^{+0.33}$ & 129.0/120 \\ % & M \\ % & XMM\\ 0 % Updated referee
NGC 7582 & 1.3 & 0 & 353.8$_{-101.8}^{+142.2}$ & 2.14$_{-0.06}^{+0.06}$ & 0.61$_{-0.42}^{+0.43}$ & 18.9/21 & 525.6$_{-130.7}^{+231.8}$ & 2.00$_{-0.04}^{+0.05}$ & 0.27$_{-0.09}^{+0.09}$ & 340.7/320  \\ % & M \\ % & XRT\\ 0 % Updated referee
\hline
\hline
\end{tabular}}\caption{\normalsize Best fit properties for the 26 candidate CT-AGN analyzed in this work, without and with the inclusion of the \nustar\ data to the fit. $N_{\rm H, gal}$ is the Galactic absorption, from \citet{kalberla05}, in units of 10$^{20}$\,cm$^{-2}$; $\theta_{\rm obs, AS, AL}$ is the reflection component viewing angle; $N_{\rm H, z}$ is the intrinsic AGN absorption, in units of 10$^{22}$\,cm$^{-2}$; $\Gamma$ is the power law photon index; EW is the equivalent width of the iron K$\alpha$ line at 6.4\,keV. In NGC 4102, leaving $N_{\rm H, z, 2-10}$ free to vary with respect to $N_{\rm H, z, NuS}$ lead to a significant improvement of the fit: we report the $N_{\rm H, z, 2-10}$ value in Section \ref{sec:variable}. Parameters fixed to a given value are flagged with $^f$. 90\% confidence errors flagged with $l$ and $u$ indicate that the value is pegged at either the lower ($\Gamma$=1.4, $N_{\rm H, z}$=10$^{22}$\,cm$^{-2}$) or the upper ($\Gamma$=2.6, $N_{\rm H, z}$=10$^{25}$\,cm$^{-2}$) boundary  of the parameter in the \texttt{MyTorus} model. For these sources, the reported values should therefore be treated as lower limits on the actual 90\% confidence uncertainties.}\label{tab:results}
\end{table*}
\endgroup

\begingroup
\renewcommand*{\arraystretch}{1.5}
\begin{table*}
\centering
\scalebox{0.85}{
\begin{tabular}{ccccccccccc}
\hline
\hline
Source & $C_{NuS-2-10}$ & norm$_{\rm 1}$ & A$_{\rm S}$ & A$_{\rm L}$  & $f_{\rm scatt}$ & f$_{\rm 2-10}$ & L$_{\rm 2-10}$ & f$_{\rm 15-55}$ & L$_{\rm 15-55}$\\
%            &                               &                      &                     &   &     \%            &  \\
\hline
NGC 424$^R$                           & 0.91$_{-0.07}^{+0.07}$ & 0.37$_{-0.04}^{+0.04}$ & 282.28$_{-45.55}^{+58.96}$ & 101.95$_{-17.31}^{+21.97}$ & -- & --12.03$_{-0.01}^{+0.01}$ & -- & --11.05$_{-0.04}^{+0.03}$ & -- \\
MCG +08--03--018                  & 0.93$_{-0.14}^{+0.19}$ & 20.15$_{-7.61}^{+9.88}$ & 1.00$^f$ & =A$_{\rm S}$ & 3.7$_{-1.4}^{+2.4}$ & --11.89$_{-0.12}^{+0.06}$ & 42.69$_{-0.18}^{+0.13}$ & --11.26$_{-0.08}^{+0.02}$ & 42.98$_{-0.13}^{+0.10}$\\ 
NGC 1068                               & 1.50$_{-0.05}^{+0.05}$ & 51.87$_{-13.83}^{+21.60}$ & 1.00$^f$ & 4.02$_{-0.60}^{+0.57}$ & 11.1$_{-2.2}^{+3.2}$ & --11.42$_{-0.01}^{+0.02}$ & 41.80$_{-0.01}^{+0.01}$ & --10.64$_{-0.06}^{+0.03}$ & 41.92$_{-0.01}^{+0.01}$\\
NGC 1194                               & 0.98$_{-0.07}^{+0.08}$ & 16.63$_{-5.31}^{+8.28}$ & 1.00$^f$ & =A$_{\rm S}$ & 2.0$_{-0.5}^{+0.7}$ & --11.97$_{-0.05}^{+0.02}$ & 42.57$_{-0.06}^{+0.05}$ & --10.88$_{-0.04}^{+0.02}$ & 43.05$_{-0.12}^{+0.10}$\\
NGC 1229                               & 1.21$_{-0.17}^{+0.23}$ & 3.31$_{-0.89}^{+1.00}$ & 2.07$_{-1.66}^{+2.65}$ & =A$_{\rm S}$ & 2.3$_{-1.3}^{+1.8}$ & --12.23$_{-0.07}^{+0.07}$ & 42.77$_{-0.37}^{+0.20}$ & --11.22$_{-0.04}^{+0.03}$ & 43.33$_{-0.70}^{+0.25}$\\
ESO 201-IG 004                     & 0.62$_{-0.08}^{+0.09}$ & 5.75$_{-4.18}^{+9.07}$ & 3.25$_{-1.71}^{+3.77}$ & =A$_{\rm S}$ & 4.9$_{-2.5}^{+3.4}$ & --12.18$_{-0.27}^{+0.02}$ & 42.96$_{-0.09}^{+0.07}$  & --11.05$_{-0.18}^{+0.03}$ & 43.47$_{-0.22}^{+0.14}$\\
2MASXJ03561995--6251391 & 1.24$_{-0.17}^{+0.26}$ & 34.30$_{-16.08}^{+24.47}$ & $<$0.67 & =A$_{\rm S}$ & 0.4$_{-0.2}^{+0.4}$ & --12.13$_{-0.17}^{+0.05}$ & 44.48$_{-0.30}^{+0.18}$ & --11.11$_{-0.13}^{+0.02}$ & 44.73$_{-1.03}^{+0.28}$\\
CGCG 420--15                       & 1.34$_{-0.11}^{+0.13}$ & 18.07$_{-8.12}^{+12.03}$ & 2.70$_{-0.95}^{+1.97}$ & =A$_{\rm S}$ & 2.4$_{-0.9}^{+1.6}$ & --11.90$_{-0.05}^{+0.02}$ & 43.15$_{-0.07}^{+0.06}$ & --10.82$_{-0.05}^{+0.02}$ & 43.65$_{-0.12}^{+0.09}$\\
ESO 005--G 004                    & 0.51$_{-0.07}^{+0.09}$ & 23.91$_{-11.77}^{+27.77}$ & 1.00$^f$ & =A$_{\rm S}$ & 1.1$_{-0.5}^{+0.9}$ & --12.11$_{-0.20}^{+0.04}$ & 41.64$_{-0.33}^{+0.19}$ & --11.26$_{-0.29}^{+0.02}$ & 42.62$_{-0.49}^{+0.22}$\\
MCG +06--16--028                & 1.08$_{-0.14}^{+0.19}$ & 12.11$_{-7.08}^{+10.32}$ & 1.43$_{-1.13}^{+2.54}$ & =A$_{\rm S}$ & 2.5$_{-1.3}^{+2.3}$ & --12.31$_{-0.18}^{+0.06}$ & 42.37$_{-0.28}^{+0.17}$ & --11.08$_{-0.14}^{+0.03}$ & 42.93$_{-0.15}^{+0.10}$ \\ 
2MASXJ09235371--3141305& 1.46$_{-0.23}^{+0.30}$ & 30.27$_{-16.09}^{+4.27}$ & $<$0.44 & =A$_{\rm S}$ & 0.2$_{-0.1}^{+0.3}$ & --12.01$_{-0.19}^{+0.05}$ & 43.53$_{-0.72}^{+0.37}$ & --10.92$_{-0.20}^{+0.02}$ & 44.65$_{-0.63}^{+0.25}$ \\
NGC 3079                             & 1.17$_{-0.10}^{+0.11}$ & 184.84$_{-70.52}^{+117.64}$ & 1.00$^f$ & =A$_{\rm S}$ & 0.3$_{-0.2}^{+0.1}$ & --12.49$_{-0.11}^{+0.03}$ & 42.18$_{-0.03}^{+0.03}$ & --10.77$_{-0.08}^{+0.02}$ & 42.43$_{-0.11}^{+0.09}$ \\
NGC 3393                             & 1.60$_{-0.21}^{+0.23}$ & 57.13$_{-26.04}^{+11.74}$ & 1.00$^f$ & =A$_{\rm S}$ & 0.5$_{-0.3}^{+0.3}$ & --12.03$_{-0.01}^{+0.01}$ & 42.84$_{-0.09}^{+0.07}$ & --11.05$_{-0.04}^{+0.03}$ & 43.05$_{-0.28}^{+0.17}$ \\
RBS 1037                             & 1.08$_{-0.08}^{+0.09}$ & 7.18$_{-0.54}^{+0.81}$ & -- & -- & -- & --11.50$_{-0.02}^{+0.03}$ & 43.66$_{-0.02}^{+0.02}$ & --11.39$_{-0.04}^{+0.04}$ & 43.74$_{-0.02}^{+0.02}$ \\
NGC 4102                             & 0.99$_{-0.20}^{+0.20}$ & 39.72$_{-18.53}^{+34.99}$ & 0.58$_{-0.42}^{+0.63}$ & =A$_{\rm S}$ & 1.3$_{-0.5}^{+0.8}$ & --12.30$_{-0.13}^{+0.05}$ & 41.45$_{-0.07}^{+0.06}$ & --10.80$_{-0.13}^{+0.02}$ & 41.80$_{-0.16}^{+0.12}$ \\
NGC 4945                             & 16.14$_{-4.35}^{+7.23}$ & 783.02$_{-284.05}^{+385.06}$ & 0.019$_{-0.003}^{+0.003}$ & 0.009$_{-0.002}^{+0.003}$ & 0.3$_{-0.1}^{+0.1}$ & --11.77$_{-0.11}^{+0.02}$ & 42.33$_{-0.12}^{+0.10}$  & --9.88$_{-0.11}^{+0.02}$ & 43.36$_{-0.08}^{+0.04}$ \\
IGR J14175-4641                  & 0.63$_{-0.10}^{+0.12}$ & 35.86$_{-20.73}^{+49.04}$ & 0.55$_{-0.55}^{+1.56}$ &  =A$_{\rm S}$ & $<$0.5 & --12.01$_{-0.22}^{+0.06}$ & 44.17$_{-0.69}^{+0.51}$ & --11.24$_{-0.31}^{+0.02}$ & 44.82$_{-0.73}^{+0.50}$ \\
NGC 5643                            & 0.93$_{-0.09}^{+0.10}$ & 16.10$_{-0.66}^{+6.76}$ & 3.72$_{-1.32}^{+3.87}$ & =A$_{\rm S}$ & 1.0$^f$ & --12.15$_{-0.03}^{+0.03}$ & 41.19$_{-0.20}^{+0.14}$ & --11.28$_{-0.06}^{+0.04}$ & 41.82$_{-0.24}^{+0.14}$ \\
NGC 5728                             & 1.00$_{-0.05}^{+0.06}$ & 185.02$_{-53.29}^{+69.66}$ & 1.82$_{-0.50}^{+0.72}$ & =A$_{\rm S}$ & 0.2$_{-0.1}^{+0.1}$ & --11.80$_{-0.04}^{+0.02}$ & 42.98$_{-0.12}^{+0.10}$ & --10.44$_{-0.02}^{+0.01}$ & 43.45$_{-0.23}^{+0.15}$ \\
CGCG 164--019                   & 0.82$_{-0.19}^{+0.25}$ & 18.36$_{-7.30}^{+8.36}$ & 0.56$_{-0.37}^{+0.52}$ & =A$_{\rm S}$ & 3.8$_{-2.3}^{+5.2}$ & --12.27$_{-0.50}^{+0.07}$ & 43.17$_{-0.18}^{+0.13}$ & --11.45$_{-0.48}^{+0.03}$ & 43.18$_{-0.17}^{+0.12}$ \\
NGC 6232                             & 0.46$_{-0.17}^{+0.21}$ & 3.51$_{-1.12}^{+1.03}$ & -- & -- & 2.7$_{-1.8}^{+2.0}$ & --12.51$_{-0.63}^{+0.11}$ & 42.01$_{-0.29}^{+0.17}$ & --11.83$_{-0.43}^{+0.02}$ & 42,53$_{-0.72}^{+0.36}$ \\
NGC 6240                             & 1.07$_{-0.05}^{+0.05}$ & 139.50$_{-33.27}^{+43.74}$ & 1.00$_{-0.26}^{+0.32}$ & =A$_{\rm S}$ & 1.8$_{-0.4}^{+0.4}$ & --11.70$_{-0.03}^{+0.01}$ & 43.81$_{-0.02}^{+0.02}$ & --10.50$_{-0.02}^{+0.01}$ & 43.94$_{-0.04}^{+0.04}$ \\
ESO 464--G016                    & 1.20$_{-0.29}^{+0.45}$ & 20.79$_{-18.71}^{+38.57}$ & $<$1.14 & =A$_{\rm S}$ & $<$0.9 & --12.36$_{-0.48}^{+0.06}$ & 43.27$_{-0.81}^{+0.43}$ & --11.23$_{-0.51}^{+0.02}$ & 43.80$_{-0.87}^{+0.41}$ \\
NGC 7130                             & 0.95$_{-0.17}^{+0.17}$ & 8.35$_{-1.58}^{+2.34}$ & 1.10$_{-0.39}^{+0.97}$ & 5.23$_{-1.51}^{+1.40}$ & 2.4$_{-0.7}^{+0.7}$ & --12.66$_{-0.04}^{+0.03}$ & 42.51$_{-0.06}^{+0.05}$ & --11.31$_{-0.05}^{+0.02}$ & 42.83$_{-0.14}^{+0.10}$ \\
NGC 7212                            & 0.86$_{-0.10}^{+0.11}$ & 35.26$_{-8.76}^{+10.08}$ & 0.91$_{-0.44}^{+0.79}$ & =A$_{\rm S}$ & 1.4$_{-1.1}^{+2.5}$ & --12.18$_{-0.11}^{+0.03}$ & 43.21$_{-0.10}^{+0.08}$ & --11.31$_{-0.12}^{+0.04}$ & 43.24$_{-0.17}^{+0.21}$ \\
NGC 7582                           & 0.99$_{-0.05}^{+0.05}$ & 308.67$_{-56.42}^{+69.59}$ & 1.00$^f$ & =A$_{\rm S}$ &  13.6$_{-2.3}^{+2.9}$ & --11.30$_{-0.03}^{+0.02}$ & 42.06$_{-0.12}^{+0.10}$ & --10.44$_{-0.02}^{+0.02}$ & 42.57$_{-0.21}^{+0.20}$ \\
\hline
\hline
\end{tabular}}\caption{\normalsize Best fit properties for the 26 candidate CT-AGN analyzed in this work. The reported parameters have been obtained by fitting all the available data for the given source, including \nustar. $C_{NuS-2-10}$ is the cross-normalization constant between the 2--10\,keV and the \nustar\ data; norm$_{\rm 1}$ is the main power law normalization (in units of ph cm$^2$ s$^{-1}$ keV$^{-1}$$\times$10$^{-4}$), measured at 1\,keV; A$_{\rm S}$ is the intensity of the \texttt{MyTorus} reflected component with respect to the main one; $f_{\rm scatt}$ is the percentage of main power law emission scattered, rather than absorbed, by the obscuring material. f$_{\rm 2-10}$, L$_{\rm 2-10}$, f$_{\rm 15-55}$ and L$_{\rm 15-55}$ are the logarithms of the observed flux (in units of erg s$^{-1}$ cm$^{-2}$) and the intrinsic, unabsorbed luminosity (in units of erg s$^{-1}$) measured in the 2--10\,keV and in the 15--55\,keV bands, respectively. Fluxes and luminosities are obtained with XSPEC, using the \texttt{flux} and the \texttt{clumin} commands, respectively. Parameters fixed to a given value are flagged with $^f$. NGC 424, flagged with $^R$, is a reflection dominated source where the continuum is poorly constrained and it is therefore not possible to properly assess the intrinsic luminosity values.
}\label{tab:results2}%NGC 7130 needed A$_L$ different from A$_S$ equal to 5.23$_{-1.51}^{+1.40}$.
\end{table*}
\endgroup

\section{The role of \nustar\ in the characterization of CT-AGN}\label{sec:nus_role}
As already mentioned in the introduction, the excellent effective area of \nustar\ in the 5--30\,keV range represents a fundamental tool to study heavily obscured AGN, and provides more accurate results than those obtained through the joint fit of \swi\ and 2--10\,keV data only. To quantitatively validate this assumption, in this section we compare the best-fit results obtained without and with the addition of the \nustar\ data.

First of all, the significantly improved statistics allows us to better constrain the spectral parameters. For example, the mean uncertainty on the intrinsic absorption $N_{\rm H, z}$ is reduced by a factor $\sim$3, being $\sigma_{\rm NH, NoNuS}\sim$50\% without the \nustar\ data, and becoming $\sigma_{\rm NH, NuS}\sim$17\% when the \nustar\ data are added to the fit. Similarly, the average uncertainty on $\Gamma$ decreases by $\sim$30\% when the \nustar\ data are included to the fit, from 11\% to 7\%. In these computations, we do not take into account those sources whose either best-fit $N_{\rm H, z}$ or $\Gamma$ value is pegged to a \texttt{MyTorus} boundary value. In particular, while computing the $N_{\rm H, z}$ average value we do not include NGC 1068, whose intrinsic absorption best-fit value is pegged at the \texttt{MyTorus} upper boundary for this parameter, $N_{\rm H, z}$=10$^{25}$ cm$^{-2}$, RBS 1037, which we find being an unobscured AGN ($N_{\rm H, z}<$10$^{22}$ cm$^{-2}$; see Section \ref{sec:results}), and NGC 4102, which is best-fitted by a model where $N_{\rm H, z, Nus}$ is different from $N_{\rm H, z, 2-10keV}$. Similarly, we exclude three sources from the average $\Gamma$ computation: 2MASXJ09235371-3141305, whose $\Gamma_{\rm NoNuS}$ value is pegged at \texttt{MyTorus} upper boundary, $\Gamma$=2.6, NGC 1229, whose $\Gamma_{\rm NuS}$ value is pegged at \texttt{MyTorus} lower boundary, $\Gamma$=1.4, and NGC 4945, which is best-fitted by a model where $\Gamma_{\rm Nus}$ is different from $\Gamma_{\rm 2-10keV}$.

The importance of \nustar\ is even more evident when measuring the intensity of the iron K$\alpha$ line at 6.4\,keV: since all the sources in our sample are at low redshift ($z_{\rm max}$=0.11), the line energy is observed at E$\geq$5.75\,keV, where the effective area of all 2--10\,keV instruments significantly declines. Consequently, without \nustar, it is possible to constrain the line equivalent width (EW) only for 14 out of 26 objects, with average uncertainty $\sigma_{\rm EW, NoNuS}\sim$62\%. For those same sources, the addition of the \nustar\ information significantly reduces the average uncertainty, down to $\sigma_{\rm EW, NuS}\sim$41\%. 
Furthermore, with the additional \nustar\ data it is possible to measure EW at a 90\% confidence level for another eight objects which only had an upper limit: we therefore have a significant EW measurement for 22 out of 26 objects, with average uncertainty $\sigma_{\rm EW, NuS}\sim$51\%.

\begin{figure}%[!t]
  \centering
  \includegraphics[width=1.02\linewidth]{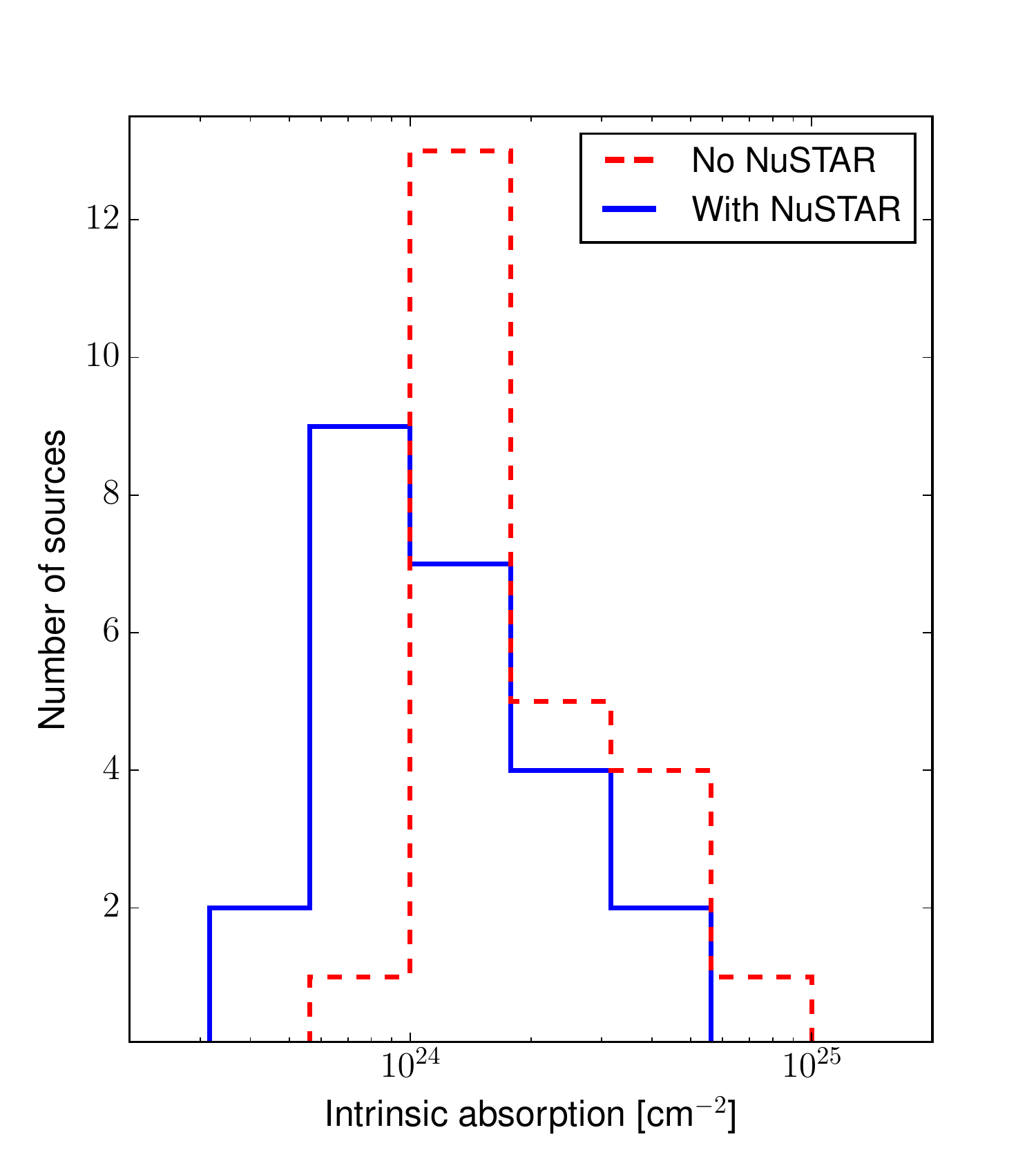}
\caption{\normalsize Histogram of the best-fit $N_{\rm H, z}$ value measured without (red dashed line) and with (blue solid line) the addition of \nustar\ data to the fit. As can be seen, a large fraction of objects moved from the log($N_{\rm H, z}$)=[24-24.25] bin to the log($N_{\rm H, z}$)$<$24 ones.}\label{fig:histo_nh}
\end{figure}

In several cases, the addition of \nustar\ data allowed us to refine the measurements of the spectral parameters not only reducing the uncertainties on the measurements, but also finding significantly different values. Particularly, nine candidate CT-AGN on the basis of the 2--10\,keV and BAT information have $N_{\rm H, z}<$10$^{24}$ cm$^{-2}$ at a $>$3$\sigma$ confidence level when the \nustar\ data are added to the fit. As already discussed in Section \ref{sec:results}, one of these objects, RBS 1037, is actually an unobscured AGN. Furthermore, another three sources have best-fit intrinsic absorption $N_{\rm H, z}<$10$^{24}$ cm$^{-2}$, but with the 3$\sigma$ confidence upper limit being $N_{\rm H, z}>$10$^{24}$ cm$^{-2}$. Finally, four objects have best-fit $N_{\rm H, z}>$10$^{24}$ cm$^{-2}$, but 3$\sigma$ confidence lower limit  $N_{\rm H, z}<$10$^{24}$ cm$^{-2}$. We show the distribution of $N_{\rm H, z}$ measured without (red dashed line) and with (blue solid line) the contribution of \nustar\ in Figure \ref{fig:histo_nh}. 

Consequently, taking into account also the four sources which are not confirmed as Compton thick on the basis of the 2--10\,keV+BAT data only (see Appendix \ref{sec:app_no-ct}),  only 14$^{+3}_{-4}$ out of 30 sources ($\sim$47$^{+10}_{-13}$\%) are confirmed as bona-fide CT-AGN in our analysis. Such an outcome significantly affects the overall observed fraction of CT-AGN in the nearby Universe: for example, based on our results the observed CT-AGN fraction in the 70-month \swi\ catalog, which was reported to be 7.6$^{+1.1}_{-2.1}$\% in \citet{ricci15}, decreases to 6.0$^{+0.4}_{-0.5}$\% (i.e., from 55 to 44\footnote{Five candidate CT-AGN reported in our sample, namely the four objects with $N_{\rm H, z}$$<$10$^{24}$ cm$^{-2}$ both without and with the addition of the \nustar\ data (see Section \ref{sec:results} and Appendix A) and RBS 1037, are not presented as CT in \citet{ricci15} work.} CT-AGN out of 728 sources). 
We also point out that this is in fact a very conservative upper limit, since it is computed under the assumption that all the 30 remaining candidate CT-AGN with no \nustar\ data available reported in \citet{ricci15} sample are indeed CT-AGN, an assumption that we deem unlikely based on the results of this work. If we instead assume for these 30 objects an outcome similar to the one found in our analysis, i.e., that only $\sim$50\% of the candidate CT-AGN are actual CT-AGN, the CT-AGN fraction would drop to $\sim$4\%, in agreement with the observed fraction measured by \citet{burlon11} in a sample of 199 \swi-selected AGN ($f$=4.6$^{+2.1}_{-1.5}$\%). It is worth noticing however that \citet{burlon11} showed that even hard X-ray instruments such \swi\ are biased against detecting heavily obscured sources. Once this bias is taken into account,  the intrinsic CT-AGN fraction they derived is $f_{\rm int}$=20$^{+9}_{-6}$\%.

\subsection{A systematic shift in the $\Gamma$ and $N_{\rm H, z}$ measurements}\label{sec:gamma_nh_shift}
As already pointed out in the previous section, analyzing the average properties of our sample we find evidence of a systematic offset in the measurements of $\Gamma$ and $N_{\rm H, z}$ computed without the \nustar\ data with respect with those computed adding \nustar\ information to the fit. More specifically, we find that 23 out of 25 objects with significant $N_{\rm H, z}$ measurement have a harder photon index when the \nustar\ data are taken into account (i.e., $\Gamma_{\rm NuS}<\Gamma_{\rm NoNuS}$), the average decrease in $\Gamma$ being $\langle(\Gamma_{\rm NoNuS}-\Gamma_{\rm NuS}$)/$\Gamma_{\rm NoNuS}\rangle\sim$13\%. It is worth noticing that the average photon index value of the sources fitted with the additional \nustar\ data is $\langle\Gamma_{\rm NuS}\rangle$=1.74$^{+0.13}_{-0.11}$, a value in excellent agreement with the typical photon index values measured in large samples of unobscured and obscured AGN in the 0.5--10\,keV band \citep[see, e.g.,][]{marchesi16c}.

In a similar way, for 20 out of 25 sources we measure a decrease in $N_{\rm H, z}$ when \nustar\ is added to the fit, the systematic reduction being on average by $\langle(N_{\rm H,z,NoNuS}-N_{\rm H,z,NuS}$)/max($N_{\rm H,z,NoNuS}$, $N_{\rm H,z,NuS}$)$\rangle\sim$32\%.

\begin{figure*}
\begin{minipage}[b]{.5\textwidth}
  \centering
  \includegraphics[width=1.14\textwidth]{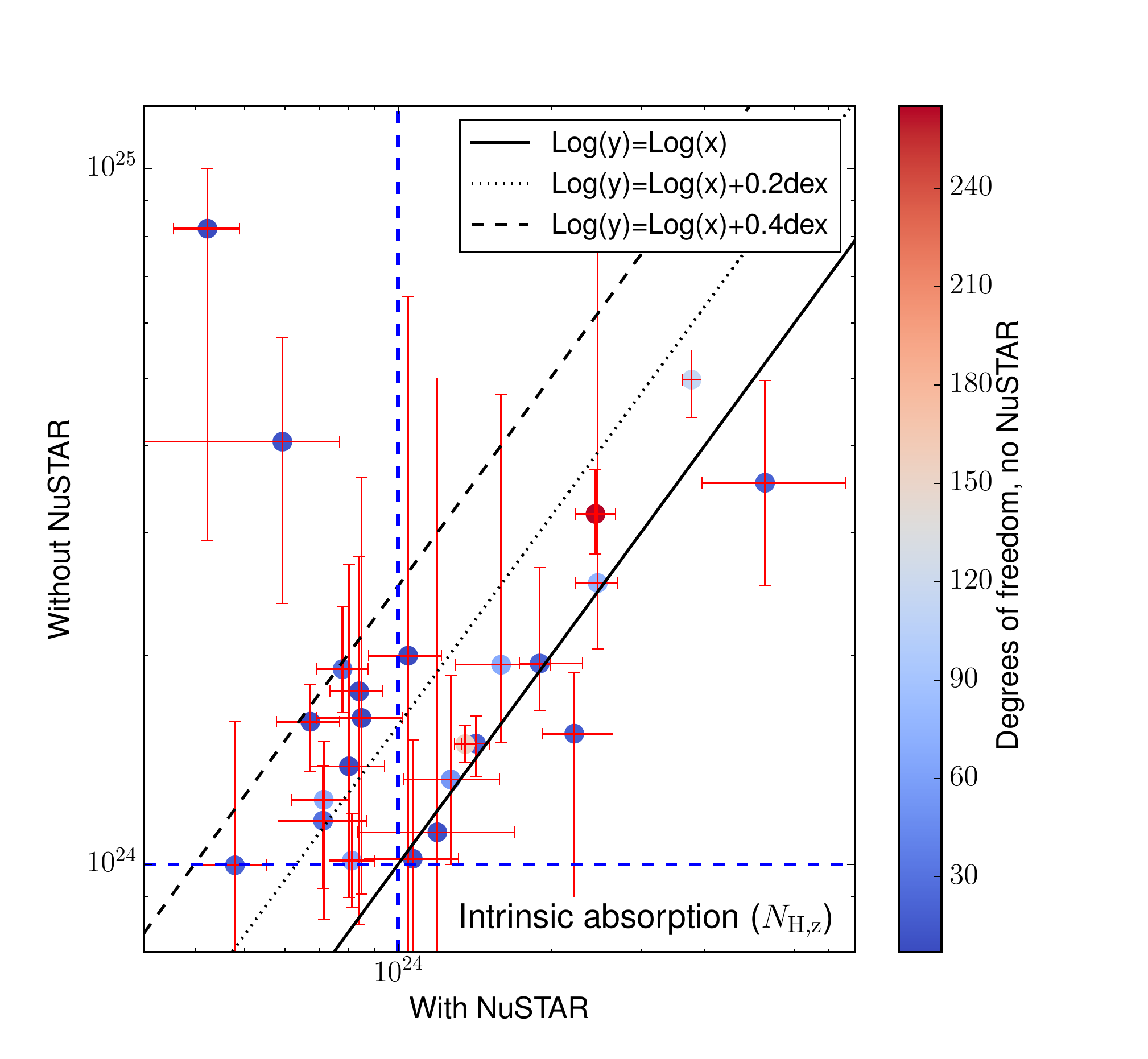}
  \end{minipage}
\begin{minipage}[b]{.5\textwidth}
  \centering
  \includegraphics[width=1.1\textwidth]{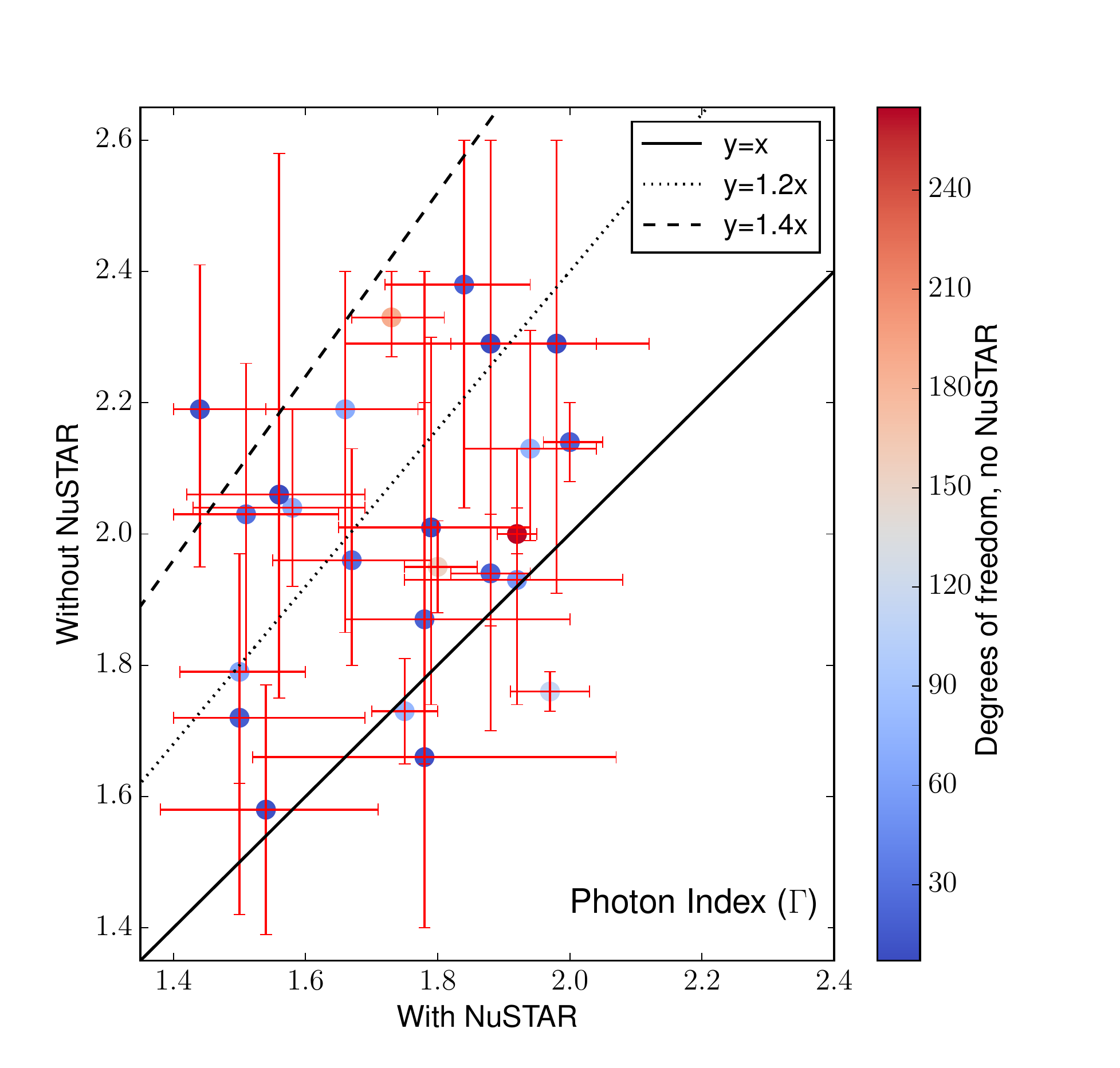}
  \end{minipage}
\caption{\normalsize \textit{Left}: Distribution of the intrinsic absorption measured without the \nustar\ contribution, $N_{\rm H, z, NoNuS}$, as a function of the same parameter measured including the \nustar\ data to the fit $N_{\rm H, z, NuS}$. NGC 1068, which has $N_{\rm H, z}>$10$^{25}$ cm$^{-2}$, and RBS 1037, which we find to be an unobscured AGN, are not shown in the plot.  The Log($y$)=Log($x$), Log($y$)=Log($x$)+0.2\,dex and Log($y$)=Log($x$)+0.4\,dex relations are plotted as a black solid, dotted and dashed line, respectively. The blue dashed horizontal and vertical lines mark the CT threshold, $N_{\rm H, z}$=10$^{24}$\,cm$^{-2}$.  \textit{Right}: same as left, but for the photon index $\Gamma$. Here, we do not plot NGC 1229, whose $\Gamma_{\rm NuS}$ value is pegged to $\Gamma$=1.4,   the \texttt{MyTorus} model's lower boundary, and 2MASXJ09235371-3141305, whose $\Gamma_{\rm NoNuS}$ value is pegged at  the \texttt{MyTorus} model's upper boundary, $\Gamma$=2.6. The $y$=\,$x$, $y$=1.2\,$x$ and $y$=1.4\,$x$ relations are plotted as a black solid, dotted and dashed line, respectively.}\label{fig:trends}
\end{figure*}

We report a summary of the average $\Gamma$ and $N_{\rm H, z}$ values in Table \ref{tab:nus_change}, while in Figure \ref{fig:trends} we show the distribution of $N_{\rm H, z, NoNuS}$ as a function of $N_{\rm H, z, NuS}$ (left) and the distribution of $\Gamma_{\rm NoNuS}$ as a function of $\Gamma_{\rm NuS}$ (right). As can be seen, the majority of the sources lie significantly above the 1--1 relation.

While a $\Gamma$-$N_{\rm H, z}$ degeneracy may be expected in low-quality X-ray data, a similar systematic offset has never been reported so far and represents an important result for future analyses of heavily obscured objects. Therefore, it is interesting to understand what is the origin of this behaviour. First, we study if the technical limitations of the instruments used in this analysis can explain the observed offset.

In a few objects, the photon index best-fit value is strongly driven by the \swi\ data, since the 2--10\,keV contribution is less significant. Particularly, three sources (MCG +08--03--018, 2MASXJ03561995-6251391 and 2MASXJ09235371-3141305), all of which with less than 21 degrees of freedom in their best-fit without \nustar, have photon index computed without \nustar, $\Gamma_{\rm NoNuS}\simeq\Gamma_{\rm BAT}>$2.1 and CT $N_{\rm H, z}$ values. When \nustar\ data are taken into account, all objects are found to have $\Gamma_{\rm NoNuS}$ and Compton thin obscuration. We point out that the \swi\ spectra are grouped in order to have only 8 data points, and in relatively faint sources the flux uncertainties in the first and last bins, where the the \swi\ effective area is smaller, can significantly affect the spectral fitting results.

Since the \swi\ data seem to be responsible for the offset only in a minority of objects, we checked for possible instrumental effects caused by low-quality 2--10\,keV spectra. To do so, we used the \texttt{MyTorus} model to simulate 1000 15\,ks \xrt\ observations of a Compton thin AGN with $\Gamma$=1.75 and $N_{\rm H, z}$=7.5$\times$10$^{23}$ cm$^{-2}$; we assumed both A$_{\rm S}$ and A$_{\rm L}$ to be equal to 1. Such an observation leads to a detection of 30--50\,cts in the 2--10\,keV band. We then fitted each simulated spectrum together with a simulated BAT spectrum, and we computed the best-fit $\Gamma$ and $N_{\rm H, z}$. Interestingly, we find that the median photon index and intrinsic absorption of the simulated population are in good agreement with the input values, being $\Gamma$=1.78 and $N_{\rm H, z}$=8$\times$10$^{23}$ cm$^{-2}$. However, both distributions also have very large dispersion values ($\sigma_\Gamma$=0.34 and  $\sigma_{\rm NH}$=8.5$\times$10$^{23}$ cm$^{-2}$), and 34\% of the simulated spectra are wrongly found to have a CT  $N_{\rm H, z}$ best-fit value. The median photon index of this ``wrongly-CT'' population is $\Gamma$=2.05. Consequently, these simulations show that the use of 2--10\,keV spectra with poor counts statistics can easily lead to an improper measurement of the basic X-ray spectral parameters, and, in a significant fraction of cases, to a wrong CT classification. We note that in the heavily obscured regime this effect is more likely to over-estimate the number of candidate CT-AGN rather than to wrongly classify as Compton thin actual CT sources,  since actual CT-AGN would have even lower count-rates and would therefore be missed by typical (in terms of exposure) \xrt\ observations.

A possible solution to this issue, in future works, can be fitting low-quality 2--10\,keV spectra (particularly \xrt\ ones, or those extracted from short \cha\ observations) fixing the photon index to a typical AGN value and measuring only $N_{\rm H, z}$. In this work, we find that heavily obscured AGN have an average photon index $\Gamma$$\sim$1.7--1.8, consistent with the typical photon index of the typical unobscured AGN. Consequently, a fit with photon index fixed to, e.g., $\Gamma$=1.8 is likely to produce better measurements of $N_{\rm H, z}$ than a fit where both parameters are free to vary and the photon index best-fit value is soft ($\Gamma$$>$2), even if this implies obtaining a higher reduced $\chi^2$ value for the fit.

We test this assumption on the seven sources in our sample having less than 25 degrees of freedom in the 2--10\,keV spectrum and best-fit $\Gamma_{\rm NoNuS}$$>$2, namely MCG+08-03-018, 2MASXJ03561995--6251391, MCG+06--16--028, 2MASXJ09235371--3141305, IGR J14175--4641, NGC 6232 and ESO 464--G016; we do not include in this subsample NGC 7582, due to its spectral complexity (see Section \ref{sec:peculiar}). We find that this method is indeed effective, since the average $N_{\rm H, z}$ value of the sample decreases by 42\%; more importantly, for all sources but NGC 6232 the new $N_{\rm H, z, NoNuS}$ value is in excellent agreement, with the $N_{\rm H, z, NuS}$ one, the discrepancy between the two quantities always being smaller than 25\%. Notably, for four objects (MCG+08-03-018, 2MASXJ03561995--6251391, MCG+06--16--028, 2MASXJ09235371--3141305) the agreement is even closer, with only a $\leq$5\% discrepancy. As a comparison, for the sources in this subsample the average offset between $N_{\rm H, z, NuS}$ and $N_{\rm H, z, NoNuS}$, when $\Gamma$ is left free to vary, is 55\%. Finally, we point out that NGC 6232 is the source in our sample with the worst \nustar\ data and both $N_{\rm H, z, NuS}$ and $\Gamma_{\rm Nus}$ are therefore quite poorly constrained (see Table \ref{tab:results}), therefore the larger discrepancy between $N_{\rm H, z, NoNuS}$ and $N_{\rm H, z, NuS}$ is not surprising.

While the above mentioned results confirm that low-quality spectra can explain the measured offset, a scenario where the sources underwent an intrinsic change in $N_{\rm H, z}$ is much less supported by our data. In fact, only nine out of 26 objects in our sample show evidence of variability in either flux or $N_{\rm H, z}$ (see Section \ref{sec:variable}), while in the remaining ones allowing $N_{\rm H, z, NoNuS}$ to vary with respect to $N_{\rm H, z, NuS}$ does not significantly improve the fit.

%\item If the he offset between BAT and NuSTAR could be explained in terms of a sort of Eddington bias?  Is it possible that for very high NH, such as those measured by BAT, it is more likely to MEASURE a fluctuation towards lower NH rather than higher NH due to the difficulty to measure higher NH which depress the spectrum at all energies? 

\begingroup
\renewcommand*{\arraystretch}{1.8}
\begin{table}
\centering
\scalebox{0.88}{
\begin{tabular}{ccccc}
\hline
\hline
Sample & $\langle\Gamma_{\rm NoNuS}\rangle$ & $\langle\Gamma_{\rm NuS}\rangle$ & $\langle N_{\rm H, z, NoNuS}\rangle$ & $\langle N_{\rm H, z, NuS}\rangle$\\
 & & 10$^{22}$ cm$^{-2}$ & 10$^{22}$ cm$^{-2}$\\
\hline
All & 2.00$^{+0.22}_{-0.21}$ & 1.74$^{+0.13}_{-0.11}$ & 222.4$^{+132.9}_{-75.6}$ & 147.4$^{+29.9}_{-21.4}$ \\ 
dof$_{\rm NoNuS}<$30 & 2.00$^{+0.28}_{-0.26}$ & 1.74$^{+0.16}_{-0.13}$ & 230.6$^{+143.9}_{-101.8}$ &  134.0$^{+36.7}_{-24.2}$\\ 
\hline
\hline
\end{tabular}}\caption{\normalsize Average photon index $\Gamma$ and intrinsic absorption $N_{\rm H, z}$ values, computed with and without the \nustar\ data, for our whole sample and using only a subsample of 15 sources with low statistics in the 2--10\,keV band, having less than 30 degrees of freedom in the fit. We do not include in the N$_{\rm H, z}$ computation NGC 1068, for which we are only able to measure a lower limit N$_{\rm H, z}>$10$^{25}$ cm$^{-2}$, RBS 1037, because we find it to be unobscured, and NGC 4102, which is best-fitted by a model where N$_{\rm H, z, NuS}$ and N$_{\rm 2-10keV}$ have different values (see Table \ref{tab:results2}). We do not include in the $\Gamma$ computation NGC 1229, whose $\Gamma_{\rm NuS}$ value is pegged to $\Gamma$=1.4,  the \texttt{MyTorus} model's lower boundary, 2MASXJ09235371-3141305, whose $\Gamma_{\rm NoNuS}$ value is pegged at the \texttt{MyTorus} model's upper boundary, $\Gamma$=2.6, and NGC 4945, which is best-fitted by a model where $\Gamma_{\rm Nus}$ is different from $\Gamma_{\rm 2-10keV}$.}
\label{tab:nus_change}
\end{table}
\endgroup

\section{Testing the Spectral Curvature technique}\label{sec:sc}
The Spectral Curvature (SC) technique has been recently developed by \citet{koss16} to identify candiate CT-AGN among sources having available \swi\ or \nustar\ data. 
The SC technique is based on sampling at different energy ranges the curvature observed above $>$10\,keV in heavily obscured AGN spectra. Particularly, for sources with \nustar\ data, SC is parameterized as follows:
\begin{equation}
SC=\frac{-0.46 \times A+ 0.64 \times B+ 2.33 \times C}{Tot},
\end{equation} 

where $A$, $B$, $C$ and $Tot$ are the count rates measured with \nustar\ in the 8--14\,keV, 14--20\,keV, 20--30\,keV and 8--30\,keV bands, respectively. SC=0.4 is the CT-AGN selection threshold: in \citet{koss16} work, seven of the nine sources with $SC_{BAT}>$0.4 in their sample have $N_{\rm H, z}>$ 10$^{24}$ cm$^{-2}$, and the remaining two are significantly obscured ($N_{\rm H, z}>$5$\times$10$^{23}$\,cm$^{-2}$).

\begin{figure}%[!t]
  \centering
  \includegraphics[width=1.04\linewidth]{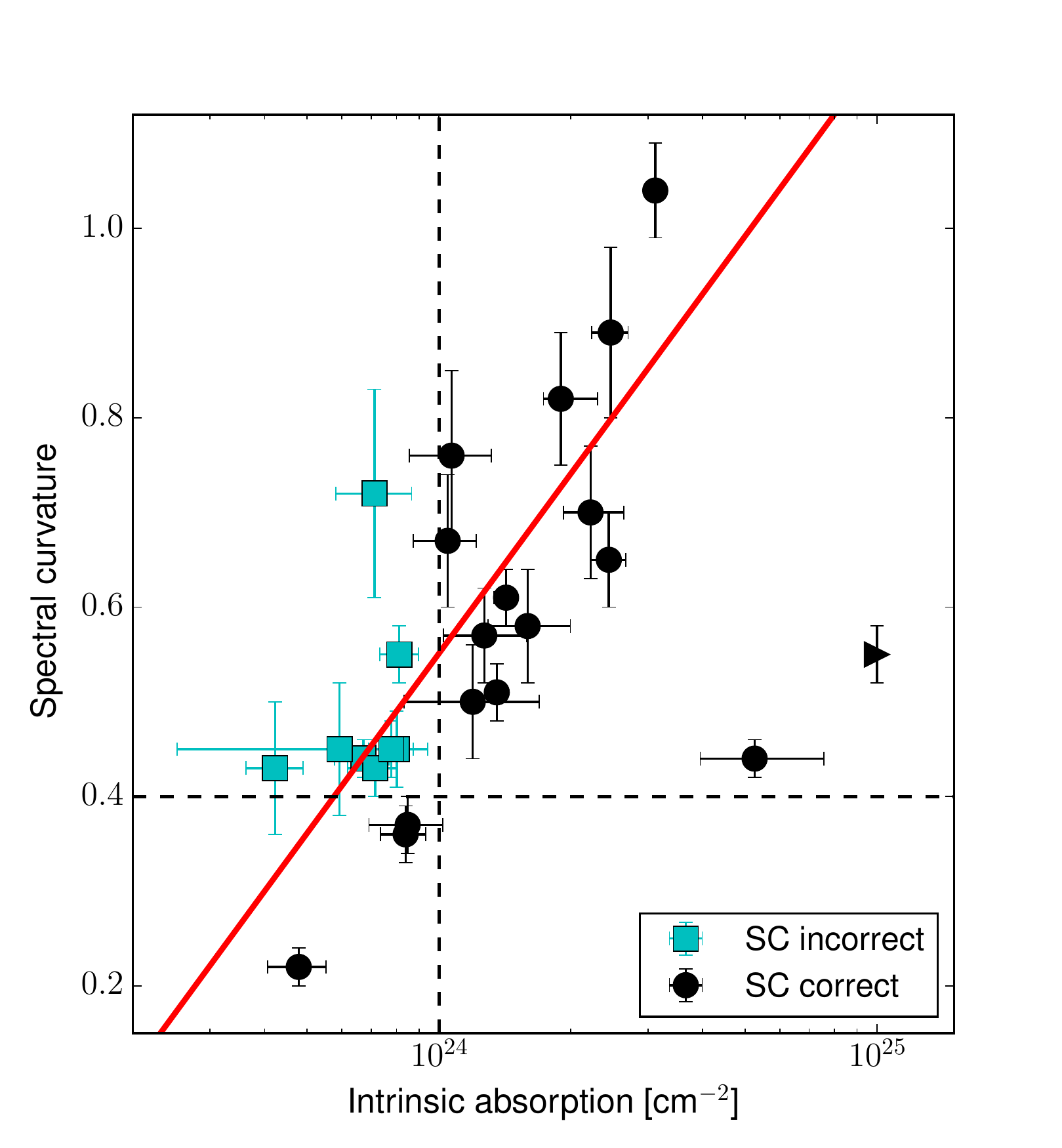}
\caption{\normalsize Spectral curvature parameter as a function of $N_{\rm H, z}$ for the 25 sources with $N_{\rm H, z}\geq$5$\times$10$^{23}$\,cm$^{-2}$ in our sample. Sources for which the SC threshold (SC$>$0.4, horizontal black dashed line) correctly predicted the CT origin of the source are plotted as black circles, while sources having SC$>$0.4 but  $N_{\rm H, z}<$10$^{24}$\,cm$^{-2}$ (vertical black dashed line) are plotted as cyan squares. NGC 1068, for which only a lower limit on $N_{\rm H, z}$ is available, is plotted as a black rightwards triangle. The best fit to the distribution (computed without taking into account the sources with $N_{\rm H, z}>$5$\times$10$^{24}$\,cm$^{-2}$) is shown with a red solid line.}\label{fig:nh_sc}
\end{figure}

In Figure \ref{fig:nh_sc} we plot the SC values as a function of $N_{\rm H, z}$ for the 25 objects with $N_{\rm H, z}\gtrsim$4$\times$10$^{23}$\,cm$^{-2}$ in our sample. As can be seen, there is a clear linear trend between SC and log($N_{\rm H, z}$), sources with higher SC values being also the most obscured ones: the Spearman rank order correlation coefficient of the distribution is $\rho$=0.72, and the p-value for such a $\rho$ value to be derived by an uncorrelated population is $p$=8.3$\times$10$^{-5}$. Notably, the two most heavily obscured objects in our sample, NGC 1068 and NGC 7582, do not follow this trend: such a behaviour is not unexpected, since \citet{koss16} reported that the SC technique is biased against objects with $N_{\rm H, z}>$5$\times$10$^{24}$\,cm$^{-2}$, due to their significant reduction in count rates. Furthermore, in Figure \ref{fig:nh_sc} we do not show RBS 1037, since we found it to be an unobscured AGN (see Section \ref{sec:results}): this is supported also by the SC technique, since the spectral curvature value of RBS 1037 is SC=0.13$\pm$0.01, i.e., the smallest SC value among the objects analyzed in this work.

In Figure \ref{fig:nh_sc} we plot as cyan squares the eight objects having SC above the CT threshold proposed by \citet{koss16}, SC=0.4, and best-fit intrinsic absorption value $N_{\rm H, z}<$10$^{24}$\,cm$^{-2}$. These eight sources can be divided in two classes: ($i$) six objects have SC$\leq$0.45 and their 3\,$\sigma$ lower limit is $<$0.4; these sources, having SC value close to the CT threshold one, all have $N_{\rm H, z}<$10$^{24}$\,cm$^{-2}$ at a $>$3$\sigma$ confidence level. ($ii$) The remaining two objects have SC$\geq$0.55: both sources, i.e., NGC 1194 and ESO 201-IG 004, are consistent with being CT within a $\sim$3.5\,$\sigma$ uncertainty.

In conclusion, we confirm that the SC technique is a reliable method to select candidate heavily obscured AGN, and the SC=0.4 threshold allows one to select a population of heavily obscured objects without missing a significant fraction of actual CT sources.

\section{Conclusions}\label{sec:concl}
In this work, we analyzed the combined 2--100\,keV spectra of 30 candidate CT-AGN. These objects have been selected among those candidate CT sources in the 100-month BAT catalog having an archival \nustar\ observation. 2--10\,keV data have been obtained using archival \xmm\ (14 sources), \cha\ (2 sources) and \xrt\ (14 sources). For 17 out of 30 objects, this is the first time when the \nustar\ data are analyzed and discussed. 

The additional \nustar\ data allows us to significantly improve the constraints on the main spectral parameters. The mean uncertainty on $N_{\rm H, z}$ is reduced by a factor $\sim$3, from 50\% to 17\%. Similarly, the average uncertainty on $\Gamma$ decreases by $\sim$40\%, from 11\% to 7\%. Finally, with \nustar\ we get an iron K$\alpha$ EW measurement for 22 out of 26 objects, 8 of which only had an EW upper limit when the \nustar\ data was not added to the fit.

The main result of our analysis is the discovery that a significant fraction of candidate CT-AGN are actually Compton thin based on the fitting of the high-quality \nustar\ spectra. We find evidence of a systematic offset between the spectral parameters measured without and with the \nustar\ data, i.e., a trend to artificially over-estimating the intrinsic absorption and the steepness of the spectrum when only the 2--10\,keV and the \swi\ data are included in the fit (see Figure \ref{fig:trends}). On average, the reduction in the photon index value is $\sim$13\% and the one in the intrinsic absorption is $\sim$32\%. 

As a consequence, only 14$^{+3}_{-4}$ out of 30 sources ($\sim$47$^{+10}_{-13}$\%) are confirmed as bona-fide CT-AGN in our analysis. This result strongly indicates that the analysis of 2--10\,keV+BAT spectra, while useful to detect new candidate heavily obscured sources, can lead to overestimating the CT-AGN fraction. For example, based on our results the observed CT-AGN fraction in the 70-month \swi\ catalog, which was reported to be 7.6$^{+1.1}_{-2.1}$\% in \citet{ricci15}, decreases to 6.0$^{+0.4}_{-0.5}$\% and potentially even down to $\sim$4\%, extrapolating the results of our work to the population of candidate CT-AGN with no public \nustar\ data available. 

We tested different possible explanations to this systematic offset and we favour a scenario where low-quality data in the 2--10\,keV band and/or from \swi\ can produce an artificial steepening in the intrinsic spectral shape of heavily obscured AGN. Such a steepening is then erroneously fitted by models with an excessively soft photon index ($\Gamma>$2) and an overestimated $N_{\rm H, z}$ value. A possible workaround for this fitting issue is to fit low-quality 2--10\,keV data fixing the photon index to a typical AGN value, $\Gamma$=1.7--1.8, and measuring only $N_{\rm H, z}$. In our sample of seven low-quality objects this approach is found to be very effective, the discrepancy between $N_{\rm H, z, Nus}$ and $N_{\rm H, z, NoNuS}$ being reduced to less than 25\% in six out of seven objects.

Finally, we tested the Spectral Curvature (SC) technique developed by \citet{koss16} to detect new candidate CT-AGN on the basis of their hard X-ray spectral shape. We confirm that the SC parameter directly correlates with $N_{\rm H, z}$, with  Spearman rank order correlation coefficient $\rho$=0.72, and that the SC=0.4 threshold is effective in selecting heavily obscured AGN.

\subsection*{Acknowledgements}
This work made use of data supplied by the UK Swift Science Data Centre at the University of Leicester, as well as of the TOPCAT software \citep{taylor05} for the analysis of data tables.

We thank the anonymous referee for the interesting discussion and for their comments, that significantly improved the paper. S.M. thanks Francesca Civano, Alberto Masini and Gianni Zamorani for the useful comments and suggestions. A.C., G.L. and C.V. acknowledge support from the ASI/INAF grant I/037/12/0 - 011/13.

%\newpage
\appendix
\section{A. Spectra of the 26 candidate CT-AGN studied in this work}\label{sec:app_spectra}

\begin{figure*}
\begin{minipage}[b]{.5\textwidth}
  \centering
  \includegraphics[width=0.78\textwidth,angle=-90]{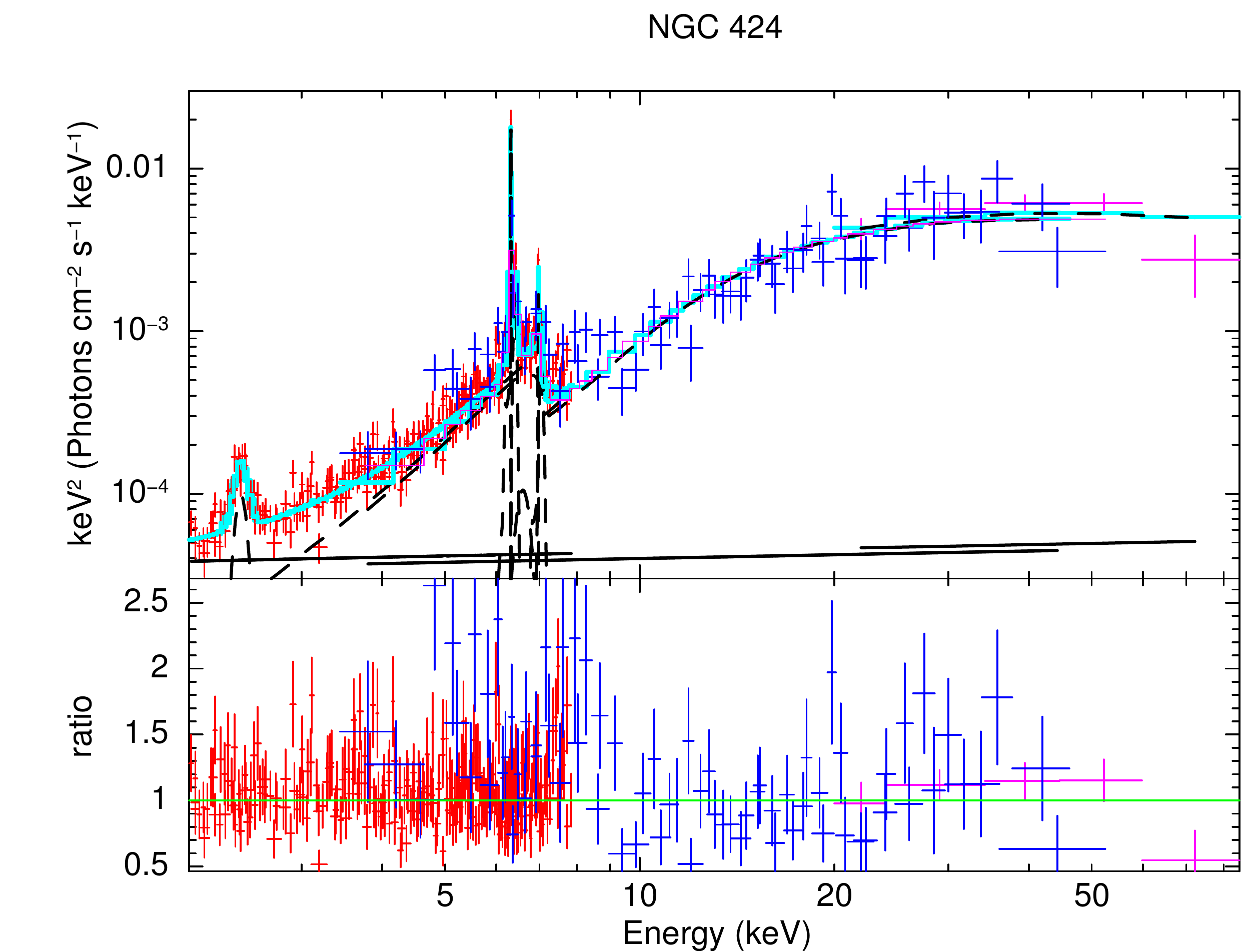}
  \end{minipage}
\begin{minipage}[b]{.5\textwidth}
  \centering
  \includegraphics[width=0.78\textwidth,angle=-90]{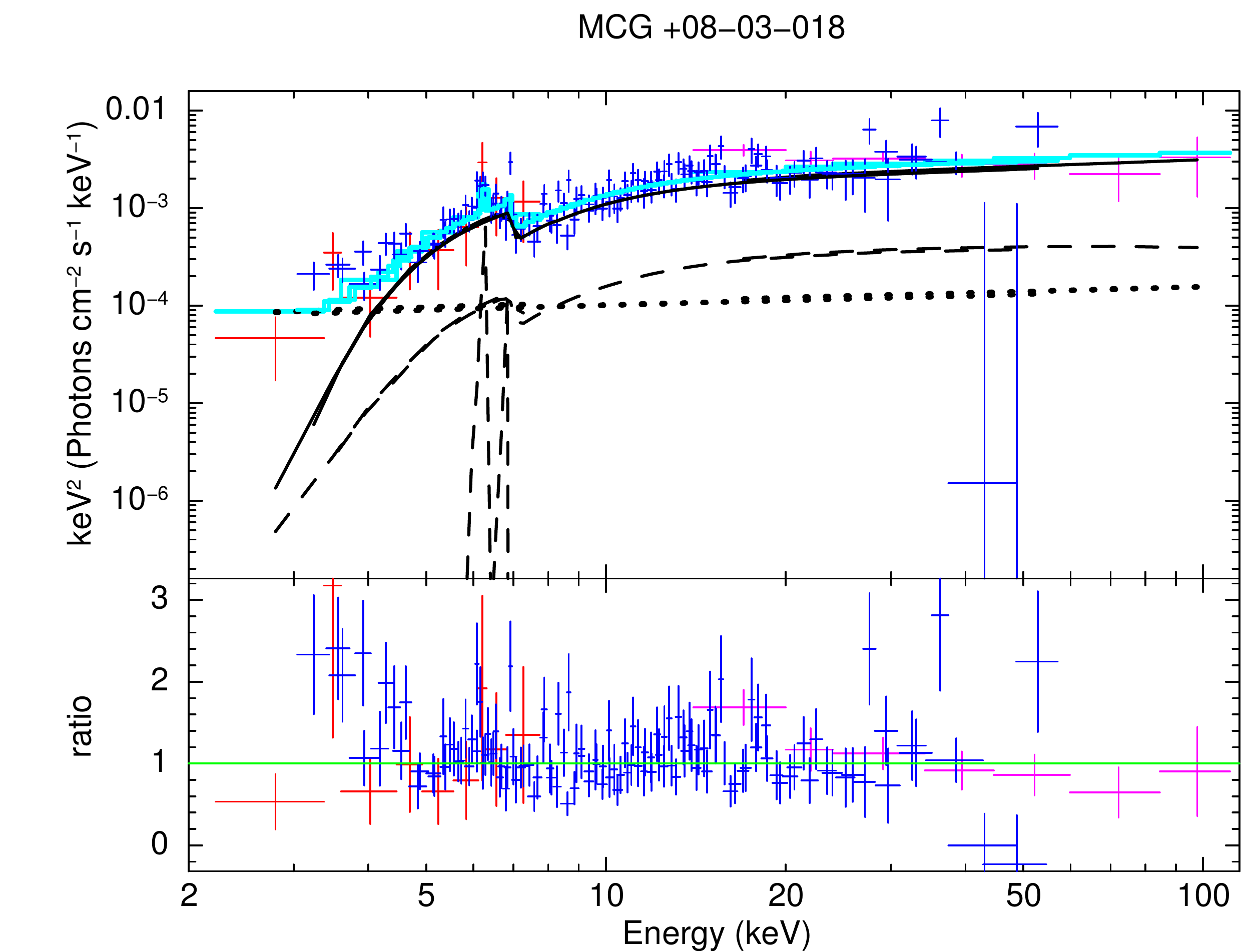}
  \end{minipage}
\begin{minipage}[b]{.5\textwidth}
  \centering
  \includegraphics[width=0.78\textwidth,angle=-90]{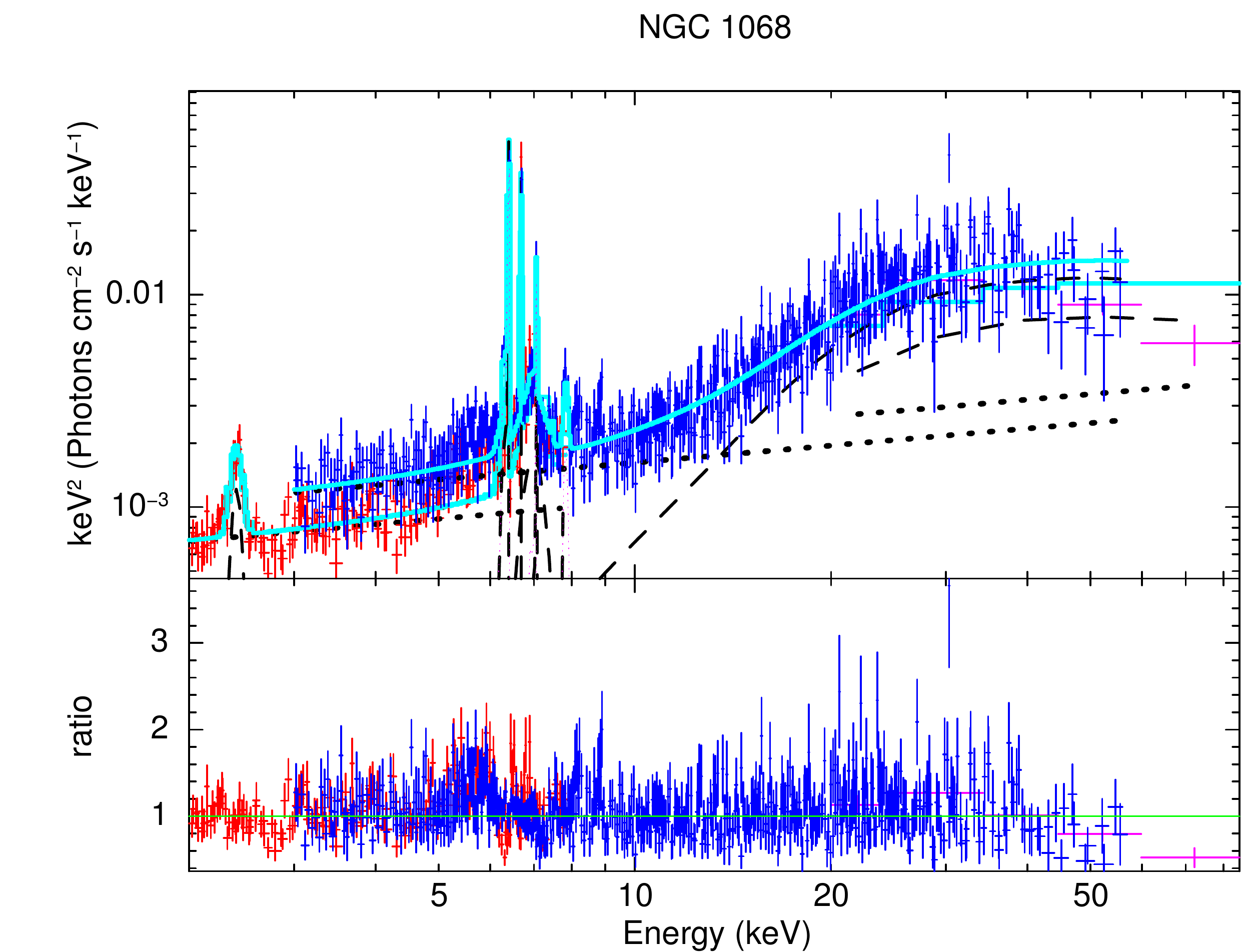}
  \end{minipage}
\begin{minipage}[b]{.5\textwidth}
  \centering
  \includegraphics[width=0.78\textwidth,angle=-90]{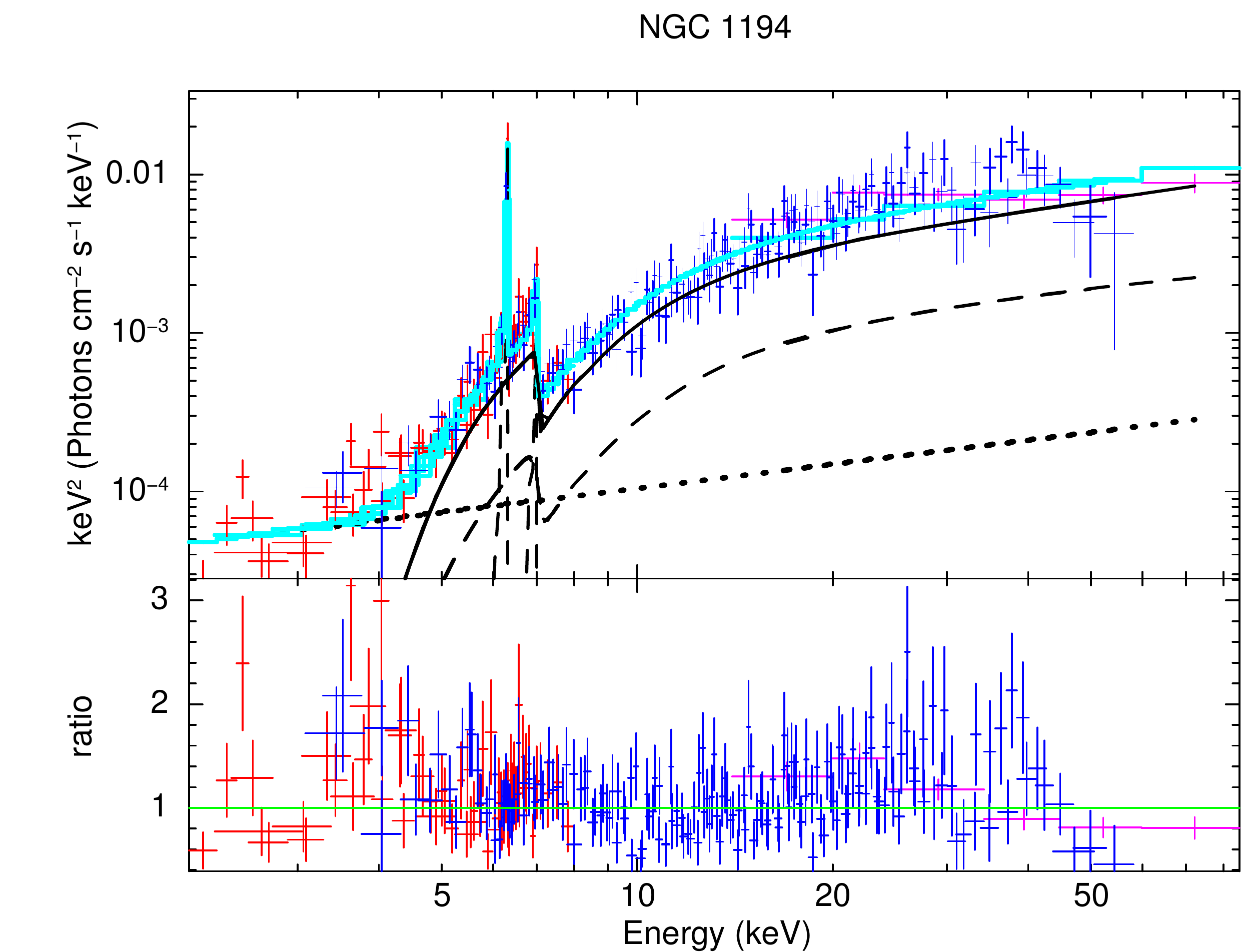}
  \end{minipage}
\begin{minipage}[b]{.5\textwidth}
  \centering
  \includegraphics[width=0.78\textwidth,angle=-90]{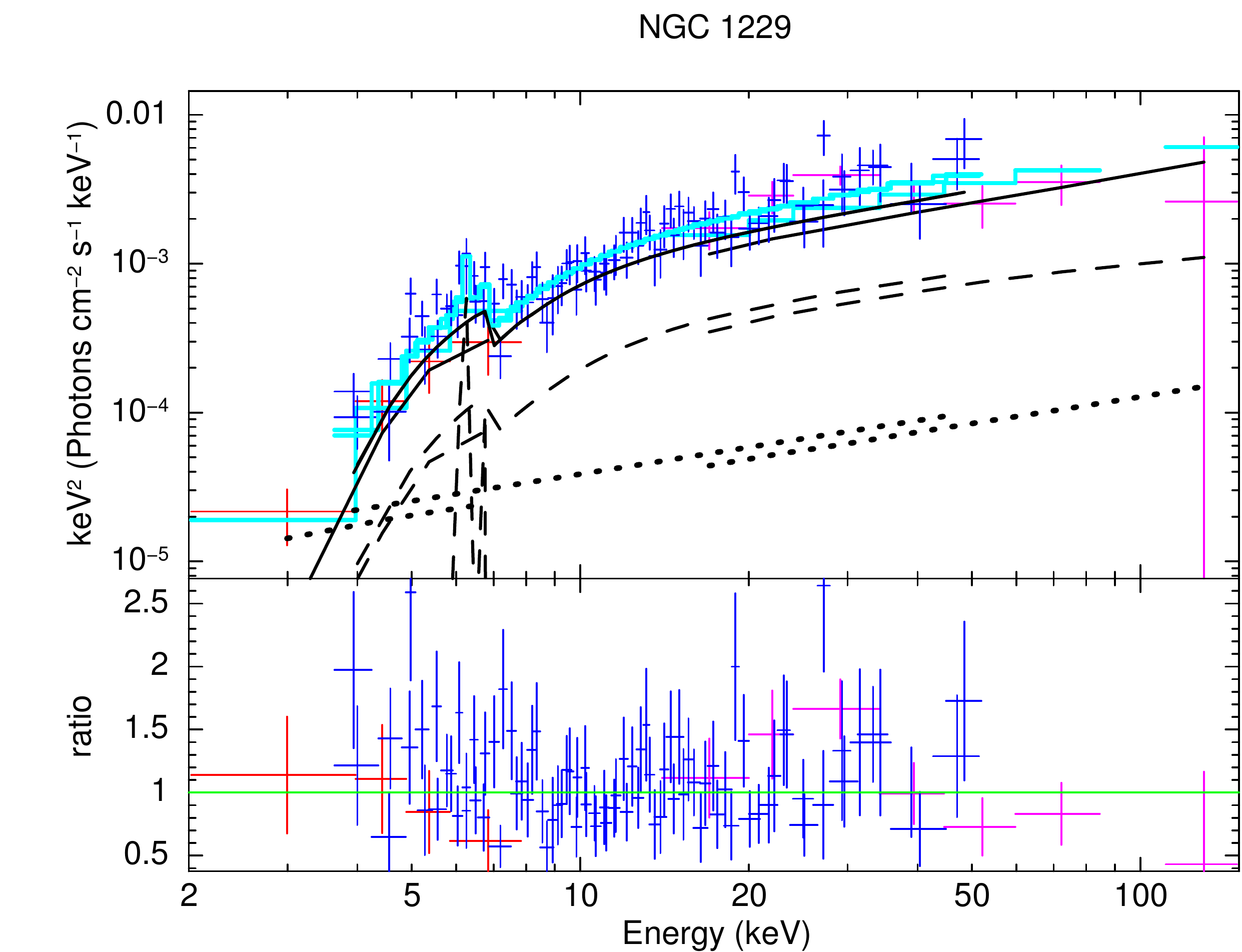}
  \end{minipage}
  \begin{minipage}[b]{.5\textwidth}
  \centering
  \includegraphics[width=0.78\textwidth,angle=-90]{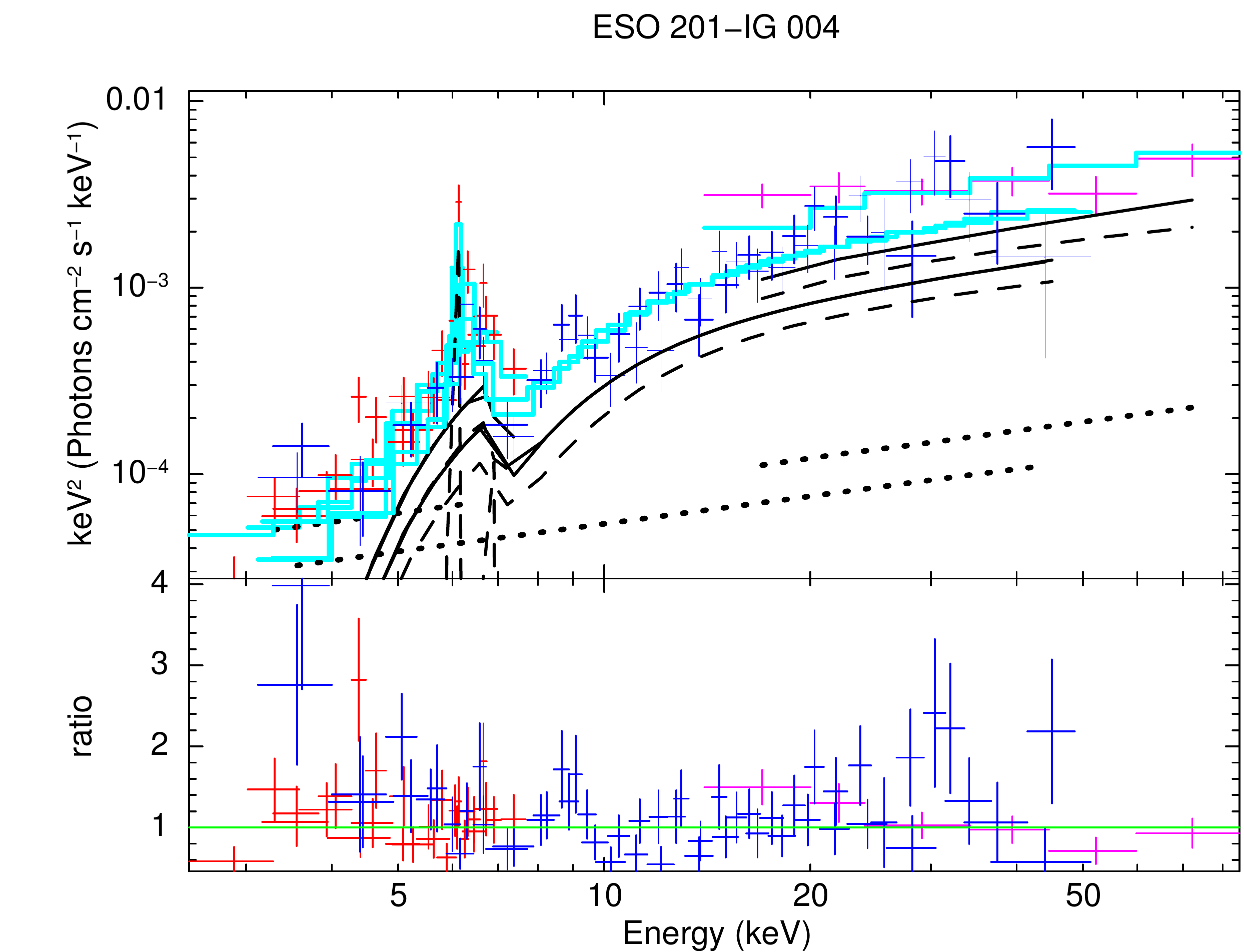}
  \end{minipage}
\caption{\normalsize Spectra (top panel) and data-to-model ratio (bottom) of the CT-AGN analyzed in this work. 2--10\,keV data are plotted in red, \nustar\ data in blue and \swi\ data in magenta. The best-fitting model is plotted as a cyan solid line, while the single \texttt{MyTorus} components are plotted as black solid (zeroth-order continuum) and dashed (reflected component and emission lines) lines. Finally, the main power law component scattered, rather than absorbed, by the torus is plotted as a black dotted line.}\label{fig:spectra}
\end{figure*}

\begin{figure*}%\ContinuedFloat
\begin{minipage}[b]{.5\textwidth}
  \centering
  \includegraphics[width=0.78\textwidth,angle=-90]{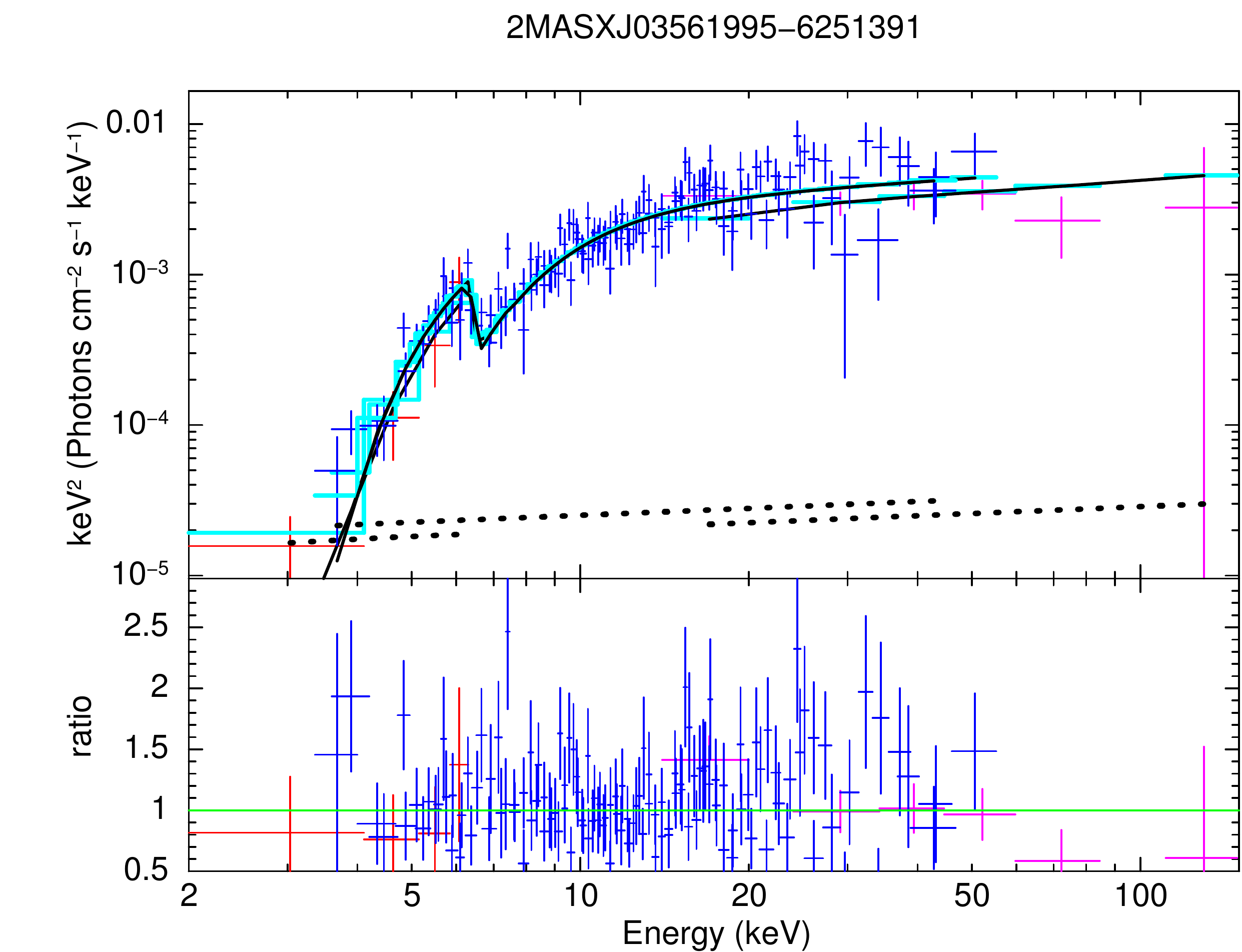}
  \end{minipage}
\begin{minipage}[b]{.5\textwidth}
  \centering
  \includegraphics[width=0.78\textwidth,angle=-90]{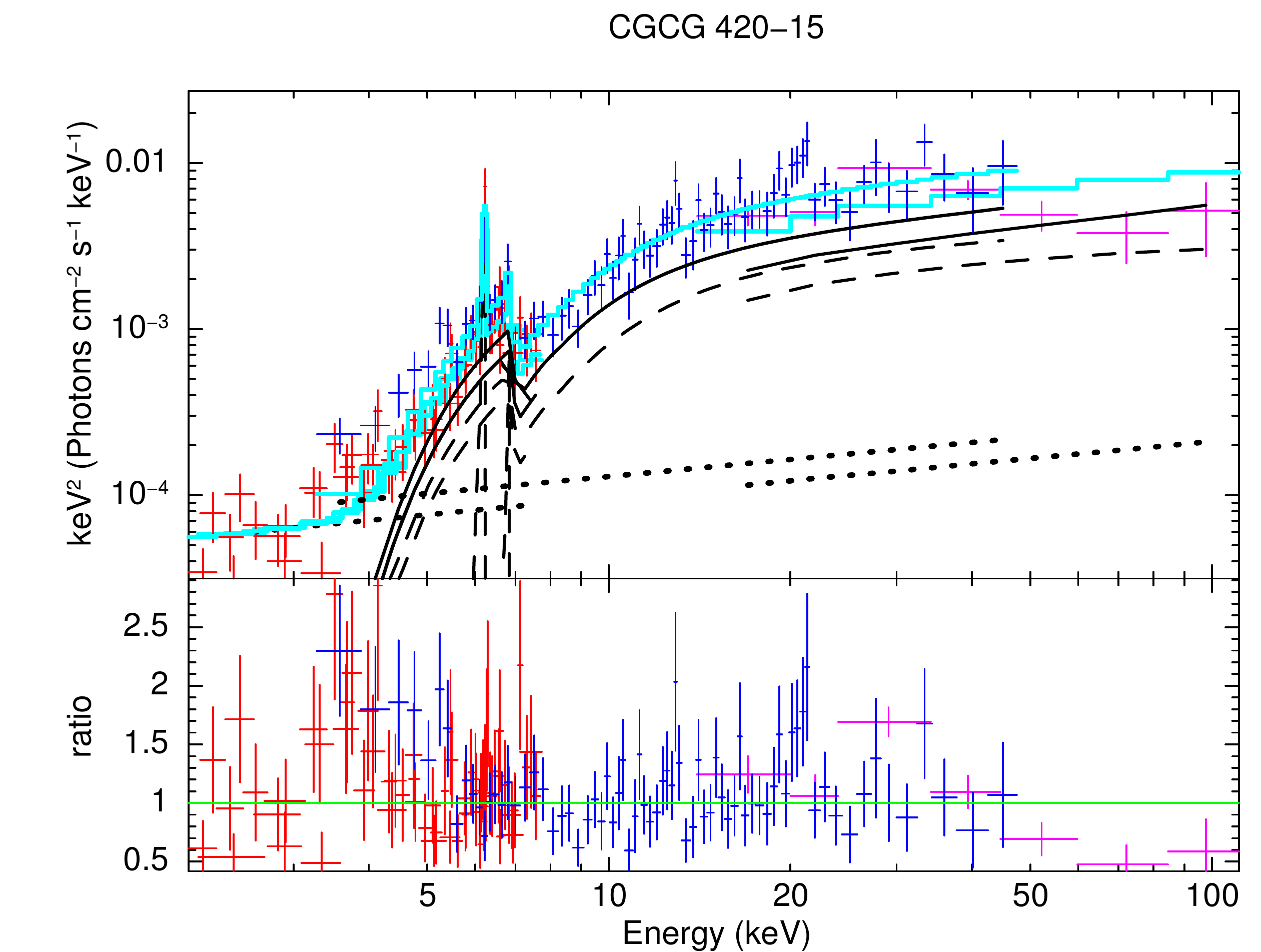} %0453
  \end{minipage}
\begin{minipage}[b]{.5\textwidth}
  \centering
  \includegraphics[width=0.78\textwidth,angle=-90]{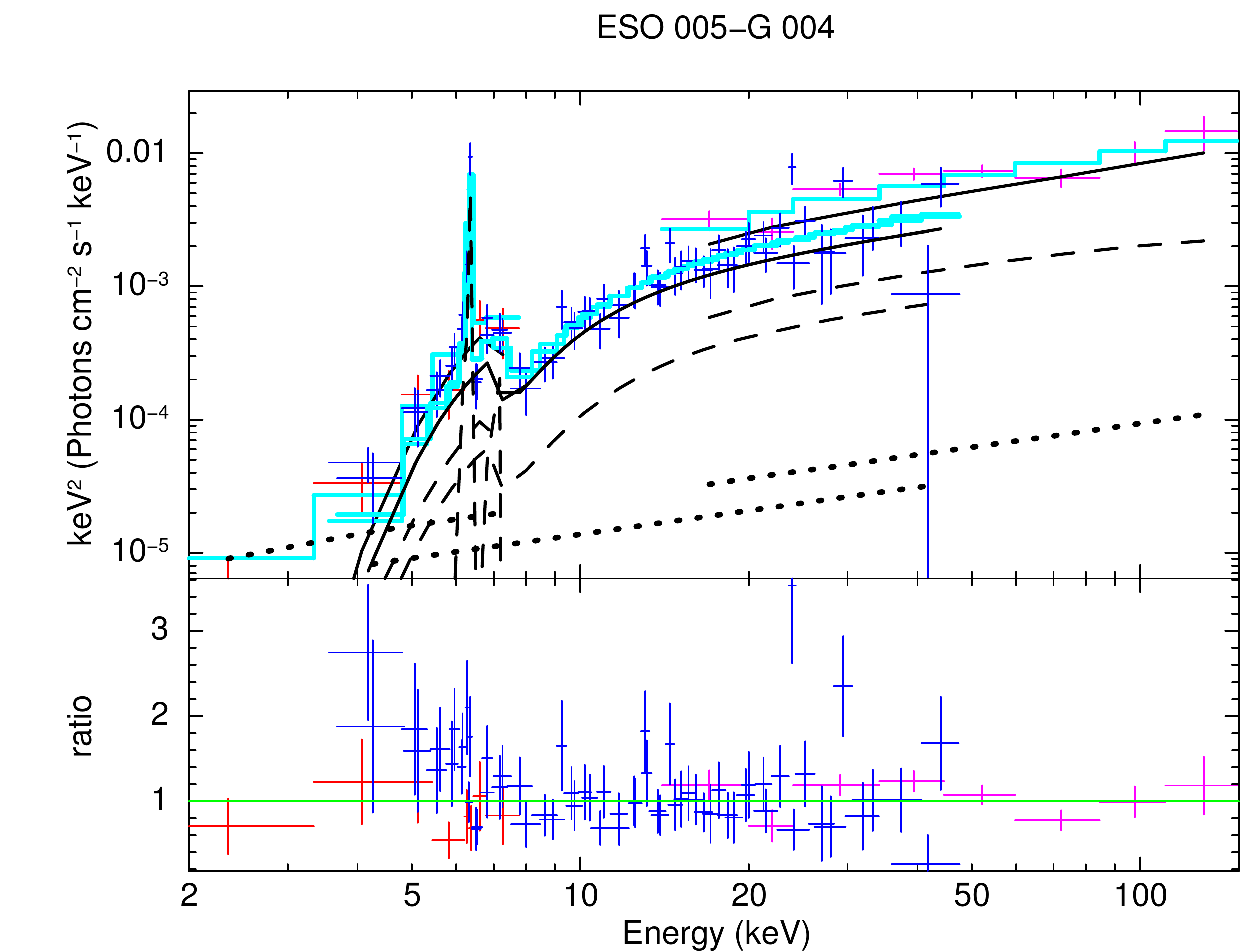}
  \end{minipage}
\begin{minipage}[b]{.5\textwidth}
  \centering
  \includegraphics[width=0.78\textwidth,angle=-90]{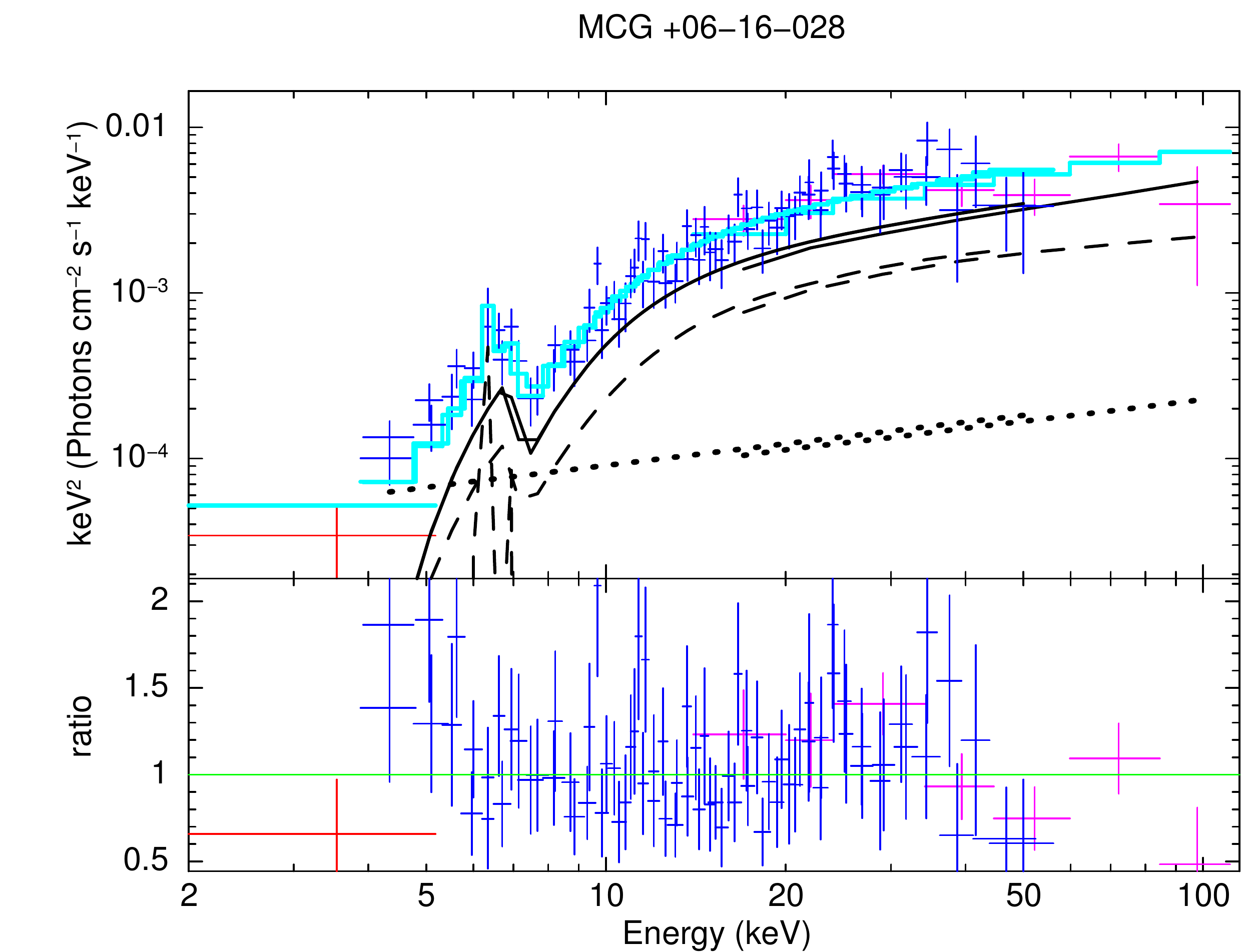}
  \end{minipage}
\begin{minipage}[b]{.5\textwidth}
  \centering
  \includegraphics[width=0.78\textwidth,angle=-90]{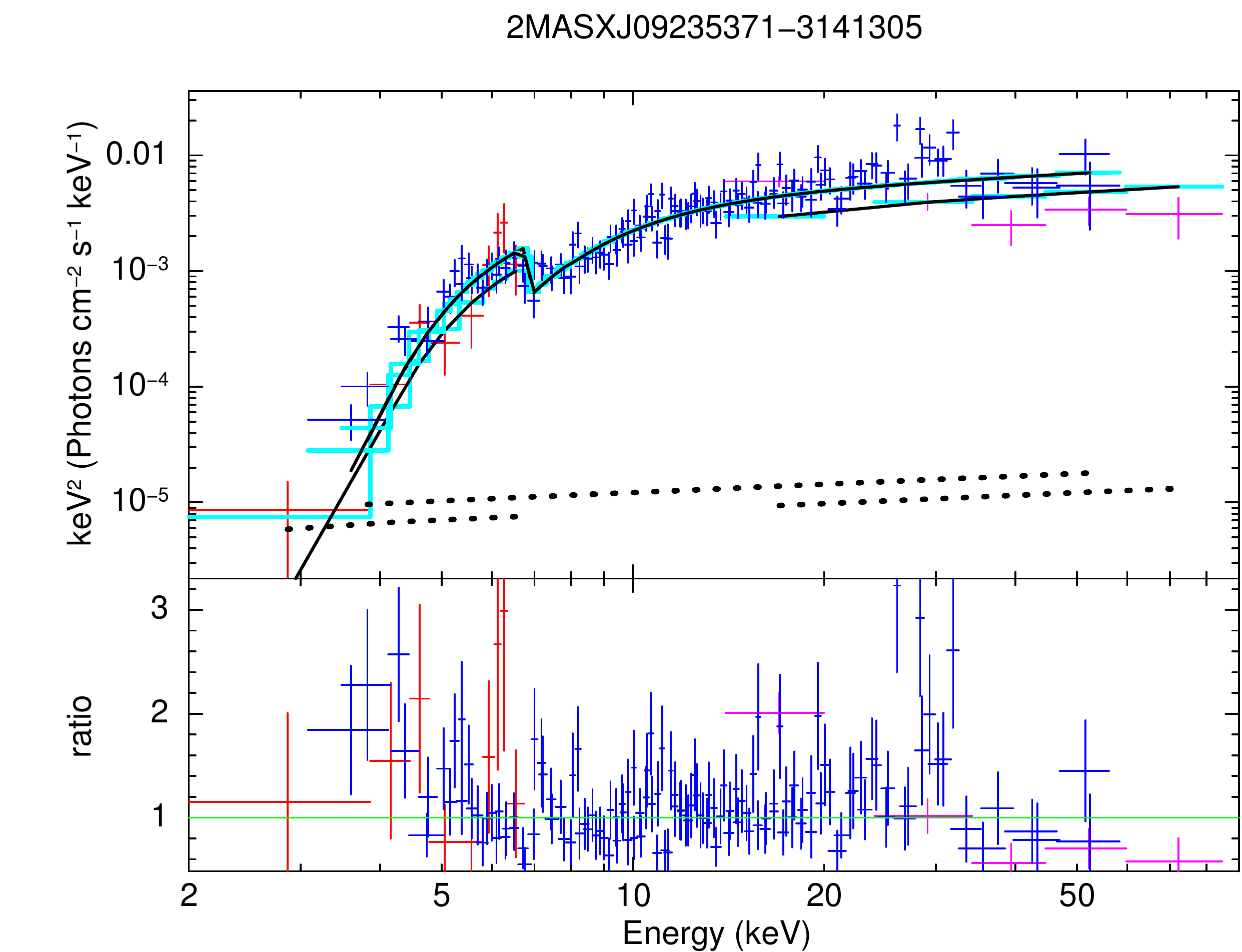}
  \end{minipage}  
\begin{minipage}[b]{.5\textwidth}
  \centering
  \includegraphics[width=0.78\textwidth,angle=-90]{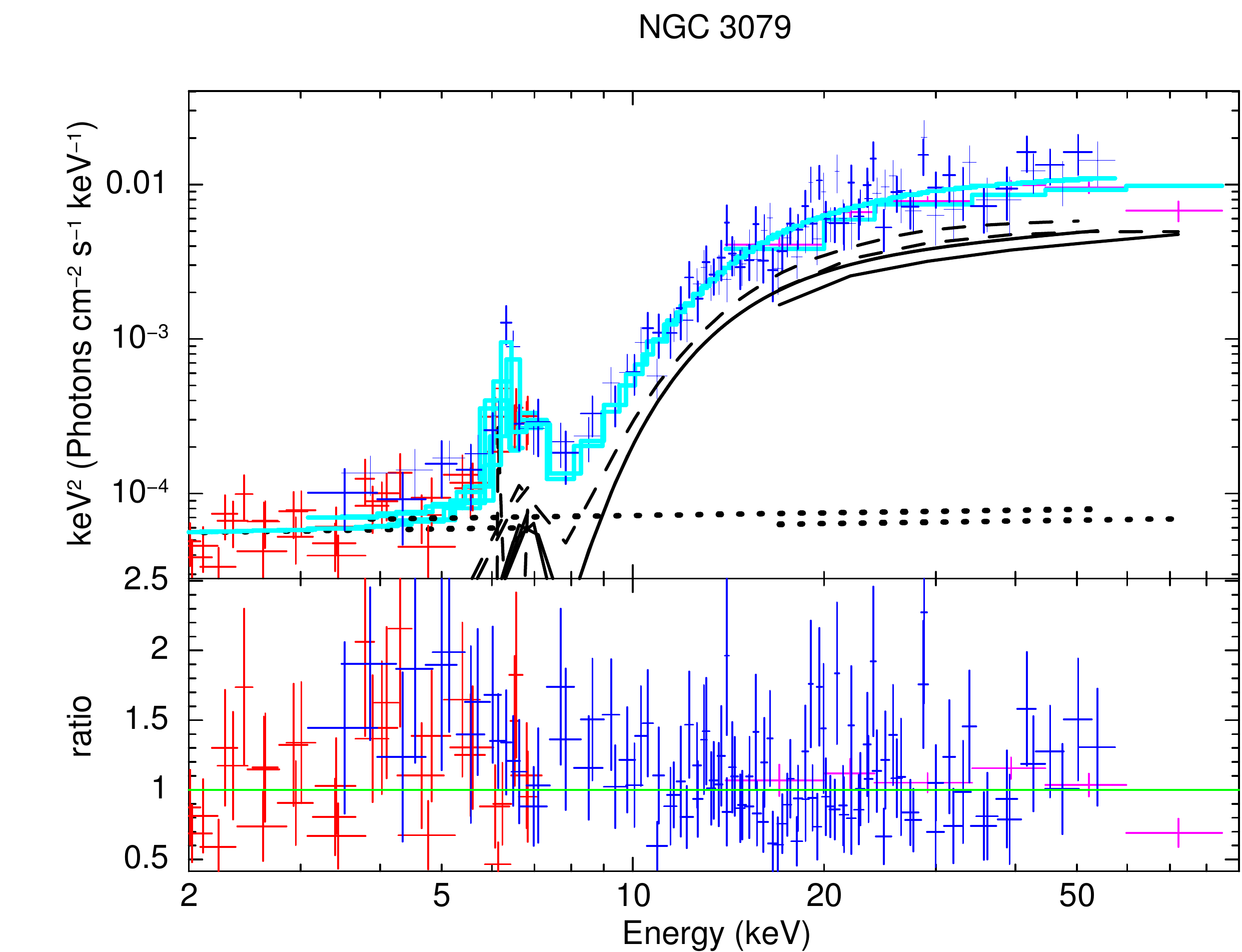}
  \end{minipage}
\caption{\normalsize Spectra (top panel) and data-to-model ratio (bottom) of the CT-AGN analyzed in this work. 2--10\,keV data are plotted in red, \nustar\ data in blue and \swi\ data in magenta. The best-fitting model is plotted as a cyan solid line, while the single \texttt{MyTorus} components are plotted as black solid (zeroth-order continuum) and dashed (reflected component and emission lines) lines. Finally, the main power law component scattered, rather than absorbed, by the torus is plotted as a black dotted line.} %\label{fig:spectra}
\end{figure*}

\begin{figure*}%\ContinuedFloat
\begin{minipage}[b]{.5\textwidth}
  \centering
  \includegraphics[width=0.78\textwidth,angle=-90]{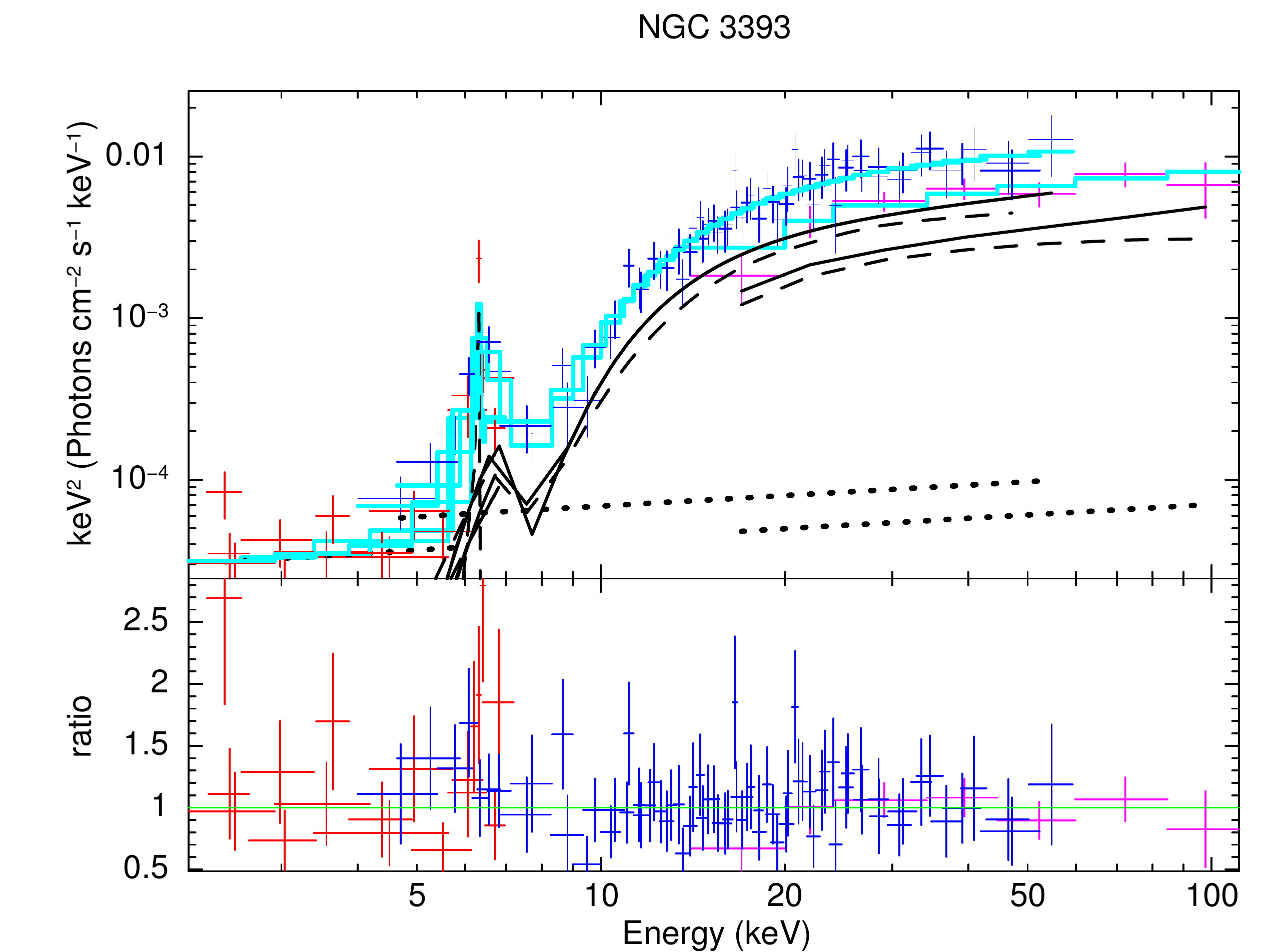}
  \end{minipage}
\begin{minipage}[b]{.5\textwidth}
  \centering
  \includegraphics[width=0.78\textwidth,angle=-90]{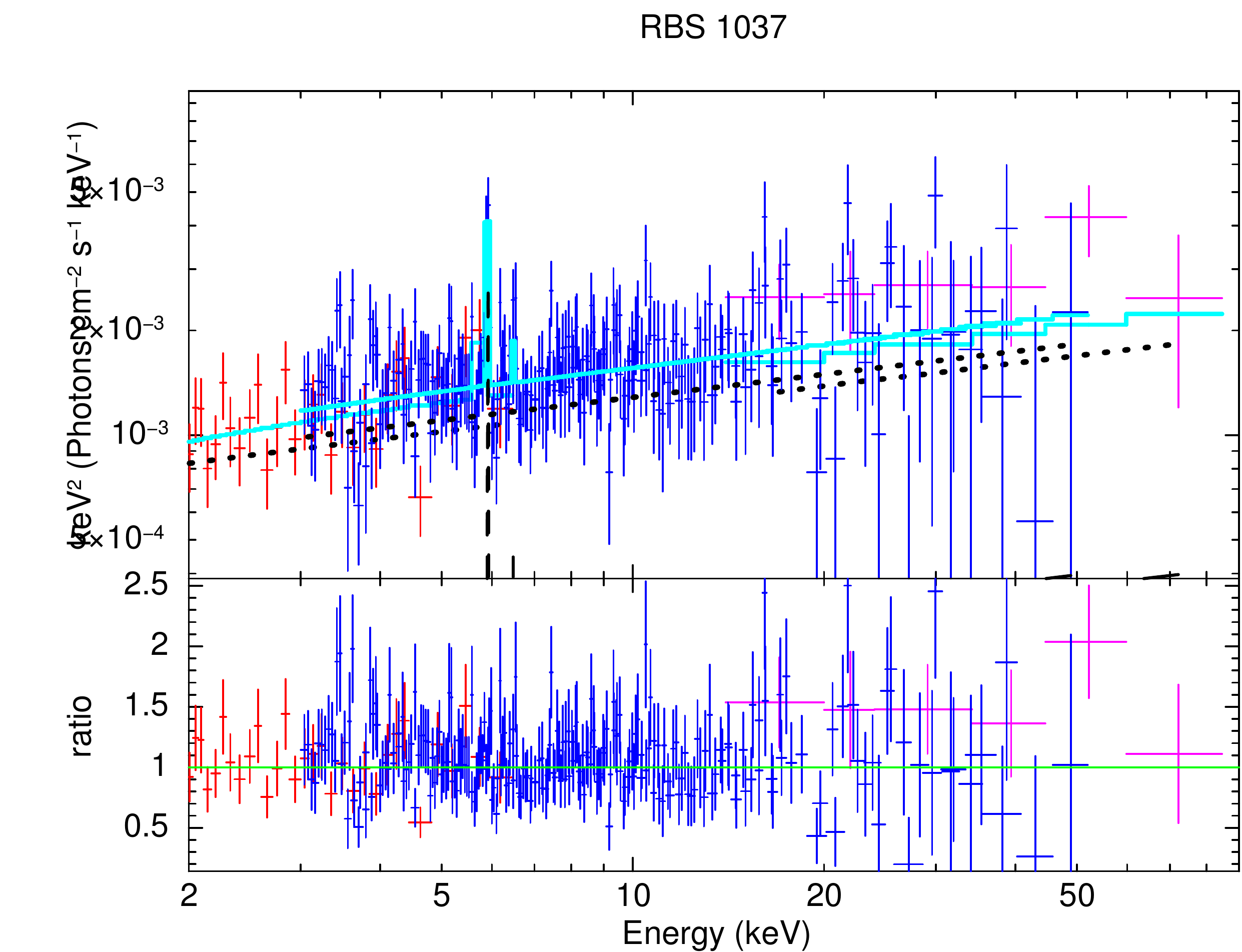}
  \end{minipage}
\begin{minipage}[b]{.5\textwidth}
  \centering
  \includegraphics[width=0.78\textwidth,angle=-90]{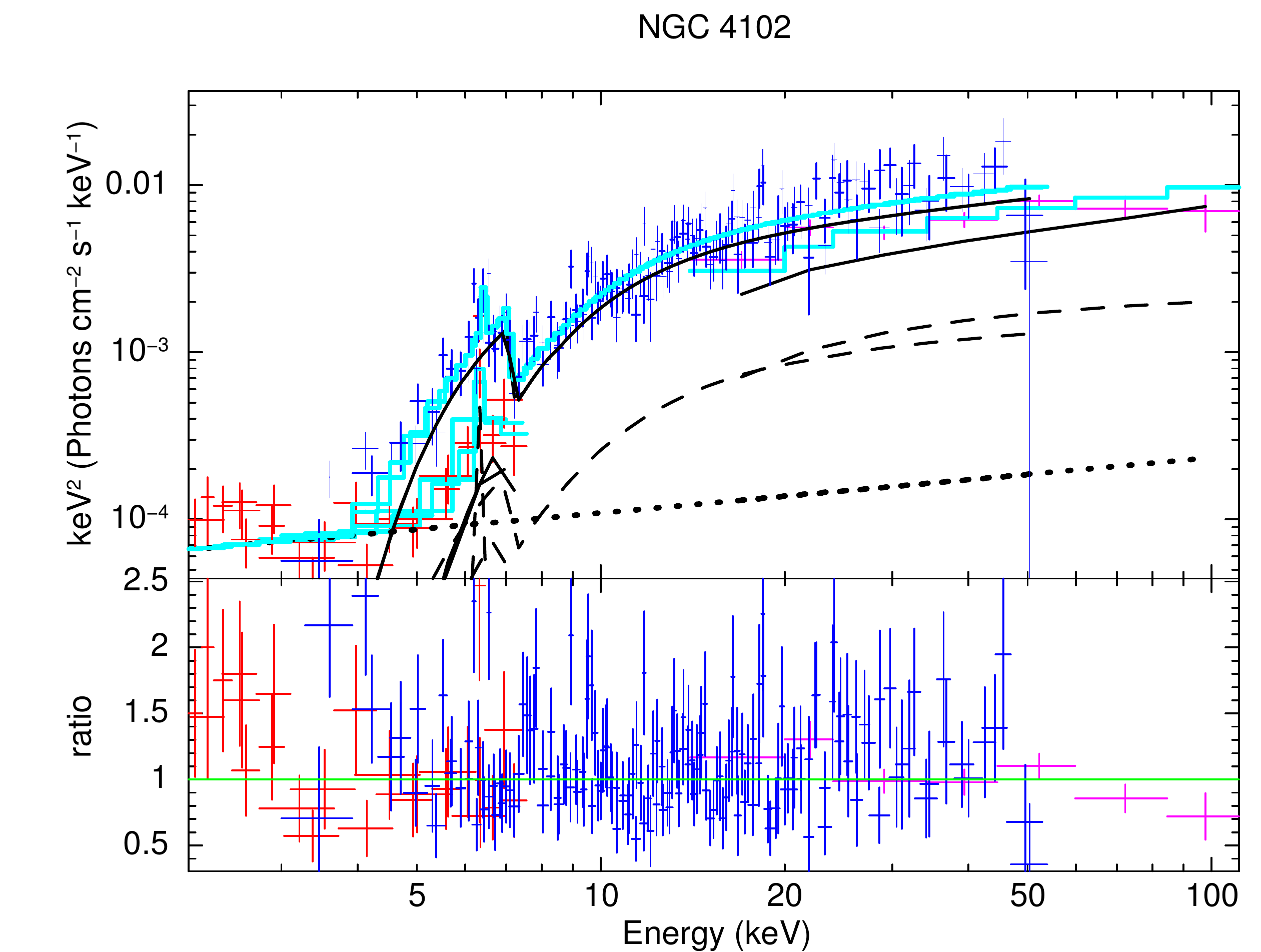}
  \end{minipage}
\begin{minipage}[b]{.5\textwidth}
  \centering
  \includegraphics[width=0.78\textwidth,angle=-90]{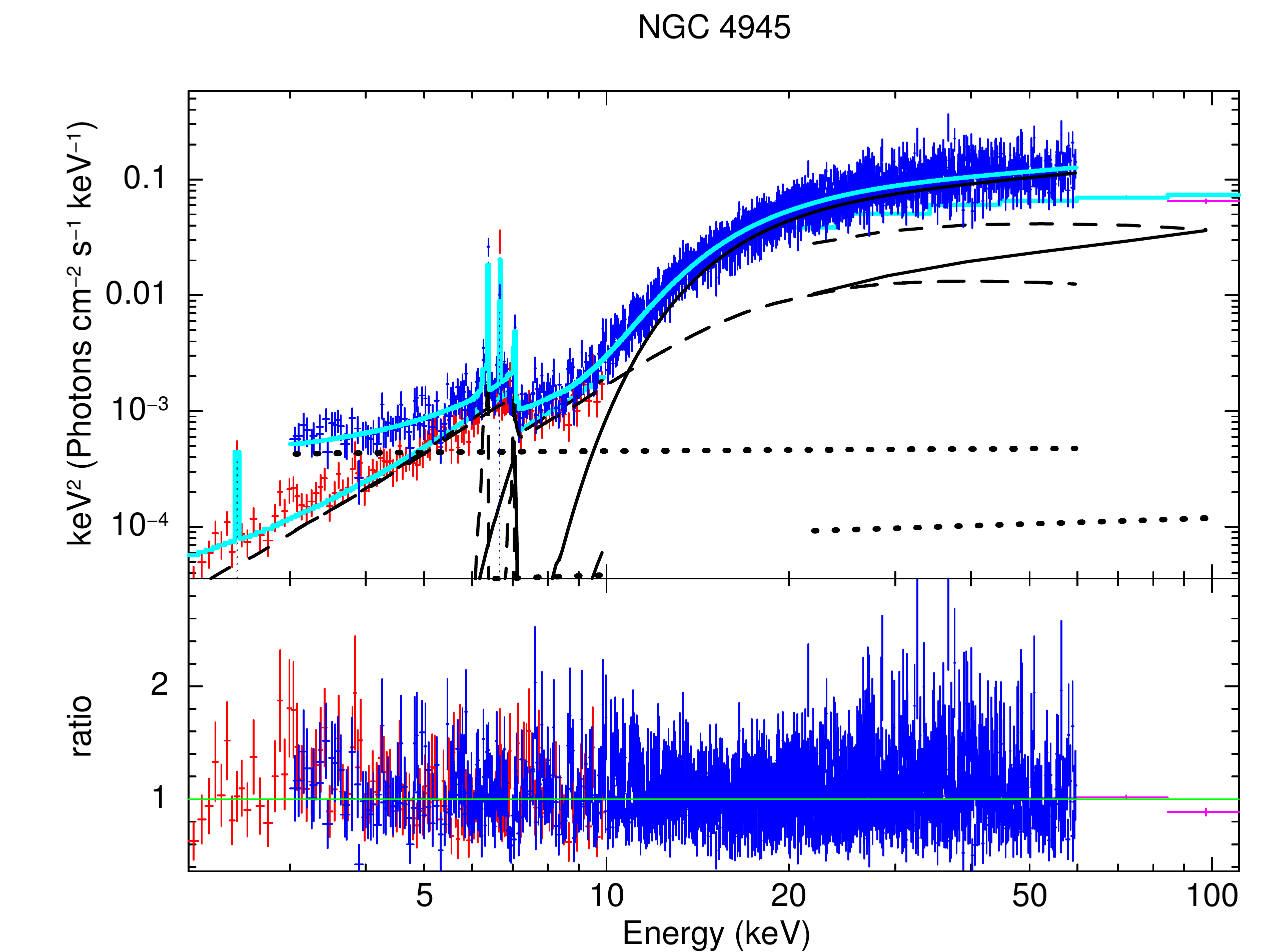}
  \end{minipage}
  \begin{minipage}[b]{.5\textwidth}
  \centering
  \includegraphics[width=0.78\textwidth,angle=-90]{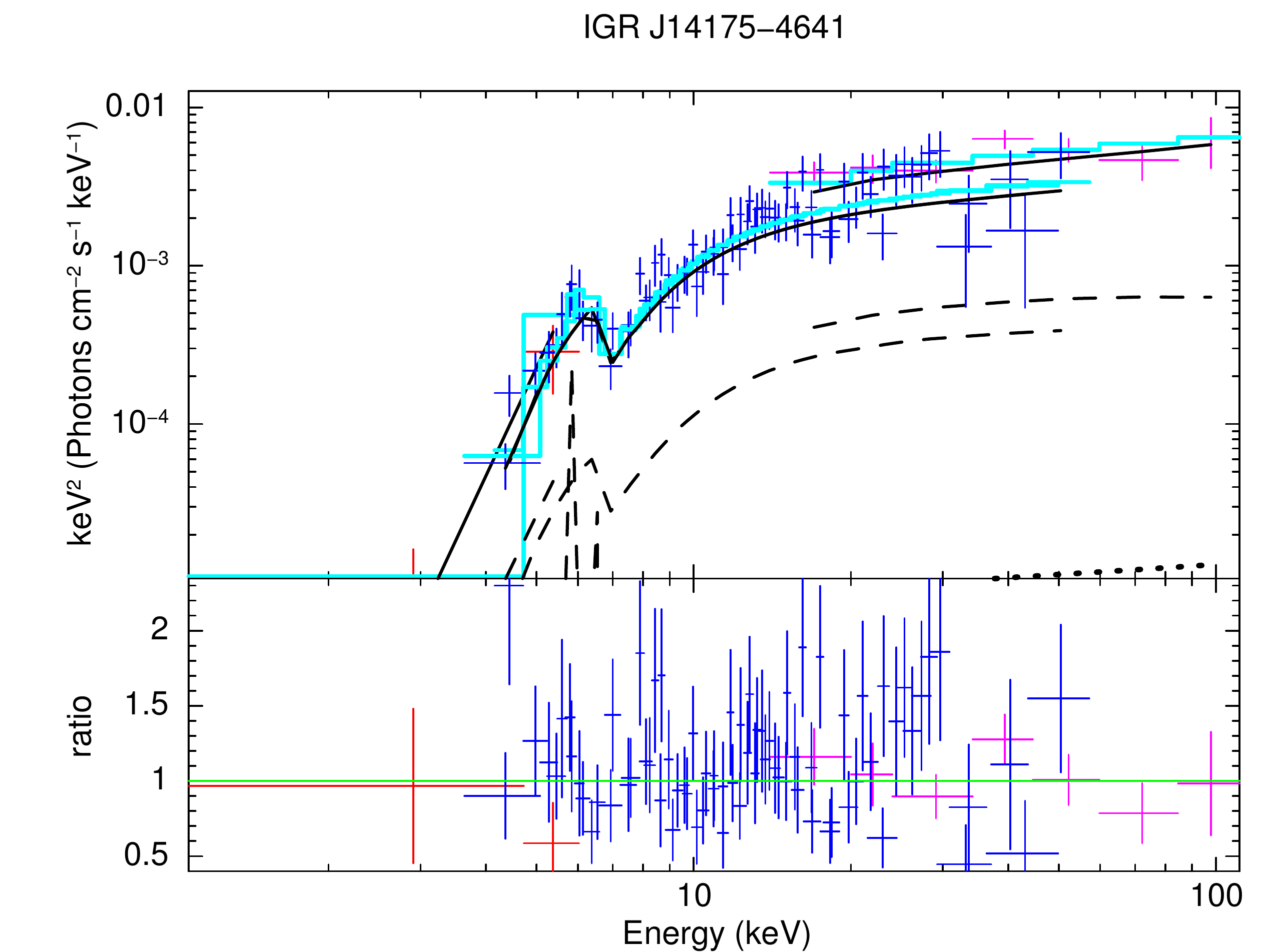}
  \end{minipage}
  \begin{minipage}[b]{.5\textwidth}
  \centering
  \includegraphics[width=0.78\textwidth,angle=-90]{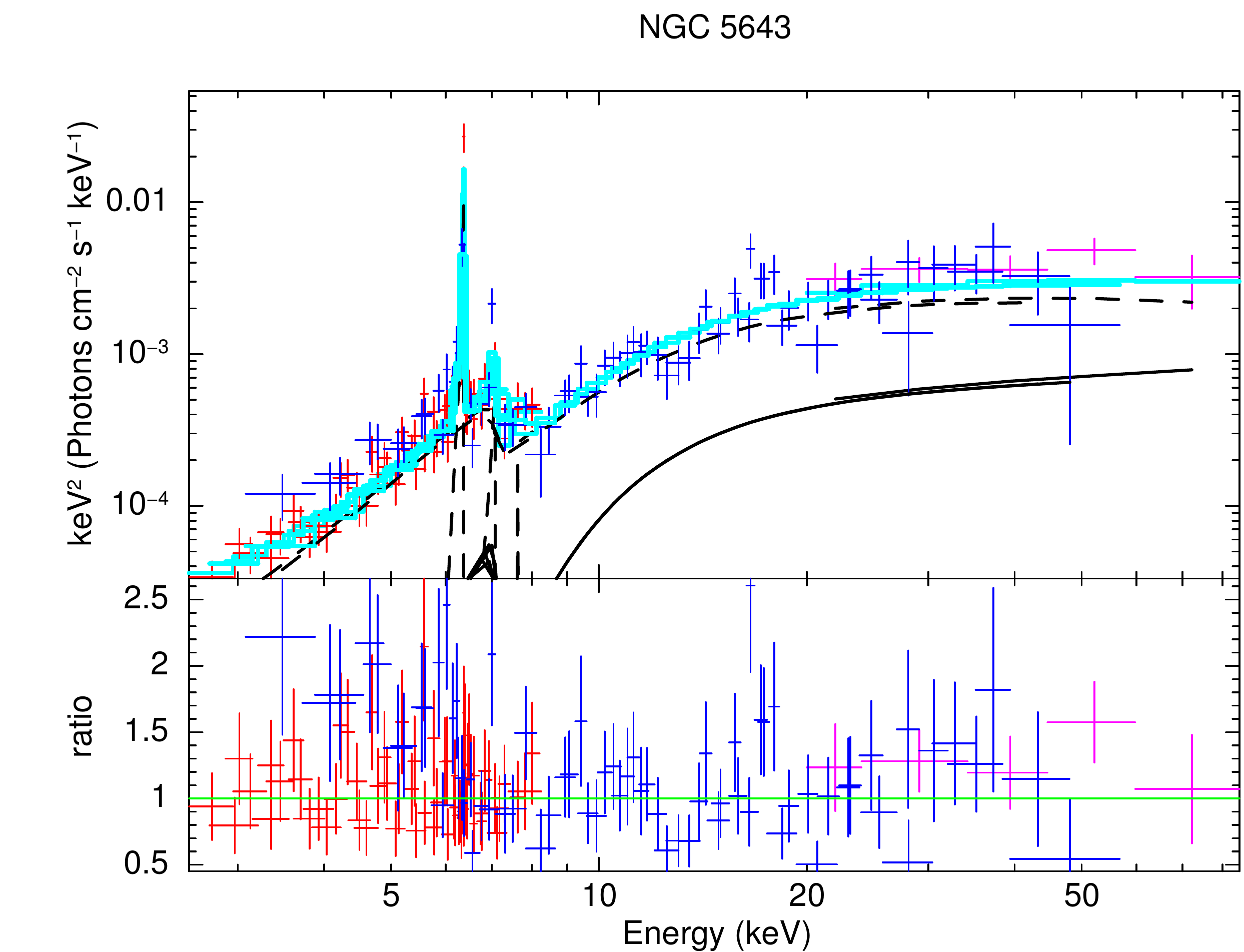}
  \end{minipage}
\caption{\normalsize Spectra (top panel) and data-to-model ratio (bottom) of the CT-AGN analyzed in this work. 2--10\,keV data are plotted in red, \nustar\ data in blue and \swi\ data in magenta. The best-fitting model is plotted as a cyan solid line, while the single \texttt{MyTorus} components are plotted as black solid (zeroth-order continuum) and dashed (reflected component and emission lines) lines. Finally, the main power law component scattered, rather than absorbed, by the torus is plotted as a black dotted line.}
\end{figure*}

\begin{figure*}%\ContinuedFloat
 \begin{minipage}[b]{.5\textwidth}
  \centering
  \includegraphics[width=0.78\textwidth,angle=-90]{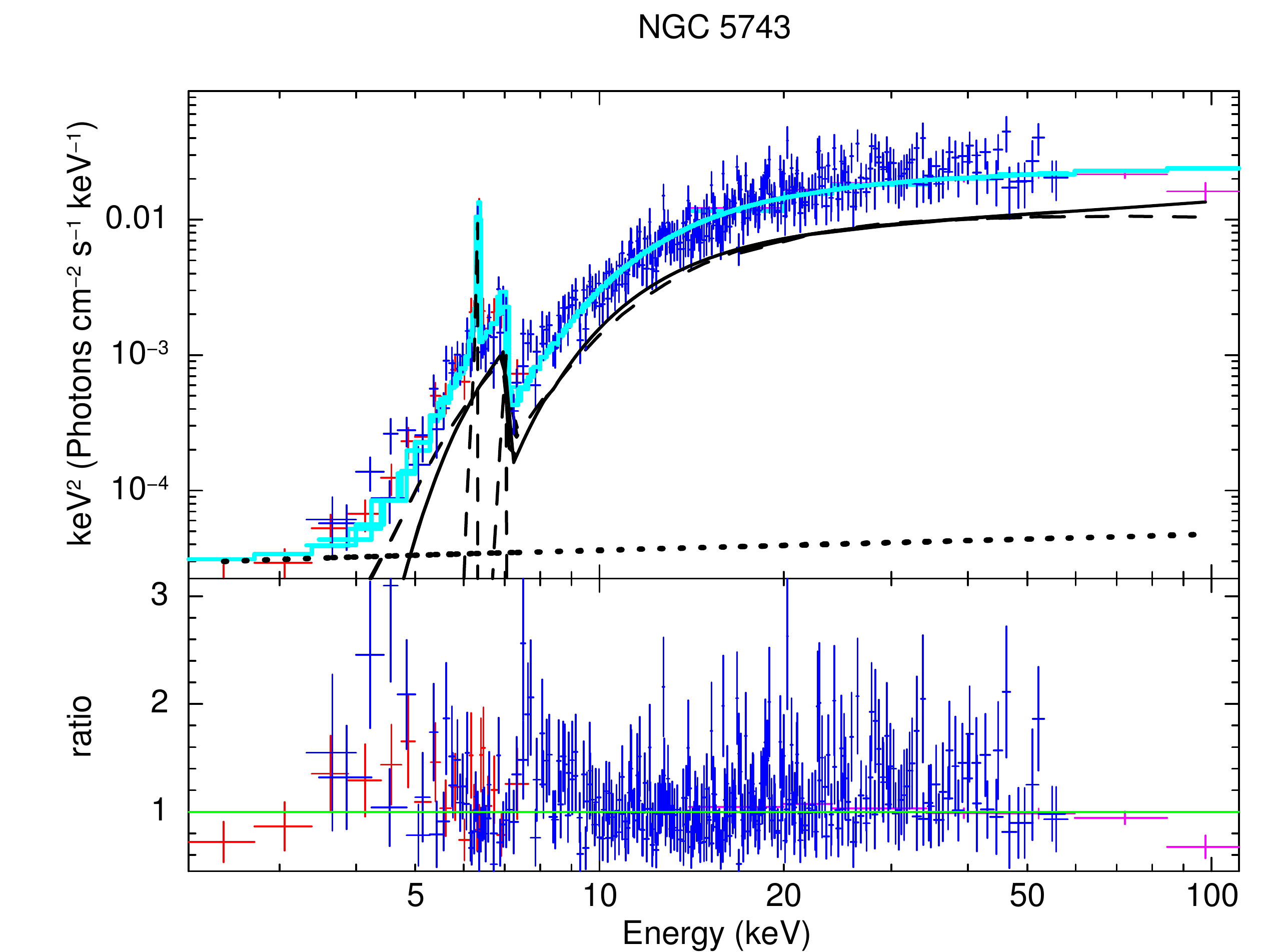}
  \end{minipage}
\begin{minipage}[b]{.5\textwidth}
  \centering
  \includegraphics[width=0.78\textwidth,angle=-90]{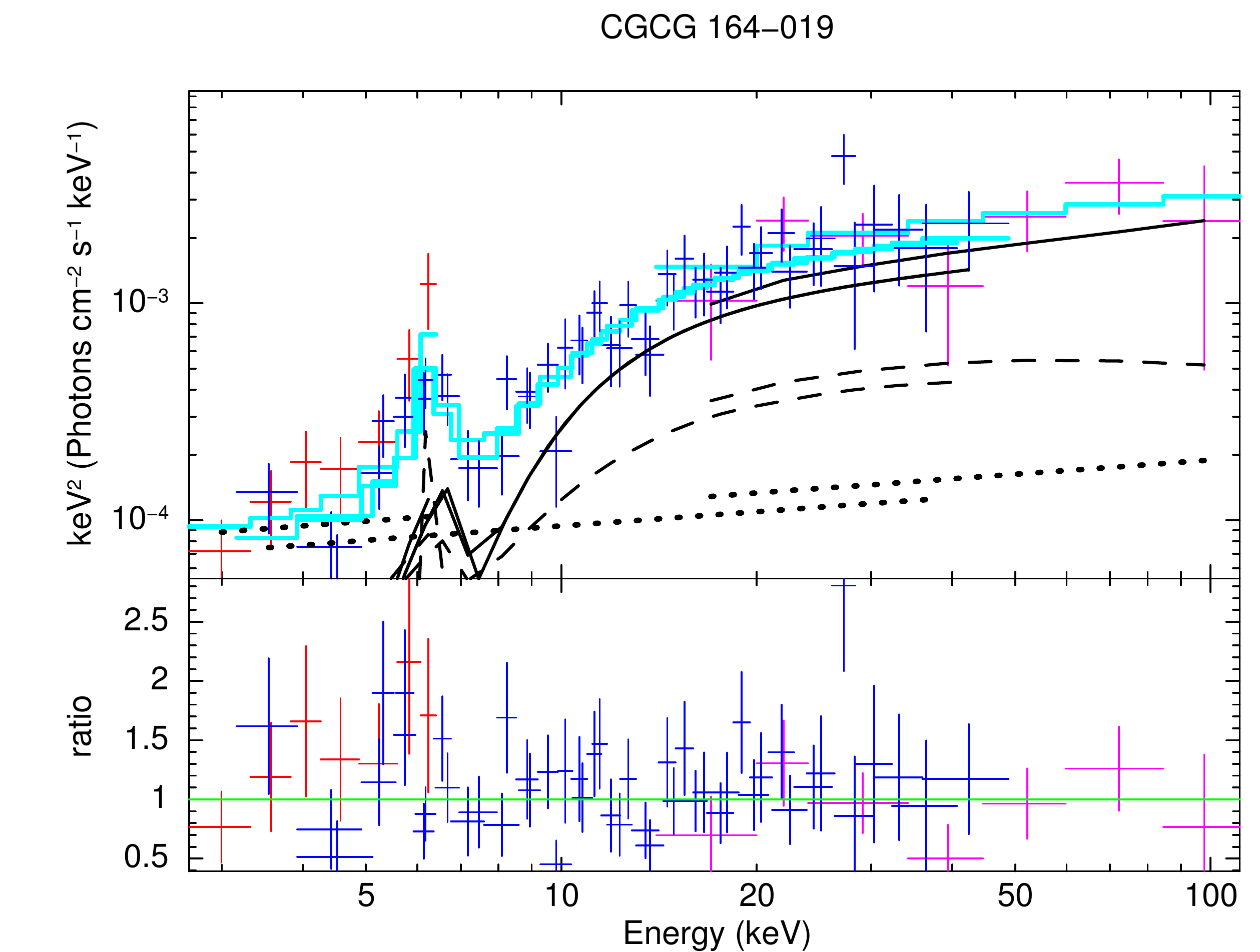}
  \end{minipage}
\begin{minipage}[b]{.5\textwidth}
  \centering
  \includegraphics[width=0.78\textwidth,angle=-90]{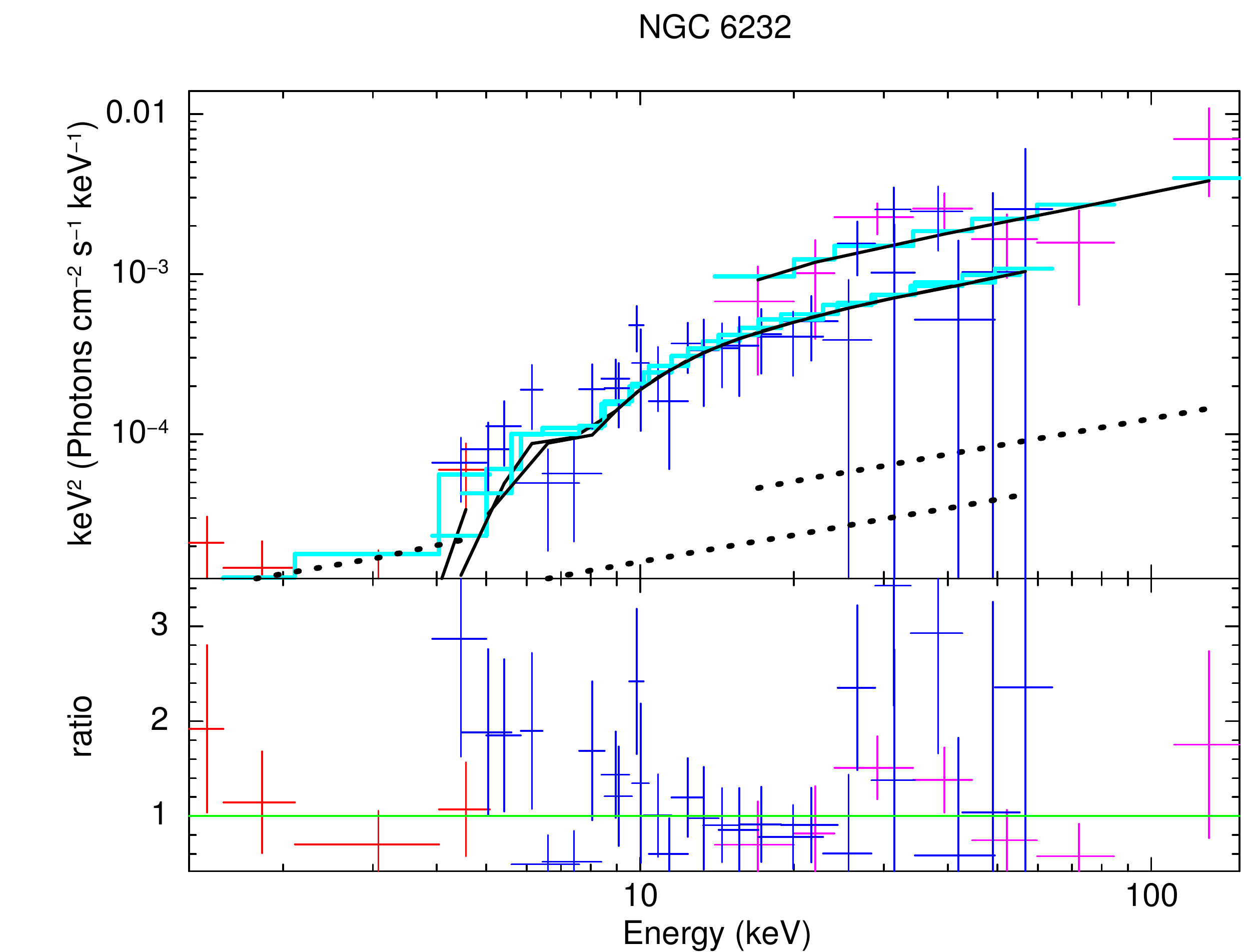}
  \end{minipage}
\begin{minipage}[b]{.5\textwidth}
  \centering
  \includegraphics[width=0.78\textwidth,angle=-90]{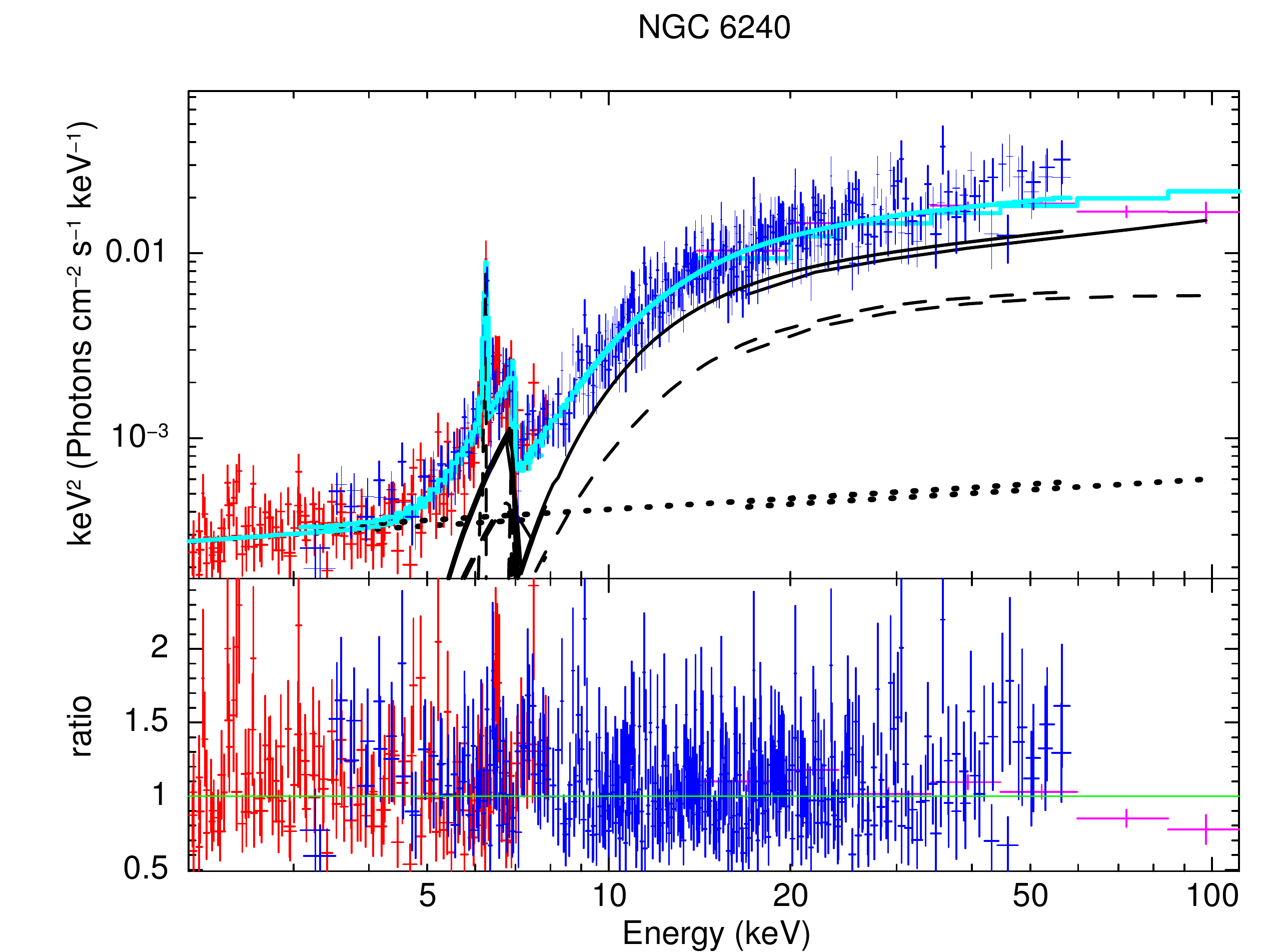}
  \end{minipage}
  \begin{minipage}[b]{.5\textwidth}
  \centering
  \includegraphics[width=0.78\textwidth,angle=-90]{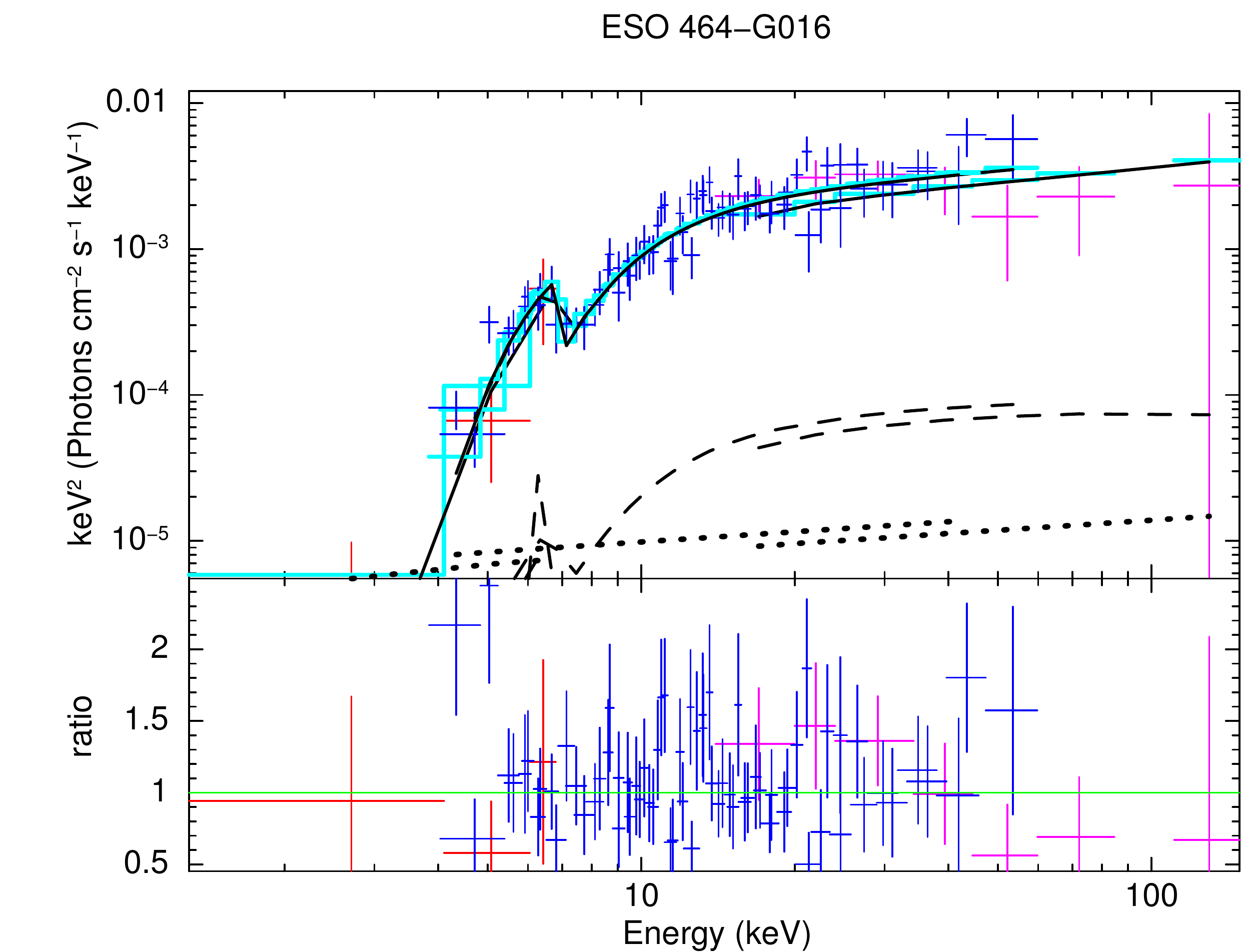}
  \end{minipage}
\begin{minipage}[b]{.5\textwidth}
  \centering
  \includegraphics[width=0.78\textwidth,angle=-90]{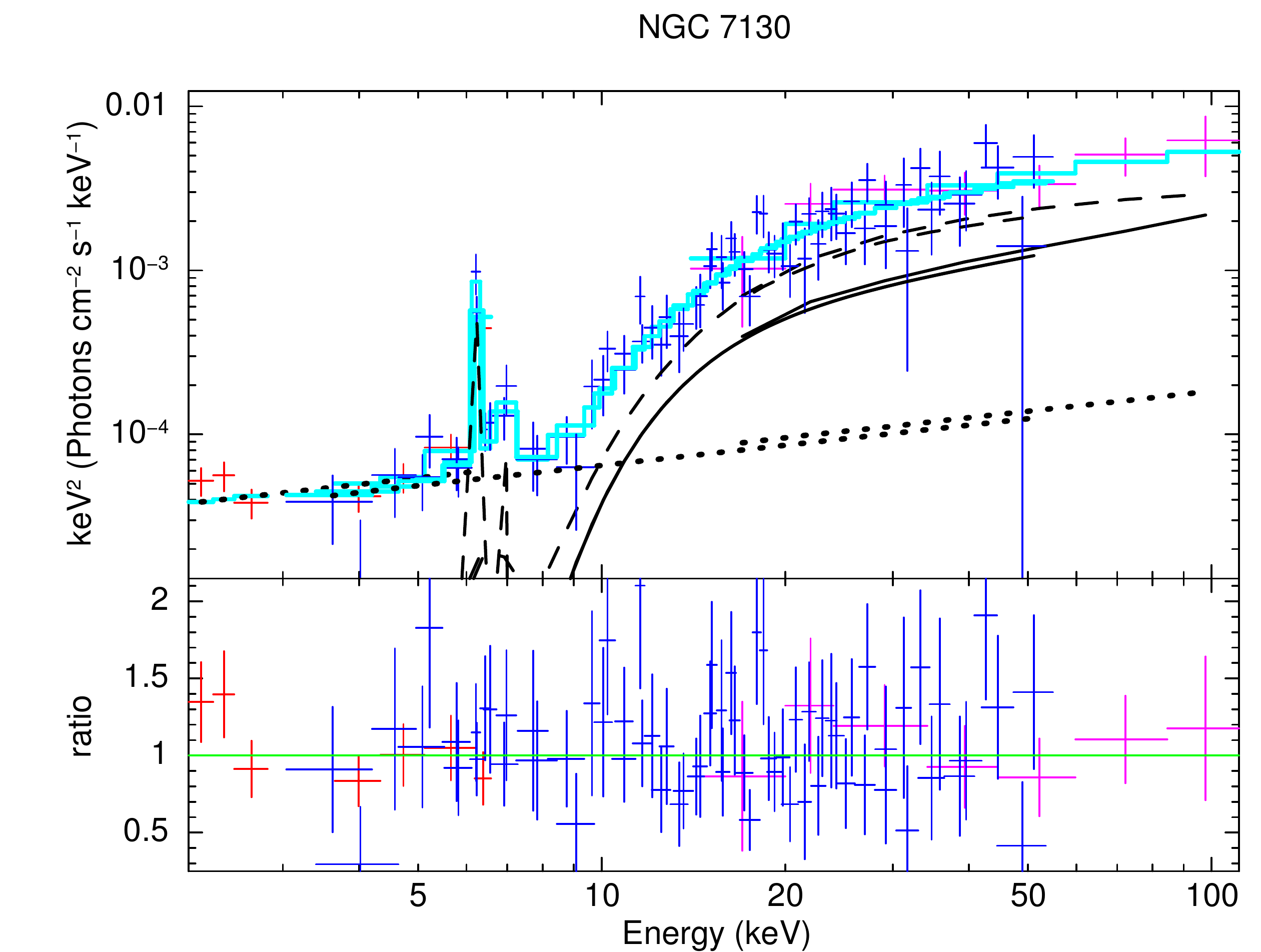}
  \end{minipage}
\caption{\normalsize Spectra (top panel) and data-to-model ratio (bottom) of the CT-AGN analyzed in this work. 2--10\,keV data are plotted in red, \nustar\ data in blue and \swi\ data in magenta. The best-fitting model is plotted as a cyan solid line, while the single \texttt{MyTorus} components are plotted as black solid (zeroth-order continuum) and dashed (reflected component and emission lines) lines. Finally, the main power law component scattered, rather than absorbed, by the torus is plotted as a black dotted line.}
\end{figure*}

\begin{figure*}%\ContinuedFloat
  \begin{minipage}[b]{.5\textwidth}
  \centering
  \includegraphics[width=0.78\textwidth,angle=-90]{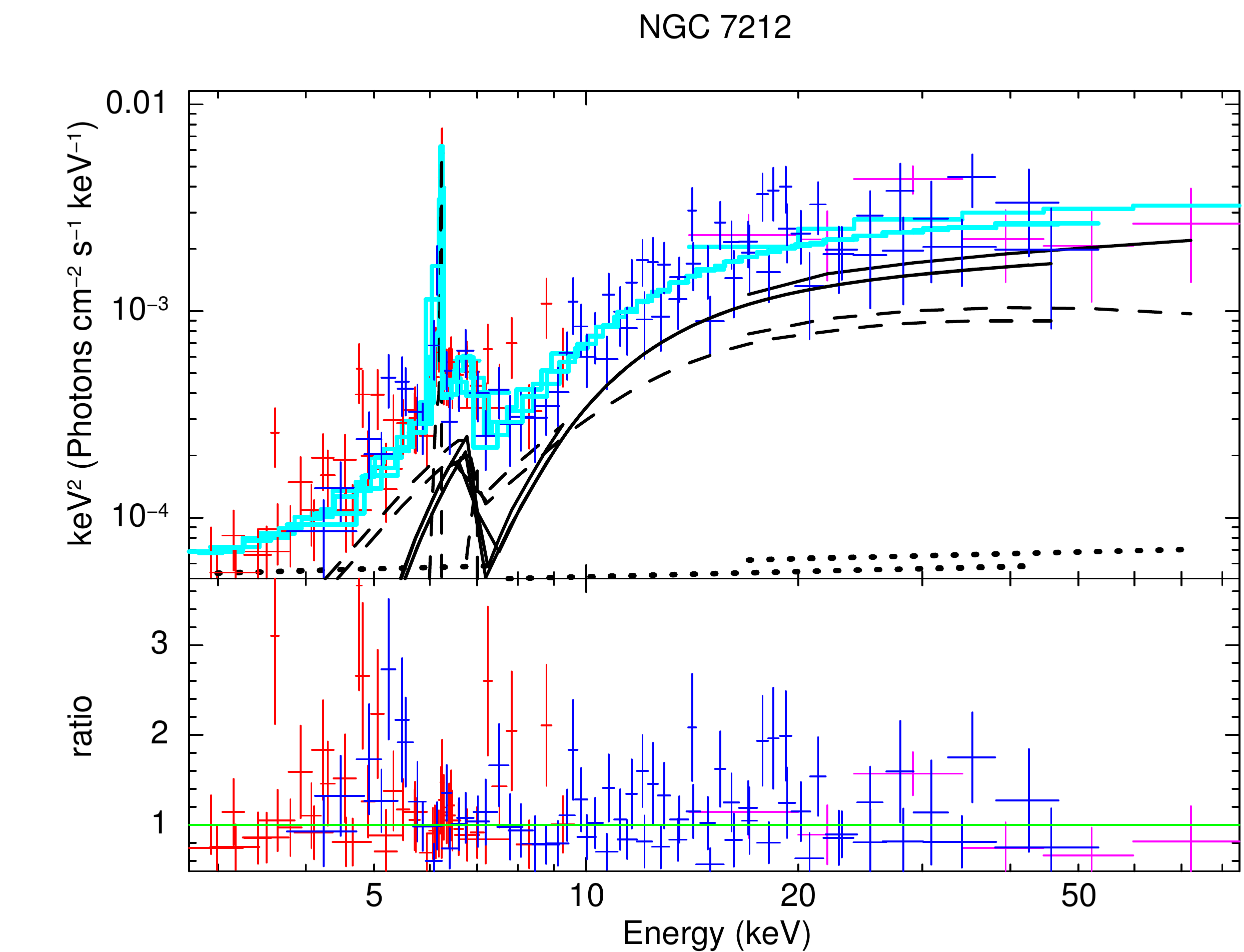}
   \end{minipage}
\begin{minipage}[b]{.5\textwidth}
  \centering
  \includegraphics[width=0.78\textwidth,angle=-90]{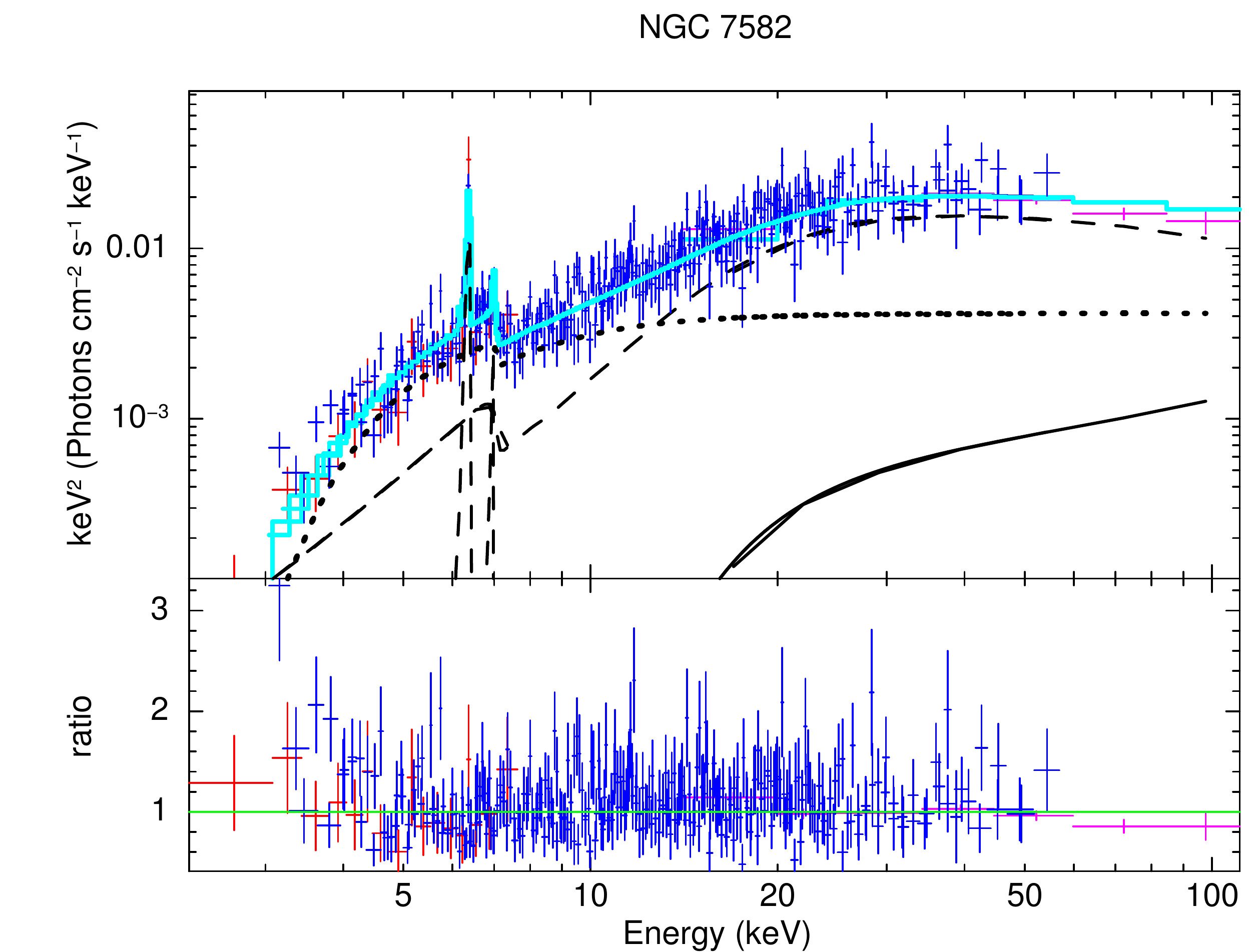}
  \end{minipage}
  \caption{\normalsize Spectra (top panel) and data-to-model ratio (bottom) of the CT-AGN analyzed in this work. 2--10\,keV data are plotted in red, \nustar\ data in blue and \swi\ data in magenta. The best-fitting model is plotted as a cyan solid line, while the single \texttt{MyTorus} components are plotted as black solid (zeroth-order continuum) and dashed (reflected component and emission lines) lines. Finally, the main power law component scattered, rather than absorbed, by the torus is plotted as a black dotted line.} \label{fig:spectra_last}
\end{figure*}
%\clearpage

\newpage
\section{B. Sources classified as non-CT-AGN based on their 0.5--10\,keV and \swi\ information only}\label{sec:app_no-ct}
The first part of the spectral analysis performed in this work makes use of the combined 2--10\,keV and \swi\ spectra, without the \nustar\ data contribution. The results of the fit to these data, and specifically the measurements of key parameters such as the photon index $\Gamma$ and the intrinsic absorption $N_{\rm H, z}$, have then been compared with those obtained adding the \nustar\ information to the fit.

\begin{figure*}
  \begin{minipage}[b]{.5\textwidth}
  \centering
  \includegraphics[width=1.\textwidth]{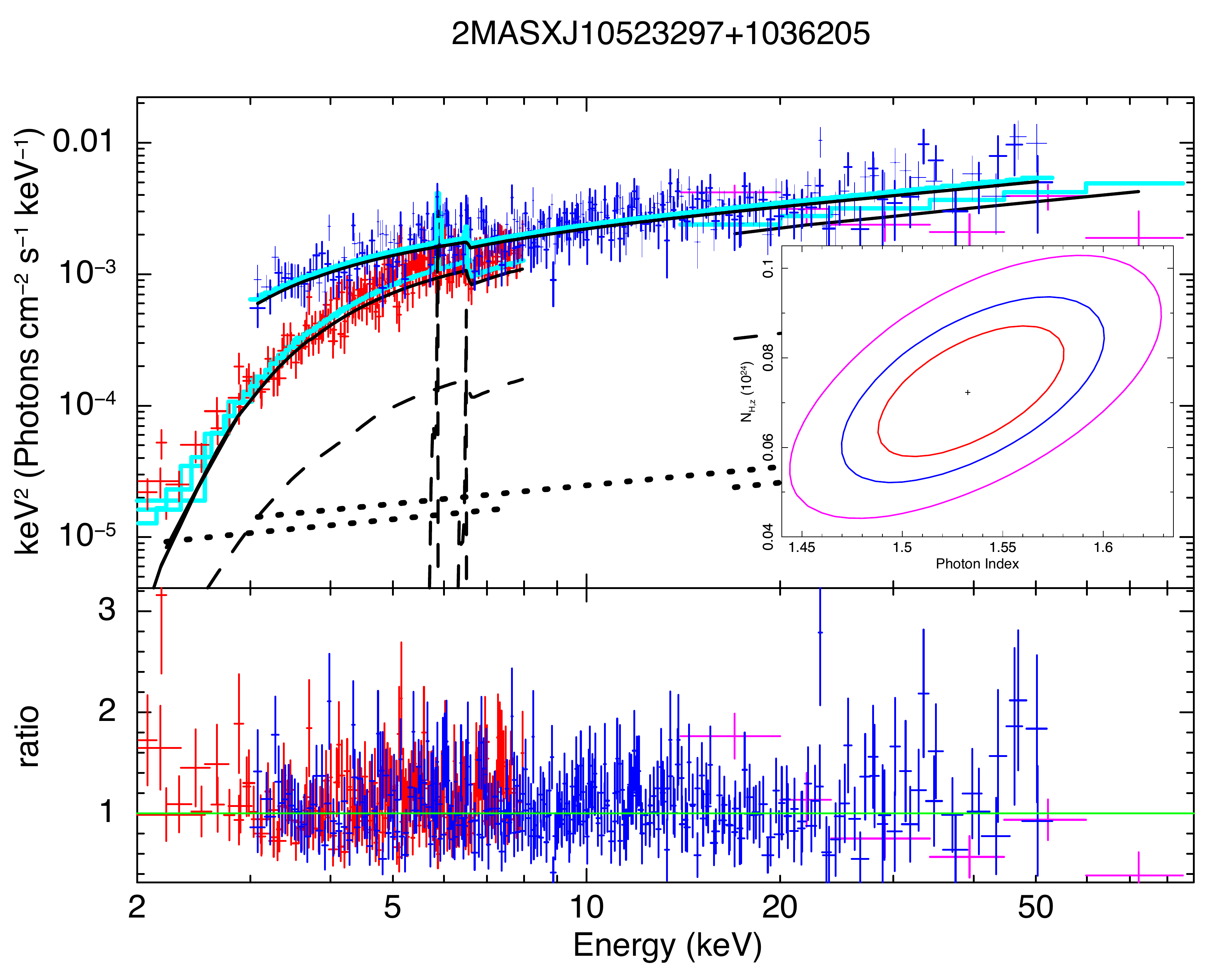}
   \end{minipage}
\begin{minipage}[b]{.5\textwidth}
  \centering
   \includegraphics[width=1.\textwidth]{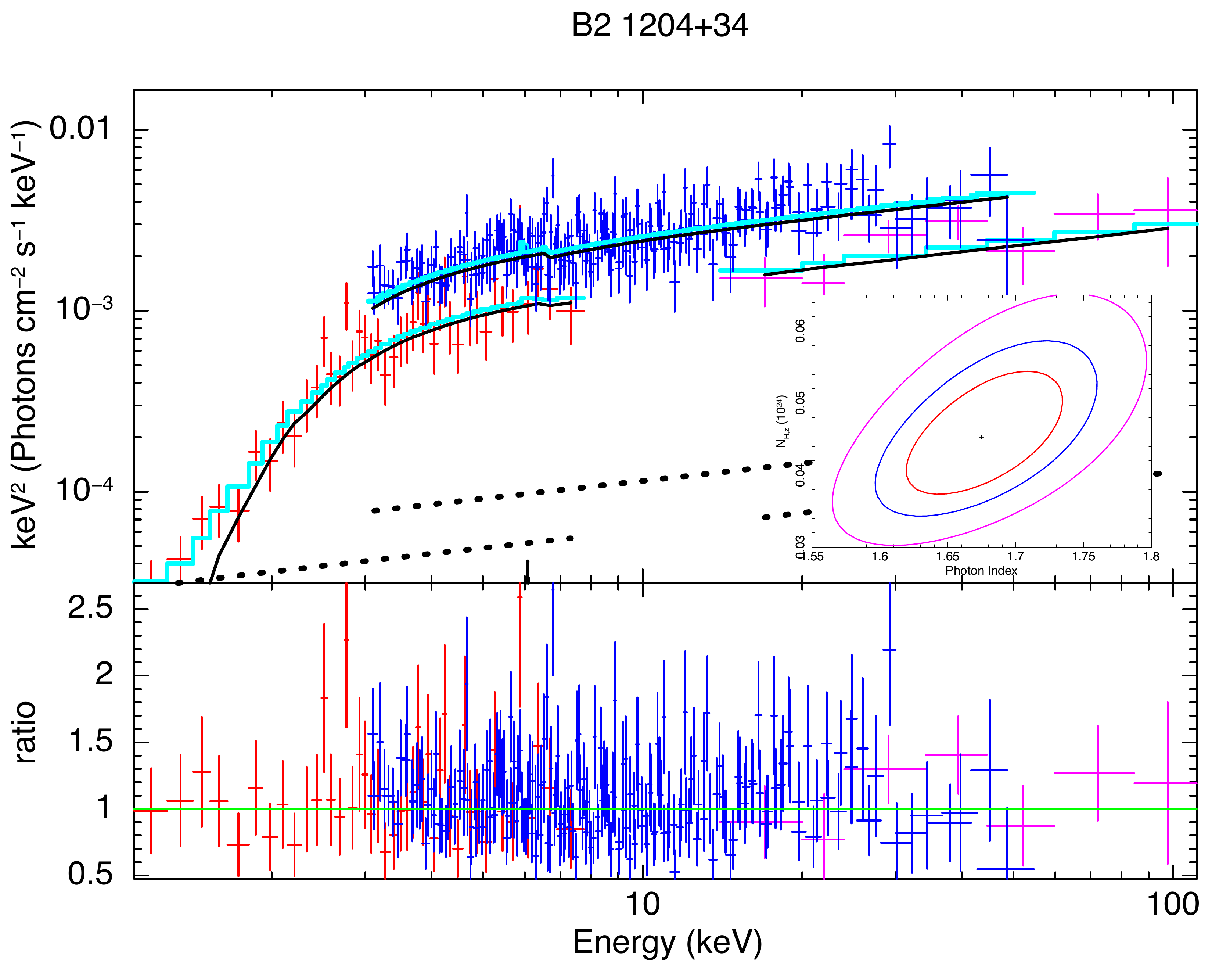}
  \end{minipage}
\begin{minipage}[b]{.5\textwidth}
  \centering
   \includegraphics[width=1.\textwidth]{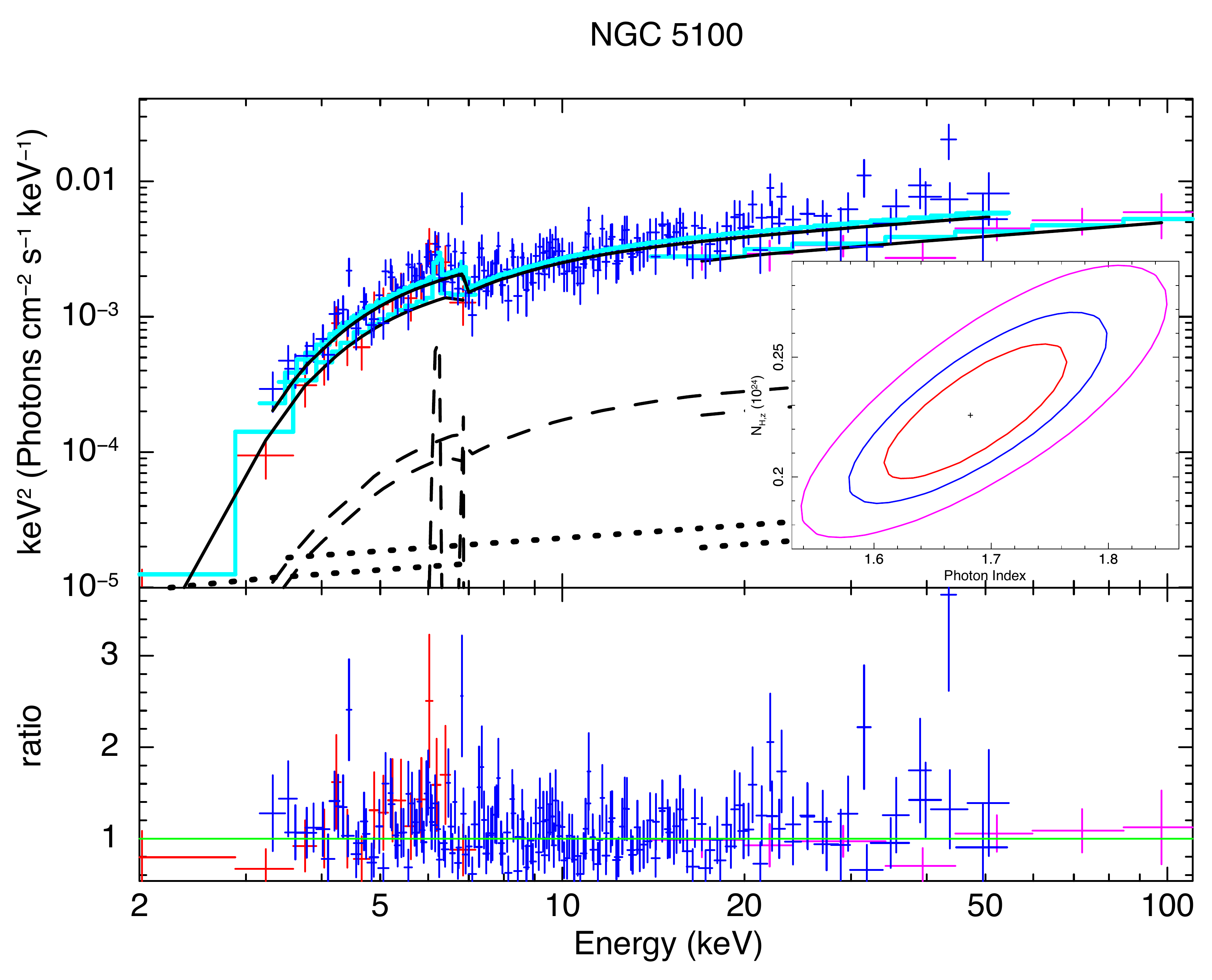}
  \end{minipage}
 \begin{minipage}[b]{.5\textwidth}
  \centering
   \includegraphics[width=1.00\textwidth]{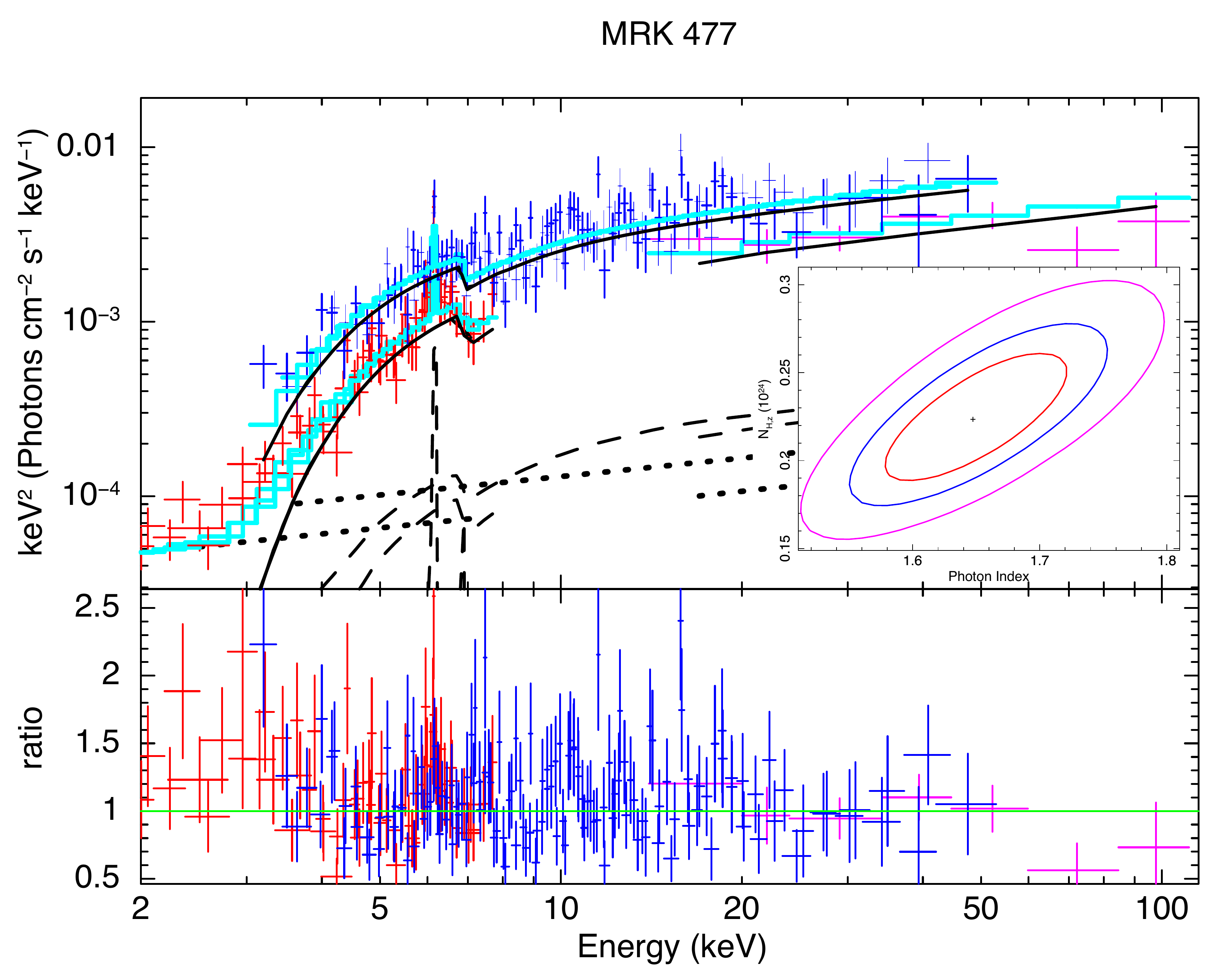}
  \end{minipage}
  \caption{\normalsize Spectra (top panel) and data-to-model ratio (bottom) of 2MASX J10523297+1036205, B2 1204+34, NGC 5100 and Mrk 477, the four sources for which we did not find CT obscuration both without and with the addition of the \nustar\ data. 2--10\,keV data are plotted in red, \nustar\ data in blue and \swi\ data in magenta. The best-fitting model is plotted as a cyan solid line, while the single \texttt{MyTorus} components are plotted as black solid (zeroth-order continuum) and dashed (emission lines and reflected component) lines. Finally, the main power law component scattered, rather than absorbed, by the torus is plotted as a black dotted line. We show in the inset the confidence contours at 68, 90 and 99\% confidence level for $\Gamma$ and $N_{\rm H, z}$ (in 10$^{22}$ cm$^{-2}$) units.} \label{fig:spectra_noct}
\end{figure*}

However, during this analysis we found that four out of 30 objects (namely 2MASXJ10523297+1036205, B2 1204+34, NGC 5100 and Mrk 477) have best-fit parameters that are not consistent with a CT-AGN origin at a $>$3$\sigma$ level; for all these objects, the non-CT identity of these sources has been confirmed adding the \nustar\ spectra to the fit. We present the spectra of these objects in Figure \ref{fig:spectra_noct}: in the inset, the confidence contours for $\Gamma$ and $N_{\rm H, z}$ are also shown. In Table \ref{tab:results_no-ct} we report the best-fit results, both without and with the \nustar\ spectra. For two out of four objects (namely, 2MASXJ10523297+1036205 and Mrk 477) we find that the fit significantly improves leaving $N_{\rm H, z, 2-10\,keV}$ free to vary from $N_{\rm H, z, NuS}$, therefore suggesting variability in the obscuring material surrounding the SMBH. Nonetheless, both $N_{\rm H, z}$ values are well below the $N_{\rm H, z}$=10$^{24}$ cm$^{-2}$ threshold.

We point out that the CT origin of Mrk 477 was first reported in \citet{bassani99} and then again in \citet{shu07}, based on these previous results. However, the \citet{bassani99} prediction is not derived from an actual X-ray spectral fitting result, which lead to a best-fit absorption value $N_{\rm H, z}$=9$_{-9}^{+12}$$\times$10$^{22}$ cm$^{-2}$, with an unphysical photon index $\Gamma$=0.2$_{-0.7}^{+0.8}$. Instead, they inferred the CT-AGN condition of the source from its location in the X-ray to optical flux versus iron K$\alpha$ line EW diagram. Therefore, it is not fully unexpected that our analysis, which makes use of a better dataset, leads to a different result; moreover, \citet{bassani99} already point out that based on their diagnostics the CT status of Mrk 477 is one of the less secure in their sample.

The other three objects have been classified as CT by \citet{vasudevan13}: for two of these objects (B2 1204+34, NGC 5100), they analyzed the same \xrt\ observations we used in our analysis, while for 2MASXJ10523297+1036205 we used a longer \xmm\ observation, with better statistics, which can contribute to the discrepancy between their measurements and ours. It is also worth mentioning that our spectral fitting has been performed using models specifically designed to treat obscured AGN spectra (see Section \ref{sec:results}), while \citet{vasudevan13} used more general XSPEC models such as \texttt{zpcfabs}, a partial covering fraction absorption model. 

Finally, we point out that 2MASXJ10523297+1036205 and Mrk 477 are optically classified Seyfert 1 galaxies: while a fraction of Seyfert 1 galaxies are expected to be obscured \citep[see, e.g.,][]{marchesi16c}, the existence of CT Seyfert 1 galaxies is yet to be confirmed. For example, none of the 55 \swi-selected CT-AGN reported in \citet{ricci15} is classified as a Seyfert 1 source.

\begingroup
\renewcommand*{\arraystretch}{1.5}
\begin{table*}
\centering
\scalebox{0.85}{
\begin{tabular}{ccc|cccc|cccc}
\hline
\hline
&  & & \multicolumn{4}{c|}{Without \nustar} & \multicolumn{4}{c}{With \nustar}\\ 
Source & $N_{\rm H, gal}$ & $\theta_{\rm obs}$ & $N_{\rm H, z}$ & $\Gamma$ & EW & $\chi^2$/DOF & $N_{\rm H, z}$ & $\Gamma$ & EW & $\chi^2$/DOF\\% & Model \\ % & Facility\\
& 10$^{20}$ cm$^{-2}$ & \degree &10$^{22}$ cm$^{-2}$ & & keV & & 10$^{22}$ cm$^{-2}$ & & keV &\\
\hline
2MASXJ10523297+1036205 & 2.3 & 90 & 18.2$_{-1.5}^{+1.6}$ &  1.56$_{-0.07}^{+0.08}$ & 0.24$_{-0.10}^{+0.12}$ &  237.0/205 &  5.2$_{-1.6}^{+1.6}$$^*$ &  1.51$_{-0.05}^{+0.05}$ & 0.11$_{-0.04}^{+0.05}$ &  536.0/513 \\ % & T \\ % & XMM\\  90
B2 1204+34 & 1.4  & 90 & 4.6$_{-1.0}^{+1.2}$ & 1.61$_{-0.09}^{+0.11}$ & $<$0.47 & 37.7/49 & 4.5$_{-0.9}^{+1.0}$ & 1.68$_{-0.06}^{+0.06}$ & $<$0.15 & 222.6/248 \\ %  & T \\
NGC 5100 & 1.7 & 90 & 30.4$_{-6.0}^{+7.5}$ & 1.89$_{-0.16}^{+0.18}$ & 0.32$_{-0.30}^{+0.43}$ & 17.4/21 & 22.6$_{-2.9}^{+2.6}$ & 1.68$_{-0.10}^{+0.10}$ & 0.17$_{-0.10}^{+0.12}$ & 191.4/197 \\ % & T \\
Mrk 477 & 1.1 & 90 & 32.8$_{-4.6}^{+5.2}$ & 1.68$_{-0.09}^{+0.10}$ & 0.39$_{-0.14}^{+0.19}$ & 99.6/80 & 22.4$_{-3.9}^{+4.4}$$^*$ & 1.65$_{-0.08}^{+0.08}$ & 0.24$_{-0.08}^{+0.10}$ & 245.9/227 \\ % & T \\
\hline
\hline
\end{tabular}}\caption{\normalsize Best fit properties for the four objects in this work whose best-fit intrinsic absorption is measured to be below the CT threshold both without and with the \nustar\  data. $N_{\rm H, z}$ values flagged with a * indicate sources where fitting the 2--10\,keV data with a $N_{\rm H, z}$ value different from the \nustar\ one implies obtaining a significantly improved best-fit. For these objects, the 2--10\,keV $N_{\rm H, z}$ values are reported in Table \ref{tab:results_no-ct2}.}\label{tab:results_no-ct}
\end{table*}
\endgroup

\begingroup
\renewcommand*{\arraystretch}{1.5}
\begin{table*}
\centering
\scalebox{0.9}{
\begin{tabular}{cccccccccc}
\hline
\hline
Source & $C_{NuS-2-10}$ & norm$_{\rm 1}$ & A$_{\rm S}$ (=A$_{\rm L}$) & $f_{\rm scatt}$ & $N_{\rm H, z, 2-10}$ & f$_{\rm 2-10}$ & L$_{\rm 2-10}$ & f$_{\rm 15-55}$ & L$_{\rm 15-55}$\\
%            &                               & ph cm$^2$ s$^{-1}$ keV$^{-1}$$\times$10$^{-4}$                   &                     &        \%            &  10$^{22}$ cm$^{-2}$\\
\hline
2MASXJ10523297+1036205 & 1.31$_{-0.10}^{+0.11}$ & 6.53$_{-0.98}^{+1.17}$ & 1.00$^f$ & 1.0$_{-0.2}^{+0.3}$ & 17.74$_{-1.31}^{+1.41}$ & --11.75$_{-0.01}^{+0.01}$ & 43.81$_{-0.10}^{+0.08}$ & --11.06$_{-0.03}^{+0.02}$ & 44.11$_{-0.12}^{+0.10}$ \\
B2 1204+34 & 1.81$_{-0.16}^{+0.20}$ & 6.64$_{-1.10}^{+1.34}$ & 1.00$^f$ & 4.4$_{-2.5}^{+2.5}$ & --  & --11.65$_{-0.04}^{+0.03}$ & 43.61$_{-0.22}^{+0.15}$ & --11.10$_{-0.04}^{+0.03}$ & 43.82$_{-0.41}^{+0.21}$ \\
NGC 5100 & 1.38$_{-0.16}^{+0.19}$ & 13.33$_{-4.09}^{+6.09}$ & 1.00$^f$ &  0.6$_{-0.4}^{+0.5}$ & -- & --11.70$_{-0.08}^{+0.04}$ & 43.10$_{-0.45}^{+0.22}$ & --11.00$_{-0.06}^{+0.02}$ & 43.55$_{-0.72}^{+0.41}$ \\
Mrk 477 & 1.54$_{-0.19}^{+0.21}$ & 11.08$_{-2.74}^{+3.77}$ & 1.00$^f$ & 3.4$_{-0.8}^{+1.0}$ & 31.70$_{-4.07}^{+4.49}$ & --11.82$_{-0.04}^{+0.02}$ & 43.21$_{-0.05}^{+0.04}$ & --10.96$_{-0.05}^{+0.02}$ & 43.45$_{-0.37}^{+0.20}$ \\
\hline
\hline
\end{tabular}}\caption{\normalsize  The reported parameters have been obtained fitting all the available data for the given source, including \nustar. $C_{NuS-2-10}$ is the cross-normalization constant between the 2--10\,keV and the \nustar\ data; norm$_{\rm 1}$ is the main power law normalization (in units of ph cm$^2$ s$^{-1}$ keV$^{-1}$$\times$10$^{-4}$), measured at 1\,keV; A$_{\rm S}$ is the intensity of the \texttt{MyTorus} reflected component with respect to the main one; $f_{\rm scatt}$ is the percentage of main power law emission scattered, rather than absorbed, by the obscuring material. $N_{\rm H, z, 2-10}$ is the intrinsic absorption value (in units of 10$^{22}$ cm$^{-2}$)  measured in the 2--10\,keV band for those sources where leaving $N_{\rm H, z, 2-10}$ free to vary with respect to $N_{\rm H, z, NuS}$ lead to a significant improvement of the fit. f$_{\rm 2-10}$ and L$_{\rm 2-10}$ and f$_{\rm 15-55}$ and L$_{\rm 15-55}$ are the logarithms of the observed flux (in units of erg s$^{-1}$ cm$^{-2}$) and the intrinsic, unabsorbed luminosity (in units of erg s$^{-1}$) measured in the 2--10\,keV and in the 15--55\,keV bands, respectively. Parameters fixed to a given value are flagged with $^f$.
}\label{tab:results_no-ct2}
\end{table*}
\endgroup

\bibliographystyle{aa}
\bibliography{nustar_ctagn_paper}

%\begin{thebibliography}{1}
%\bibitem[Aird et al.(2015)]{2015MNRAS.451.1892A} Aird, J., Coil, A.~L., Georgakakis, A., et al.\ 2015, \mnras, 451, 1892 
%\end{thebibliography}

\end{document}